\documentclass[a4paper, onecolumn, noarxiv, accepted=2021-05-19]{quantumarticle}

\usepackage[T1]{fontenc}
\usepackage[utf8]{inputenc}
\usepackage{color}
\usepackage{bm}% bold math
\usepackage{xcolor}
\usepackage{amsmath}
\usepackage{amssymb}
\usepackage{graphicx}
\usepackage[numbers, sort&compress]{natbib}
\usepackage{braket}
\usepackage{psfrag}
\usepackage{tikz}
\usepackage[normalem]{ulem}
\usepackage{qcircuit}
\usepackage{ulem}
\usepackage{multirow}
\usepackage{algorithmic}
\usepackage[section]{algorithm}
\usepackage{hyperref}
\allowdisplaybreaks

\newcommand{\Pmax}{P_{\text{max}}}

\newcommand{\fdelta}{f_{\delta}}

\newcommand{\OUTPUT}{\ENSURE}
\begin{document}

\title{A Threshold for Quantum Advantage in Derivative Pricing}

\author{Shouvanik Chakrabarti}
\affiliation{Goldman, Sachs \& Co., New York, NY}
\affiliation{University of Maryland, College Park, MD}

\author{Rajiv Krishnakumar}
\affiliation{Goldman, Sachs \& Co., New York, NY}

\author{Guglielmo Mazzola}
\affiliation{IBM Quantum, IBM Research -- Zurich}

\author{Nikitas Stamatopoulos}
\affiliation{Goldman, Sachs \& Co., New York, NY}

\author{Stefan Woerner}
\affiliation{IBM Quantum, IBM Research -- Zurich}

\author{William J. Zeng}
\affiliation{Goldman, Sachs \& Co., New York, NY}

%\date{\today}

\begin{abstract}
We give an upper bound on the resources required for valuable quantum advantage in pricing derivatives. To do so, we
give the first complete resource estimates for useful quantum derivative pricing, using autocallable and
Target Accrual Redemption Forward (TARF) derivatives
as benchmark use cases. We uncover blocking challenges in known approaches and introduce a new
method for quantum derivative pricing - the \emph{re-parameterization method} - that avoids them.
This method combines pre-trained variational circuits with
fault-tolerant quantum computing to dramatically reduce resource requirements.
We find that the benchmark use cases we examine require 8k logical qubits and
a T-depth of 54 million.
We estimate that quantum advantage would require executing this program at
the order of a second.
While the resource requirements given here are out of reach of current systems,
we hope they will provide a roadmap for further improvements in
algorithms, implementations, and planned hardware architectures.
\end{abstract}

\maketitle

\section{Introduction}
A derivative contract is a financial asset whose value is based on (or \emph{derived} from) the price of one or more underlying assets. 
Examples of these underlying assets include stocks, currencies, commodities,
etc. 
A derivative contract is typically issued between an issuer and a
holder, and is valid until its \emph{expiration date}.
Each derivative defines a payoff that quantifies what the holder stands to gain.
Generically, payoffs depend on the value of the underlying asset(s) across the
duration of the contract.
Derivative contracts are ubiquitous in finance with various uses from hedging risk to speculation, 
and currently have an estimated global gross market value in the tens of trillions of dollars \cite{prabha_deriving_2013}.
A more detailed introduction to derivatives with descriptions of some of the common derivatives used by financial institutions
is given in Appendix~\ref{app:derivatives}.

The goal of derivative \emph{pricing} is to determine the value of entering a derivative contract today,
given uncertainty about future values of the underlying assets and consequently
the ultimate payoff.
In many cases, the pricing of derivative contracts uses Monte Carlo methods which consume significant computational
resources for financial institutions. Therefore, finding a quantum advantage for this application would be very valuable to the financial sector as a whole.
This work gives the first detailed resource estimates of the required conditions for quantum
advantage in derivative pricing. In doing so, it also introduces new methods for loading
stochastic processes into a quantum computer.

The rest of the paper is structured as follows: \textbf{Section
\ref{sec:derivative-intro}} gives a brief background on classical Monte Carlo
methods for pricing derivatives
and summarizes our results: the resources required for the quantum algorithms
for pricing some commonly traded derivatives.
Then, in \textbf{Section \ref{sec:core-approach}}, we present
our core approach to the quantum algorithm and discuss its error
analysis.
\textbf{Section \ref{sec:loading_dist}} covers subroutines to load a
stochastic model of the underlying assets into a quantum state along with the
resources required for these methods.
Here, we introduce the \emph{re-parametrization method}, a novel method
that plays a crucial role in the first feasible end-to-end path to quantum
advantage in derivative pricing.
In \textbf{Section \ref{sec:payoffs}} we discuss the subroutine
that applies the payoff operator to the stored stochastic process.
Finally, in \textbf{Section \ref{sec:discussion}}, we end with a
discussion on the implications of this work and potential future paths.

\section{Derivative Pricing and Summarized Results}
\label{sec:derivative-intro}

The price of the underlying asset(s) of a derivative contract is typically
modeled as a stochastic process under assumptions
like no-arbitrage.\footnote{No-arbitrage is the assumption that no specific asset is priced differently in
different marketplaces such that one can never buy an asset from one marketplace and immediately
sell it at another for a profit.}
A common model, that we will use in this work, is that the underlying assets evolve under
geometric Brownian motion~\cite{BlackScholes}.
Let $\vec{S}^t \in \mathbb{R}_+^d$ be a vector of values for $d$ underlyings at time $t$. 
Let $(\vec{S}^0, ..., \vec{S}^T) = \bar{\omega} \in \bar{\Omega}$ be a path of a discrete time multivariate stochastic process describing the values of those assets.
We use both notations for a path in the text.
The corresponding probability density function is denoted by $\bar{p}(\bar \omega)$.
Let $f(\bar \omega) = f\left(\vec{S}^0, ..., \vec{S}^T\right) \in \mathbb{R}$
be the discounted payoff of some derivative on those assets. 
To price the derivative we calculate
\begin{align}
\mathbb{E}(f) = \int_{\bar{\omega}\in\bar{\Omega}}\bar{p}(\bar{\omega})f(\bar{\omega})d\bar{\omega}.
\end{align}
The reason for having a discounted payoff is to take into account the opportunity cost of investing in a risk-free asset.
For the rest of the paper
all payoffs will be implicitly regarded as discounted, except in \textbf{Section \ref{sec:payoffs}} and Appendix \ref{app:derivatives} where we will be explicit about whether payoffs are discounted or not.
More details on the concept of discounted payoffs can be found in Appendix \ref{app:path-dependent}. 

If the underlying stochastic processes are modeled with geometric Brownian motion then
 they have transition probabilities
\begin{equation}
\label{eqn:multivariate_lognormal_pdf}
P(\vec{S}^t \mid \vec{S}^{t-1}) =
\frac{\text{exp}\left(-\frac{1}{2}
(\ln \vec{S}^t-\vec{\mu}^{t-1})^\intercal
\Sigma^{-1}
(\ln \vec{S}^t -\vec{\mu}^{t-1})\right)}
{(2\pi)^{d/2}(\text{det}\Sigma)^{1/2}\prod_{j=1}^d S_j^t}.
\end{equation}
where
\begin{eqnarray}
\ln \vec{S}^t &=& (\ln S_1^t, \ln S_2^t, \dotsc, \ln S_d^t)^\intercal \nonumber \\
\vec{\mu}^{t-1} &=& (\mu_1^{t-1}, \mu_2^{t-1}, \dotsc, \mu_d^{t-1})^\intercal \nonumber \\
\label{eqn:multivariate_lognormal_parameters}
\mu_j^{t-1} &=& \left(r-0.5\sigma_j^2\right)\Delta t + \ln S_j^{t-1}.
\end{eqnarray}
Note that Eq.~\eqref{eqn:multivariate_lognormal_pdf} at time $t$
has a dependency on the asset vector at time $t-1$ via $\ln S_j^{t-1}$ in $\mu_j^{t-1}$.
The parameters $r$ and $\sigma_j$ are the risk-free rate\footnote{\label{note:risk_free}
The \emph{risk-free} rate is the rate of return of investing in a risk-free asset.
Although such an asset is purely theoretical, we typically use treasury bonds to represent such an asset and
approximate $r$ as the yield of the treasury bond minus the current inflation rate.} and the volatility of
the $j$-th asset respectively,
$\Delta t$ is the time duration between steps of the stochastic process,
and $\Sigma$ is the $d\times d$ positive-definite covariance matrix of the $d$
underlyings
\begin{equation}
\label{eqn:covariance_matrix}
\Sigma = \Delta t \begin{bmatrix}
    \sigma_1^2 & \rho_{12} \sigma_1\sigma_2 & \dots  & \rho_{1d}\sigma_1\sigma_d \\
    \rho_{21} \sigma_2\sigma_1 & \sigma_2^2 & \dots  & \rho_{2d}\sigma_2\sigma_d\\
    \vdots & \vdots & \ddots & \vdots \\
    \rho_{d1} \sigma_d\sigma_1 & \dots & \dots  & \sigma_d^2
    \end{bmatrix}
\end{equation}
where $-1\leq \rho_{ij} \leq 1$ is the correlation between assets $i$ and $j$.
The probability of any particular path $\bar{\omega}\in\bar{\Omega}$ is then
\begin{equation}
\bar{p}(\bar \omega) = \prod_{t=1}^T P(\vec{S}^t \mid \vec{S}^{t-1}).
\end{equation}

Classically, some simple derivatives under this model are easy to price, such
as European call
options that can be priced analytically using the Black-Scholes
equation~\cite{BlackScholes}. Easy
to price derivatives are often \emph{path independent}, i.e. where the payoff
is only a function of the final
prices at exercise time $f(\bar \omega) = f(\vec{S}^T)$. This contrasts
with~\emph{path dependent} derivatives
that are more difficult to price. Path dependent derivatives are often
priced in practice with classical
Monte Carlo methods. More examples of standard and more complex derivatives 
are given in Appendix~\ref{app:derivatives}.

When using classical Monte Carlo, the accuracy of derivative pricing converges as $O
(1/\sqrt{M})$,
where $M$ is the number of paths that are sampled.
In general cases, Montanaro~\cite{montanaro2015quantum} showed that quantum algorithms
based on amplitude estimation~\cite{brassard2002quantum} can be used to improve
 this to $O(1/M)$.
Recent work has considered how to specialize this advantage to options
pricing (options are a subcategory of
derivatives)~\cite{rebentrost2018quantum,Stamatopoulos_2020,vazquez2020}
and risk analysis~\cite{egger2019credit,Woerner_2019}.

As this is only a quadratic speedup, it is important to focus on derivatives that are
complicated enough to require a large $M$ in practice. In this work we give end
to end quantum resource
estimates for two examples of such derivatives (autocallable options and TARFs) that are
both computationally expensive, path-dependent derivatives. In doing so, we detail and
optimize the loading into quantum states of the underlying distribution over asset paths.
This loading step was left open in
previous work~\cite{rebentrost2018quantum,Stamatopoulos_2020},
and we give the first account of the resources required for it.
Although autocallables and TARFs are usually not well known derivatives to those outside the financial sector,
they are very commonly traded derivatives,  particularly among financial institutions.
It is for this reason, in addition to their complexity, that we have chosen them as suitable examples for the end-to-end
quantum resource estimates. 
A more detailed description of autocallables and TARFs can be seen in Appendix \ref{app:autocallable} and Appendix \ref{app:tarf} respectively.

In addition to estimating the resources needed for path loading using known
methods (an extension of~\cite{vazquez2020} that we call the Riemann Sum method),
we introduce several optimizations, including intentional shifts from price space to return
space calculations and the new re-parameterization method. These methods reduce the required
resources significantly and are summarized in Table~\ref{tbl:resources}.
In this table, we quantify resource requirements in the fault tolerant setting
 where the
number of $T$-gates, called the \emph{T-count} dominates the computational
requirements. The \emph{T-depth} is the sequential depth that dominates the
runtime. \emph{Logical qubits} consist of physical qubits in a quantum
error correcting code of sufficient distance to support the required number of operations.

\begin{table}[h!]
  \resizebox{\textwidth}{!}{

  \begin{tabular}{r||c|c|cc|c|c|c|c|c|c}
       & \multicolumn{2}{c|}{$(d, T)$}  & \multicolumn{2}{c|}{Error}  & \multicolumn{2}{c|}{T-count} & \multicolumn{2}{c|}{T-depth} &\multicolumn{2}{c}{\# Logical Qubits}\\
       \hline
    Method & \textbf{Auto} & \textbf{TARF} & \multicolumn{1}{c|}{\textbf{Auto}} & \textbf{TARF} & \textbf{Auto} & \textbf{TARF} & \textbf{Auto}  & \textbf{TARF} & \textbf{Auto} & \textbf{TARF} \\
\hline

\multirow{1}{*}{Riemann Sum}
	&\multirow{3}{*}{$(3, 20)$} &\multirow{3}{*}{$(1, 26)$} &\multicolumn{2}{c|}{\multirow{3}{*}{$2 \times 10^{-3}$}} & $\ge10^{43}$ & $\ge10^{18}$ & $\ge10^{43}$ & $\ge10^{18}$ & - & -\\

\multirow{1}{*}{Riemann Sum (no-norm)}
	& & & & & $1.6 \times 10^{11}$ & $5.5 \times 10^{10}$ & $1.5 \times 10^{8}$ & $1.6 \times 10^{8}$ & 23k & 17k\\

\multirow{1}{*}{Re-parameterization}
	& & & & & $1.2 \times 10^{10}$ & $9.8 \times 10^{9}$ & $5.4 \times 10^7$ & $8.2 \times 10^7$ & 8k & 11.5k\\

  \end{tabular}
  }
  \caption{
Resources estimated in this work for pricing derivatives using
different methods for a target error of $2 \times 10^{-3}$.
As representative use cases of business interest with non-trivial complexity, we consider a basket
autocallable (Auto) with $3$ underlyings, $5$ payment dates and a knock-in put option with $20$ barrier dates, and a TARF with one underlying and $26$ payment dates.
Detailed definitions of these contracts and their parameters can be found in Appendix \ref{app:autocallable}.
We find that Grover-Rudolph methods~\cite{grover2002creating} are not applicable in
practice (details in Appendix~\ref{app:grover-rudolph}) and that
Riemann summation methods require normalization assumptions to avoid errors
that grow
exponentially in $T$. Even if those normalization issues were avoided, as
detailed in the Riemann Sum (no-norm) row, the re-parameterization method
still performs best. See Section~\ref{sec:riemann-sum} for a discussion of the
Riemann summation normalization.
The detailed resource estimation is discussed in
Sections~\ref{sec:resource_estimates_riemann} and \ref{sec:resource_estimates_reparam}.}
\label{tbl:resources}
\end{table}

\subsection{Discretized Derivative Pricing}
In order to map our derivative pricing problem into quantum states, we must discretize the values $\vec{S}^t$.
Classically, this is not that important as high precision is available, but, in
order to study the minimal qubit
requirements, we need to consider discretization explicitly in the quantum case.

Let each value $\vec{S}^t$ be discretized into a different $n$-qubit register, i.e., mapped to a regular grid.
We then define $\omega \in \Omega$ as the discrete space of paths.
The price expectation is now a sum
\begin{equation}
\mathbb{E}(f) = \sum_{\omega\in\Omega}p(\omega)f(\omega),
\end{equation}
where the probability $p(\omega)$ can be defined in multiple ways. For instance, one can take the midpoints of the grid cells so that
\begin{equation}
p(\omega) = \prod_{t=1}^TP(\vec{S^t}\,|\,\vec{S}^{t-1}),
\end{equation}
where the $\vec{S}^t$ are restricted to discrete midpoints.
Or $p(\omega)$ can be defined as an integral over the discrete cells.
These representations are the same in the limit of fine grids and in the following we will choose the midpoint method.

\subsection{Price Space vs. Return Space}
\label{sec:spaces}
In Section~\ref{sec:derivative-intro},
Eq.~\eqref{eqn:multivariate_lognormal_pdf} introduces geometric Brownian motion
to model the price on underlying assets. We call this the~\emph{price
space} description of the underlying stochastic process. In price space,
transition probabilities are given by a multivariate log-normal distribution.

An alternative, but equivalent representation, is to consider the stochastic
process on the log-returns of the underlying assets, and perform all calculations in \emph{return space}.
When asset prices obey a log-normal distribution, then the log-returns are distributed normally.
We define a vector of underlying log-returns for $d$ assets at time $t$ as $\vec{R}^t =
(R_1^t,R^t_2,...,R^t_d)$.  At any time $t'$ we can calculate
the price of asset $j$ from return space using
\begin{align}
\label{eqn:return-to-price}
S_j^{t'} = S_j^{0} \prod_{t=1}^{t'} e^{R_j^{t}}.
\end{align}
The transition probabilities are then given by a
multivariate normal distribution
\begin{align}
\label{eqn:multivariate_normal_pdf}
P(\vec{R}^t) = \frac{\text{exp}\left(-\frac{1}{2}(R^t -\vec{\mu})^{\intercal} \Sigma^{-1}(R^t -\vec{\mu})
\right)}{(2\pi)^{d/2}(\text{det}\Sigma)^{1/2}},
\end{align}
where,
\begin{align}
\mu &= (\mu_1, \mu_2, ... \mu_d)^{\intercal}, \\
\label{eq:brownian}
\mu_j &= \left(r-0.5\sigma_j^2\right)\Delta t,
\end{align}
and $\sigma$, $\Delta t$, $\Sigma$ and $r$ are the same Brownian motion
parameters as in price space. Note that this
is no longer conditioned on the value at the previous time step. In fact, the
path distribution in return space
consists of $dT$ independent Gaussians.

Note that we have overloaded the notation from the price space formulation as
these representations are interchangeable via Eq.~\eqref{eqn:return-to-price}.
This calculation is needed when the stochastic process has been modeled in
return space but the payoff is defined in terms of asset prices. In the
following sections, it will be made clear from the context which space we are operating in.

Switching between price space and return space changes from
log-normal distribution loading to normal distribution loading. In general,
 the loading of normals is easier since their underlying stochastic evolution
 is independent of the price at a previous time step which can be seen by
 comparing Eq.~\eqref{eqn:multivariate_lognormal_parameters}
 and~\eqref{eq:brownian}.
As such, the probability distribution $P(\vec{R}^1, \vec{R}^2, \dots, \vec{R}^T)$  across all $T$ timesteps of the stochastic process can be computed simultaneously with

\begin{equation}
\label{eqn:multivariate_normal_pdf_T}
P(\vec{R}) \equiv P(\vec{R}^1, \vec{R}^2, \dots, \vec{R}^T) = \frac{\text{exp}\left(\sum_{t=1}^T-\frac{1}{2}(\vec{R}^t -\vec{\mu})^{\intercal} \Sigma^{-1}(\vec{R}^t -\vec{\mu}) \right)}{(2\pi)^{dT/2}(\text{det}\Sigma)^{T/2}}.
\end{equation}
This advantage can compensate for the quantum
 arithmetic needed to evaluate the exponentials in
 Eq.~\eqref{eqn:return-to-price}. We will leverage this advantage
 with the re-parameterization method in Section~\ref{sec:re-parameterization}.
 Additionally,
 when
 working with derivatives
 that have payoffs defined in terms of log-returns directly and are independent
 of individual asset prices, this is another reason to work in return space.

\section{Core Approach}
\label{sec:core-approach}
Our approach to derivative pricing extends the quantum mean
estimation method from~\cite{montanaro2015quantum}. In this section we
review this approach and introduce an error analysis for the discrete case
of option pricing.

Let the normalized discounted
payoff of any path $\omega$ be given by
\begin{equation}
\tilde{f}(\omega) = \frac{f(\omega) - f_{\text{min}}}{f_{\text{max}} - f_{\text{min}}} \in [0, 1].
\label{eq:norm_disc_payoff}
\end{equation}
where $f_{\text{max}}$ and $f_{\text{min}}$ are the 
maximum and minimum possible payoffs respectively across all paths. 
The algorithm proceeds in four key phases.
First, a probability distribution is loaded in form of a superposition over all possible paths. 
Second, payoffs for all possible paths are calculated in quantum parallel.
Third, the expected payoff is stored in the amplitude of a marked state. 
Fourth, amplitude estimation is used to read out the amplitude using $\mathcal{O}(1/\epsilon)$ queries for a
given target accuracy $\epsilon >0$. This approach is detailed in Algorithm~\ref{alg:core}.

\begin{figure}
\begin{algorithm} [H]
\caption{Core approach to derivative pricing}
\begin{algorithmic}[1]
\REQUIRE Parameters $n$, $d$, and $T$ that are all positive integers.
\REQUIRE An operator $\mathcal{P}$ for loading a probabilistically
weighted superposition of paths onto a register of $ndT$-qubits.
\begin{enumerate}
\item Apply operator $\mathcal{P}$ to prepare the quantum state
\begin{align}
\mathcal{P}|0\rangle = \sum_\omega\sqrt{p(\omega)}\ket{\omega}.
\end{align}
\item Calculate $\delta(\omega)=\arcsin\sqrt{\tilde{f}(\omega)}$ into a quantum register
\begin{align}
\label{eqn:payoff_amplitude_encoding}
\sum_{\omega}\sqrt{p(\omega)}\ket{\omega}\ket{\delta(\omega)}.
\end{align}
\item Introduce an ancilla qubit and rotate the value of the $\tilde f(\omega)$ register into its amplitude:
\begin{align}
\label{eqn:payoff_ancilla_rotation}
\sum_{\omega}\sqrt{p(\omega)(1-\tilde{f}(\omega))}\ket{\omega}\ket{0}
    +\sum_{\omega}\sqrt{p(\omega)\tilde{f}(\omega)}\ket{\omega}\ket{1}.
\end{align}
\item Use amplitude estimation to extract the probability of the ancilla being $\ket{1}$.
\end{enumerate}
\OUTPUT The
(discretized) expected payoff $\mathbb{E}(\tilde{f})=\sum_{\omega}p(\omega)\tilde{f}(\omega)$. We rescale this to obtain
$\mathbb{E}(f) = (f_{\text{max}}-f_{\text{min}})\mathbb{E}(\tilde{f}) + f_{\text{min}}$.
\end{algorithmic}
\label{alg:core}
\end{algorithm}
\end{figure}

Note that Steps 1-3 in the Algorithm~\ref{alg:core} load the exact answer after a single
execution. Were it possible to read out an amplitude directly, then we could
compute the expectation over all paths in a constant number of queries. This
is, unfortunately, not possible, and so amplitude estimation introduces a
linear overhead to extract an answer to a given precision. This is a key
conceptual difference from classical Monte Carlo where samples from
paths are taken. In Algorithm~\ref{alg:core}, we compute all possible paths and
take (amplitude estimated) samples of the expected payoff.

Another important distinguishing feature of the quantum approach is that we
must normalize the payoff in order to store it in the amplitude of a state.
This normalization must be rescaled at the end and can have a critical impact
on error scaling, as errors are also scaled up. In the Riemann summation
method, discussed in Section \ref{sec:riemann-sum}, a version of this normalization rescaling can
rapidly accumulate errors.

\subsection{Amplitude Estimation for Derivative Pricing}
\label{subsec:amplitude_estimation}
Typically, path-dependent derivatives like autocallables and TARFs are priced
using Monte Carlo or quasi-Monte Carlo methods. Paths
$\omega = (\vec{S}^0, \vec{S}^1, ..., \vec{S}^T)$ are generated by modeling
the underlying stochastic process and then the expected payoff is calculated
using the estimator
\begin{align}
\mathbb{E}(f) \approx \frac{1}{M}\sum_{\omega=1}^M f(\omega).
\end{align}
This estimator converges to the true expected value with error $\epsilon = O(M^{-1/2})$
by the Central Limit Theorem~\cite{rubinstein2016simulation}.

This convergence can be quadratically accelerated to $\epsilon = O(M^{-1})$ using
quantum amplitude estimation~\cite{brassard2002quantum} for Monte
Carlo~\cite{montanaro2015quantum,rebentrost2018quantum,Stamatopoulos_2020}.
Amplitude estimation takes as input a unitary operator $\mathcal{A}$ on $n+1$
qubits such that
\begin{align}
\label{eq:ae-form}
\mathcal{A}|0\rangle_{n+1} = \sqrt{1-a}|\psi_0\rangle_n|0\rangle +
\sqrt{a}|\psi_1\rangle_n|1\rangle,
\end{align}
where the parameter $a$ is unknown. Here, the final qubit acts as a label to
distinguish $|\psi_0\rangle$ states from $|\psi_1\rangle$ states.

Amplitude estimation determines $a$ by repeated applications of the
operator\footnote{This is often called the Grover operator.}
$\mathcal{Q} = \mathcal{A}S_0\mathcal{A}^{\dagger}S _{\psi_0}$, where
$S_0 = \mathbb{I} - 2 |0\rangle_{n+1}\langle 0|_{n+1}$ and
$S_{\psi_0} = \mathbb{I} - 2|\psi_0\rangle_n|0\rangle\langle 0|\langle\psi_0|_n$
are reflection operators. By using phase estimation and the quantum Fourier
transform $a$ can be determined with accuracy
$O(M^{-1})$~\cite{brassard2002quantum}.
Unfortunately, the required depth of the resulting quantum circuits scales as
$O(1/\epsilon)$ and requires the use of a resource
expensive quantum Fourier transform. Recent developments have introduced
other approaches
\cite{suzuki2020amplitude,aaronson2020quantum,grinko2019iterative,nakaji2020faster,tanaka2020amplitude, giurgicatiron2020low} that aim to reduce the
resource requirements needed for amplitude estimation and can remove quantum phase estimation.

The most efficient variant of amplitude estimation known to date is
Iterative Quantum Amplitude Estimation (IQAE) introduced
in \cite{grinko2019iterative}.
It has been shown empirically that IQAE outperforms the other known variants.
Although it omits quantum phase estimation~\cite{mikeandike}, it achieves a
four times better performance than the canonical phase estimation approach.
Further, it has been shown that for practical considerations, the following bound holds:
\begin{equation}
\label{eqn:AE_oracle_calls}
N_{\text{oracle}}^{\text{wc}}
\leq \frac{1.4}{\epsilon}\log\left(\frac{2}{\alpha} \log_2\left(\frac{\pi}{4\epsilon}\right)\right),
\end{equation}
where $N_{\text{oracle}}^{\text{wc}}$ denotes the worst-case number of oracle calls,
i.e., applications of $\mathcal{Q}$, to achieve an estimation error of $\epsilon > 0$ with
confidence level $1 - \alpha$, $\alpha \in (0, 1)$.

We use the performance estimates from Eq.~\eqref{eqn:AE_oracle_calls} for IQAE for our resource estimates in this work.
This approach gives a full quadratic speedup, however it requires a quantum processor to successfully execute programs of oracle depth $\mathcal{O}(1/\epsilon)$.
This large depth is a demanding requirement on QPU performance and is a dominant contributor to the required resources.
Recent work~\cite{giurgicatiron2020low,bouland2020prospects} has shown that it is possible to use shorter depth quantum programs $\mathcal{O}(1/\epsilon^{1-\beta})$ in exchange for less quantum advantage in total oracle calls $N_{\text{oracle}} = \mathcal{O}(1/\epsilon^{1+\beta})$ for $\beta \in (0, 1]$.
Using shorter depth quantum programs means more tolerance to error and as such may result in less needed overhead for error correction.
While there may be settings where this tradeoff is advantageous overall, we leave this analysis to future work.

\subsection{Path Distribution Loading}
\label{sec:path-dist-loading}
In order for Algorithm~\ref{alg:core} to achieve a practical quantum advantage, the
resources needed for path loading and payoff calculation need to be taken into
account. In some cases, there is an analytic form that can simplify path loading.
For example in the case of path-independent derivatives, a distribution over
paths is not needed. All that is needed is a distribution over final
underlying prices $\vec{S}^T$, such as the log-normal distribution given by the
Black-Scholes model. This means that the distribution could be analytically
computed and then loaded either variationally or explicitly into quantum
states. Unfortunately for quantum advantage, the analytic form for this
distribution means that these derivatives are typically easy to
price classically.
Thus it is critical to focus on path dependent derivatives where a
superposition over paths needs to be computed.

While the loading of general distributions is exponentially
hard~\cite{plesch2011quantum}, several methods have been proposed.
If the distribution is efficiently integrable, then there does exist an
efficient quantum algorithm for loading, the Grover-Rudolph method~\cite{grover2002creating}.
However, the algorithm has limited applicability in practice for derivative pricing,
because the relevant probability distributions still require Monte Carlo integration (albeit quantumly)
which is precisely what we are trying to avoid by using Amplitude Estimation.
More details on the insufficiency of this method are detailed in
Appendix~\ref{app:grover-rudolph}.

An alternative method for loading the path distribution, using a qGAN~\cite{zoufal2019quantum}, was
proposed in~\cite{Stamatopoulos_2020}. While this has appeal for lower
overhead loading, it is not yet clear how to anticipate what the overhead
from training a given qGAN will be in practice.

\subsection{Error Analysis}
\label{sec:general-error}
In this section, we investigate the various elements that contribute to the
overall error in the quantum approach to derivative pricing. There are
three main components that introduce error in Algorithm~\ref{alg:core}.
Let $f_{\delta} = f^{\text{disc}}_{\text{max}} - f^{\text{disc}}_{\text{min}}$.
\begin{description}
  \item[Truncation Error] The price of a derivative is determined by
  an integral over all the possible values of the underlying price
  (or return). Given finite quantum memory, we cannot feasibly compute an
  integral over an infinite
  domain, and thus we restrict the domain of integration as follows: the
  prices/log-returns are restricted to a range $[B_l,B_u]$. This restriction
  of the domain leaves out a probability mass of $\alpha$. Given an upper
  bound of $P_{\text{max}}$ on the density functions at each step and an upper
  bound $\fdelta$ on the discounted payoff, we incur a truncation error which we denote by
  $\epsilon_{\mathrm{trunc}}=P_{\text{max}}^{T}\fdelta\alpha$.
  \item[Discretization Error] This error (denoted by
  $\epsilon_{\mathrm{disc}}$) arises from the use of a Riemann Sum over a
  finite grid of points to approximate the integral. This error can be
  reduced by increasing the number of qubits $(n)$ to approximate the sum.
  \item[Amplitude Estimation Error] Amplitude estimation incurs an
  error
  of $\epsilon_{\mathrm{amp}}$ when using $1/\epsilon_{\mathrm{amp}}$
  repetitions of the state preparation procedure and price computation.
\end{description}
We now analyze the truncation and discretization errors in more detail.

\subsubsection{Truncation Error}
\label{sec:truncation_error}
We present the truncation error in return space as it then extends to
price space straightforwardly. Denote the maximum
eigenvalue of the covariance matrix
$\Sigma$ by $\sigma_{\text{max}}$.
Using Chernoff tail bounds on Gaussians,
the probability that the log-returns for asset $i$ lie outside of the
interval $[\mu_{i}-w \sigma_{\text{max}},\mu_{i} + w \sigma_{\text{max}}]$ is upper
bounded by $2e^{-w^{2}/2}$. By the union bound the probability that
any log-return ($d$ assets over $T$ time steps) lies outside the interval
$[B_l = (r - 0.5\sigma_{\text{max}}^{2})\Delta t - w \sigma_{\text{max}},
B_u = (r - 0.5\sigma_{\text{max}}^{2})\Delta t + w \sigma_{\text{max}}]$
is upper bounded
by $2dTe^{-w^{2}/2}$.
Let the initial asset prices lie in the range
$[S^{0}_{\text{min}}, S^{0}_{\text{max}}]$.
Then the corresponding interval in price space is given by
$[S^{0}_{\text{min}}e^{B_lT},S^{0}_{\text{max}}e^{B_uT}]$.

We can then define the truncated
window of values for our $dT$ different $n$-qubit registers that are $w$
standard deviations around the mean for each time step.
The truncation error of the integral already normalized by $P_{\text{max}}^T\fdelta$ is then given by
\begin{equation}
  \label{eqn:trunc_error}
\epsilon_{trunc} \le 2dTe^{-w^2 / 2}.
\end{equation}

\subsubsection{Discretization Error} 
The final output of the amplitude estimation algorithm represents a
Riemann Sum that approximates the truncated multidimensional integral. The
integral is over
$dT$ variables corresponding to $d$ assets over $T$ time steps. We assume
that each underlying asset/return is restricted to an interval $[B_l,B_u]$.
To compute the discretization error, we apply a multidimensional variation
of the midpoint rule as follows: let there be $n$ qubits used to represent
each underlying asset, the domain is divided into $2^{ndT}$ cells, and
corresponding to each value of the register we associate the value of the
integrand at the midpoint of the corresponding cell. Assume that $\beta$
provides an upper bound on the second derivatives of the integrand (this can
be restated as saying that the deviation from linearity over a range of
length $l$ is bounded by $\beta l^{2}/2$).

We consider the error from discretization that accumulates over a single cell.
Each cell has side length $(B_u - B_l)/2^{n}$ and is a hypercube of dimension
$l$.
Note by symmetry that the linear component of the deviation from the value at
the center of the cell integrates to $0$ over the cell.
The error in each cell
can thus be bounded by the integral of the term $\beta x^{2}/2$ over a
$dT$-hypercube of side length $l = (B_u-B_l)/2^{n}$ centered at the origin.
\begin{align}
  \label{eq:deviation-per-cell}
  \underbrace{\int_{l/2}^{l/2}\dots\int_{l/2}^{l/2}}_{dT}
  \beta
  x^{2}/2
  = l^{dT-1}\beta \left[\beta x^{3}/6\right]^{l/2}_{l/2} =
  \frac{\beta l^{dT + 2}}{24} = \frac{\beta (B_u-B_l)^{dT + 2}}{24 \cdot 2^{n
  (dT + 2)}}.
\end{align}
Aggregating the error over all the cells, we have
\begin{align}
  \label{eqn:disc-error}
  \epsilon_{\mathrm{disc}} = \frac{\beta (B_u-B_l)^{dT + 2}}{24\cdot 2^{2n}}.
\end{align}
In terms of the number of standard deviations used in the discretization and the largest eigenvalue of the covariance matrix $\sigma_{\text{max}}$, the total discretization error is bounded by
\begin{equation}
  \label{eqn:disc_error}
\epsilon_{\text{disc}} \le \frac{\beta(2w\sigma_{\text{max}})
^{dT+2}}{24\cdot 2^{2n}}.
\end{equation}
For a target discretization error, Eq.~\eqref{eqn:disc_error} also gives us the
 total number of qubits required to represent $d$ assets for $T$ timesteps, given by
\begin{align}
  ndT = dT\lceil \frac{1}{2}\left(\log_2(\beta/24) -
  \log_2(\epsilon_{\mathrm{disc}})+
  (dT+2)\log_2(2w\sigma_{\text{max}})\right)\rceil.
\end{align}

The truncation and discretization errors apply in general to the methods we
introduce, though each method has additional method-specific error sources
which are discussed separately.

\section{Methods for Advantage in Quantum Derivative Pricing}
\label{sec:loading_dist}
In the following sections we introduce two approaches that can work effectively
 for quantum derivative pricing in practice: Riemann summation and
re-parameterization. Riemann summation was introduced in~\cite{vazquez2020},
and we present
the first resource analysis for its application for quantum advantage. This
analysis uncovers limitations in error scaling due to normalization. We then
introduce a new method called~\emph{re-parameterization} that avoids the
downsides of other methods and offers the first end-to-end path to quantum
advantage in practice.

\subsection{Riemann Summation}
\label{sec:riemann-sum}

The Riemann summation method~\cite{vazquez2020} gives an approach to construct
the  $\mathcal{P}$ path loading operator in Algorithm~\ref{alg:core}. Let $N=2^{ndT}$
be the size of the Hilbert space that contains all
possible paths. Let $\tilde{P}_{\text{max}}$ be the
maximum value of the $d$-asset multivariate transition probabilities from Eq.~\eqref{eqn:multivariate_lognormal_pdf}.
Then $\tilde{P}(\vec{S^t}\,|\,\vec{S}^{t-1}) = P(\vec{S^t}\,|\,\vec{S}^{t-1}) /
\tilde{P}_{\text{max}} \in [0, 1]$ are the normalized
transition probabilities over all choices of $\vec{S}^{t}$ and $\vec{S}^{t-1}$.
Let the asset price for each asset at each timestep be discretized in the interval $[0, S_{\text{max}}]$.
The steps of the method summarized in Algorithm~\ref{alg:riemann-sum} calculate the price of the derivative with a normalization factor $1/P_{\text{max}}^T$, with $P_{\text{max}} = \tilde{P}_{\text{max}}S_{\text{max}}^d$.
Critically, we note that the normalization factor in the final step scales exponentially in $T$.
If $P_{\text{max}} < 1$ no normalization is needed,  but this factor can be used to improve the performance.
However, if $P_{\text{max}} > 1$, the error increases exponentially, which renders this approach impractical.

\begin{figure}
\begin{algorithm} [H]
\caption{Riemann summation pricing}
\begin{algorithmic}[1]
\REQUIRE Parameters $n$, $d$, and $T$ that are all positive integers.
\REQUIRE Access to operators $\mathcal{W}_{t}, t=1, \ldots, T$ that apply the transition probabilities of the stochastic process
into an ancilla via
\begin{align}
\mathcal{W}_t\ket{\vec{S}^t}_n\ket{\vec{S}^{t-1}}_n\ket{0} \mapsto
\ket{\vec{S}^t}_n\ket{\vec{S}^{t-1}}_n\left[
\sqrt{1-\tilde{P}(\vec{S}^t|\vec{S}^{t-1}}\ket{0}
+ \sqrt{\tilde{P}(\vec{S}^t|\vec{S}^{t-1}}\ket{1}
\right]
\end{align}
\begin{enumerate}
\item Apply Hadamards to $ndT$ qubits to prepare an equal superposition of all paths.
\item Load the initial prices $\vec{S}^0$ into the zero-th $nd$-qubit register.
\item Apply each of the $T$ transition operators $\mathcal{W}_t$ to construct
\begin{multline}
    \label{eqn:riemann_transition_operator}
\frac{1}{\sqrt{N}}\sum_{\omega}|\vec{S}^0 \ldots \vec{S}^T\rangle
\left[\ldots + \sqrt{\prod_{t=1}^T
\tilde{P}(\vec{S}^t \mid \vec{S}^{t-1})}|1 \ldots 1\rangle_{T}\right] \\
=\frac{1}{\sqrt{P_{\text{max}}^T}}\frac{1}{\sqrt{N}}\sum_{\omega}|\vec{S}^0 \ldots \vec{S}^T\rangle
\left[\ldots + \sqrt{p(\omega)}|1 \ldots 1\rangle_{T}\right],
\end{multline}
where $N=2^{ndT}$.
\item Calculate $\delta(\omega)=\arcsin\sqrt{\tilde{f}(\omega)}$ into a quantum
 register, obtaining
\begin{align}
\label{eqn:riemann_payoff_computation}
\frac{1}{\sqrt{P_{\text{max}}^T}}\frac{1}{\sqrt{N}}\sum_{\omega}|\vec{S}^0...\vec{S}^T\rangle
\left[...+\sqrt{p(\omega)}|1...1\rangle_{T}\right]\ket{\delta(\omega)}.
\end{align}
\item Introduce an ancilla qubit and rotate the value of the $\tilde f(\omega)$ register into its amplitude:
\begin{align}
\label{eqn:riemann_payoff_rotation}
%\sum_{\omega}\sqrt{p(\omega)(1-\tilde{f}(\omega))}\ket{\omega}\ket{0}
\ldots + \frac{1}{\sqrt{P_{\text{max}}^T}} \frac{1}{\sqrt{N}}\sum_{\omega}\sqrt{p(\omega)\tilde{f}(\omega)}\ket{\omega}\ket{1 \ldots 1}_T\ket{1}.
\end{align}
\item Use amplitude estimation to extract the probability of the ancilla being $\ket{1}$.
\end{enumerate}
\OUTPUT The
(discretized) expected payoff $\mathbb{E}(\tilde{f}(\omega)/P_{\text{max}}^{T}) = 1/(P_{\text{max}}^T N)\sum_{\omega}{p(\omega)\tilde{f}(\omega)}$.
We rescale this to obtain
$\mathbb{E}(f) = P_{\text{max}}^{T}\left(\fdelta\mathbb{E}(\tilde{f}) + f_{\text{min}}\right)$.
\end{algorithmic}
\label{alg:riemann-sum}
\end{algorithm}
\end{figure}
The normalization factor $P_{\text{max}}$ is easier to handle in return space where the probability density function is given by Eq.~\eqref{eqn:multivariate_normal_pdf}.
If we discretize the log-returns at each timestep for each asset to $\pm w$ times the asset's volatility $\sigma_j$, we have

\begin{equation}
    P_{\text{max}}=\frac{(2w)^d\prod_{j=1}^{d}\sigma_j}{(2\pi)^{d/2}(\text{det}\Sigma)^{1/2}}.
\end{equation}
When the $d$ assets are uncorrelated, we have
\begin{equation}
\label{eqn:P_max}
    P_{\text{max}}=\left(\frac{2w}{\sqrt{2\pi}}\right)^d,
\end{equation}
and therefore we need to choose $w \le \pi/\sqrt{2}$ for $P_{\text{max}} \le 1$.
However, choosing a small discretization window $w$ increases the truncation error discussed in Section~\ref{sec:truncation_error},
and for $w \le \pi/\sqrt{2}$ we have $\epsilon_{\text{trunc}} \ge 2e^{-\pi^2/4} \sim 0.17$, which increases proportionally to the number of assets and timesteps in the computation.

\subsubsection{Riemann Summation Error Analysis}
In addition to the truncation and discretization errors from
Section~\ref{sec:general-error}, the Riemann summation approach includes errors
due to scaling considerations and quantum arithmetic.

When working in return space, we only need one transition operator which computes 
Eq.~\eqref{eqn:multivariate_normal_pdf_T} and performs the amplitude encoding 
of $\sqrt{p(\omega)}$ in Eq.~\eqref{eqn:riemann_transition_operator}.
Assuming the transition operator introduces a maximum additive error $\epsilon_{\text{dens}}$ and the payoff operator computing Eq.~\eqref{eqn:riemann_payoff_computation} and Eq.~\eqref{eqn:riemann_payoff_rotation} introduces payoff error $\epsilon_f$, the total arithmetic error of the quantity we estimate using amplitude estimation is

\begin{equation}
   \epsilon_{\mathrm{arith}} = \frac{1}{N}\sum_{\omega}[(p(\omega) + \epsilon_{\text{dens}})(f(\omega) + \epsilon_f) - p(\omega)f(\omega)].
\end{equation}
Ignoring quadratic error terms, we have

\begin{equation}
   \epsilon_{\mathrm{arith}} \approx \frac{1}{N}\sum_{\omega}p(\omega)\epsilon_f + \frac{1}{N}\sum_{\omega}f(\omega)\epsilon_{\text{dens}} \le \frac{\epsilon_f}{(2w\sigma_{\text{max}})^{dT}} + \epsilon_{\text{dens}},
\end{equation}
where we assume the payoff has been normalized to lie in $[0, 1]$ and the log-returns for each asset and each timestep have been constructed to discretize the domain $[-w\sigma_{\text{max}}, w\sigma_{\text{max}}]$.

The probability density error $\epsilon_{\text{dens}}$ arises from the computation of $\ket{\arcsin\sqrt{P(\vec{R})}}$ with $P(\vec{R})$ given by Eq.~\eqref{eqn:multivariate_normal_pdf_T}, and the ancilla rotation in Eq.~\eqref{eqn:riemann_transition_operator}.
The term inside the exponential in
Eq.~\eqref{eqn:multivariate_normal_pdf_T} can be written as
\begin{equation}
\label{eqn:normal_pdf_numerator_expansion}
-\frac{1}{2}\sum_{t=1}^T(\vec{R}^t -\vec{\mu})^{\intercal} \Sigma^{-1}(\vec{R}^t -\vec{\mu}) = -\frac{1}{2}\sum_{t=1}^T\sum_{i=1}^d \sum_{j=i}^d C_{ij}  \bar{R_i^t}\bar{R_j^t},
\end{equation}
where $\bar{R} = R-\mu$ and $C_{ij}$ are classical variables containing volatility and correlation parameters from the correlation matrix $\Sigma$.
In Eq.~\eqref{eqn:normal_pdf_numerator_expansion}, each calculation of $\bar{R}$ thus incurs an error of $\epsilon_A$ and there are $(d+{d \choose 2}) \cdot T$ multiplications in total.
Each $\bar{R}$ term is bounded by $|w|\sigma_{\text{max}}$ by construction, where each quantum register representing a log-return $R$ is constructed to represent values in the window $[-w\sigma_{\text{max}}, w\sigma_{\text{max}}]$.
Using the error analysis for addition and multiplication in Appendix~\ref{app:arithmetic_error_analysis}, the total error in computing Eq.~\eqref{eqn:normal_pdf_numerator_expansion} is

\begin{equation}
\label{eqn:error_sum}
\epsilon_{\text{sum}} = \left( \frac{2w\sigma_{\text{max}} + n}{2^{n-p}} + \frac{1}{4^{n-p}} \right) \left(d+{d \choose 2}\right) \cdot T.
\end{equation}
Then using the error propagation analysis in Appendix~\ref{app:arithmetic_error_analysis} for computing the exponential, square root, arcsine and sine functions on quantum registers which already contain arithmetic errors, we can bound $\epsilon_{\text{dens}}$ by
\begin{equation}
\label{eqn:riemann-dens-error}
    \epsilon_{\text{dens}} \le \epsilon_{\text{sin}} +
    \epsilon_{\text{arcsin}} - \arcsin(0.5 - (\epsilon_{\text{sq}} +
    \sqrt{\epsilon_{\text{exp}} + \epsilon_{\text{sum}}})) + \arcsin(0.5).
\end{equation}

Each rescaling we perform to the input variables introduces a corresponding rescaling error.
In addition to the the $\Pmax$ rescaling discussed in the previous section, we also need to scale the payoff by $1/\fdelta$ to lie in $[0, 1]$.
The final answer thus needs to be multiplied by $\Pmax^{T}\fdelta$ to account
for these rescalings, and
the error in the estimate of the \emph{truncated} integral by amplitude
estimation is therefore scaled by $\Pmax^{T}\fdelta$.
We then can bound the error in the Riemann Summation approach
\begin{equation}
\epsilon_{\text{total}} \le
P_{\text{max}}^{T}f_{\delta}
(\epsilon_{\text{trunc}} + \epsilon_{\text{disc}} + \epsilon_{\text{arith}} +
\epsilon_{\text{amp}}),
\end{equation}
where $\epsilon_{\text{trunc}}, \epsilon_{\text{disc}},$ and
$\epsilon_{\text{amp}}$ are defined as in Section~\ref{sec:general-error}.

\subsubsection{Resource Estimates}
\label{sec:resource_estimates_riemann}
As an example, we consider a basket autocallable with $5$ autocall dates and parameters
$T=20, d=3$, and target an error of $\epsilon_{\text{total}}/f_{\delta} \le 2 \times 10^{-3}$.
We need to choose $w \sim 5$ for the truncation error in Eq.~\eqref{eqn:trunc_error} to be within the total target error, and Eq.~\eqref{eqn:P_max} gives $P_{\text{max}} \approx 4^3$.
This makes the scaling factor prohibitively large with $P_{\text{max}}^T \approx 10^{40}$.
However, there may be some methods to deal with this normalization issue, such as a
method inspired by importance sampling and discussed in
Appendix~\ref{app:importance}.
For the sake of argument, we continue the resource analysis assuming that some
method could be invented
to deal with the normalization, and set $P_{\text{max}} = 1$.

Then, using resource calculations as discussed in Appendix~\ref{app:riemann-resource},
we can bound
$\epsilon_{\text{arith}} \le 2 \times 10^{-3}$ with $n=34$ and $p=2$.
Here $p$ is the integer part of the fixed point representation as defined in
Eq.~\eqref{eqn:fixed_point_repr}.
The $\mathcal{Q}$ operator in this case requires $23$k qubits and has a T-depth of $26$k, including the resources required to compute prices from log-returns using Eq.~\eqref{eqn:return-to-price}.
For a choice of $\Delta t = 1/20$ and $\sigma_{\text{min}}=0.1$ we compute that $\beta \approx 17$.
Choose $\sigma_{\text{max}}=0.4$ and $w=5$.
Thus for the choice of $n$, $\epsilon_{\text{disc}} \approx \fdelta 10^{-5}$ and $\epsilon_{\text{trunc}} \le f_{\delta}\cdot 10^{-4}$.
When the derivative is priced classically with Monte Carlo, we define the pricing error as the standard deviation of the calculated derivative prices over all simulated paths.
We therefore pick $1-\alpha = 0.68$ in Eq.~\eqref{eqn:AE_oracle_calls} to calculate the number of oracle calls needed for a given target error at the same confidence level as classical Monte Carlo.
Choosing a target $\epsilon_{\text{amp}}$ for the amplitude estimation of $10^{-3}$, we then obtain $N^{\text{wc}}_{\text{oracle}} \le 6$k.
This means that the total T-depth is about $1.5\times 10^8$.

Using the same analysis, for a TARF contract (see Section~\ref{sec:TARF}) with $d=1$, $T=26$ and $\Delta t=1/26$, assuming the underlying has annualized volatility $\sigma = 0.4$, a target error of $\epsilon_{\text{total}}/f_{\delta} \le 2 \times 10^{-3}$ can be achieved with a total T-depth of $1.6 \times 10^8$ and $17$k qubits.
These resource estimates are summarized in Table~\ref{tbl:resources}.

\subsection{Re-parameterization Method}
\label{sec:re-parameterization}
The limitations of normalization in Riemann summation motivate the need for a
new method for loading stochastic processes. In the re-parameterization method,
we shift to modeling assets in return space. As described in
Section~\ref{sec:spaces}, in return space underlying assets consist of
uncorrelated normal distributions. We recognize that these different
distributions can be loaded by preparing, in parallel, many standard normals and
then applying affine transformations to obtain the required means and standard
deviations. This approach extracts a specific subroutine - loading a
standard normal into a quantum state - and uses it as a resource to load the
full distribution of underlying paths. The normal loading subroutine itself
can then be precomputed and optimized using variational methods. This is an
advantageous combination of fault-tolerant quantum computing with variational
compilation and will be discussed in Section~\ref{sec:var_gaussian}.
Overall the re-parameterization method avoids the normalization issues in Riemann summation
and reduces the computational requirements.

The steps in re-parameterization pricing are described in
Algorithm~\ref{alg:reparam}. We note that a path $\omega_R\in\Omega_R$ in this
context refers to a series of log-returns $\vec{R}^1,\ldots,\vec{R}^T$.
The re-parameterization method removes the problematic dependence on
$P_{\text{max}}$, and the operators $\mathcal{G}_j$ can be implemented with
relatively few resources using variationally trained circuits as discussed in the following.

\begin{figure}
\begin{algorithm}[H]
\caption{Re-parameterization method pricing}
\begin{algorithmic}[1]

\REQUIRE Parameters $n$, $d$, and $T$ that are all positive integers.
\REQUIRE Access to an operator $\mathcal{G}$ that loads a standard Gaussian
distribution $\sum_i\sqrt{g_i}\ket{i}$ into an
$n$-qubit register. Let $g_i$ be the value of the probability mass function for
 a standard Gaussian distribution discretized
into $2^n$-bins.
\begin{enumerate}
\item Apply $dT$ Gaussian operators $\mathcal{G}$, to $ndT$ qubits. This constructs
\begin{align}
\label{eqn:gaussian_preparation}
%\bigotimes_{j=0}^{dT-1}\mathcal{G}_j|0\rangle =
%\sum_{i=1}^{2^{ndT}}\prod_j^{dT}\sqrt{g_j}\ket{\omega_R},
\bigotimes_{t=1}^T \bigotimes_{j=1}^{d}\mathcal{G}|0\rangle_n 
&= \sum_{\omega_{\bar{R}}} \sqrt{p(\omega_{\bar{R}})} \ket{\omega_{\bar{R}}}_{ndT},
\end{align}
where $\omega_{\bar{R}}$ runs over all $2^{ndT}$ different realizations of this multivariate standard Gaussian, and $p(\omega_{\bar{R}})$ denotes the corresponding probabilities.

\item Let $\Sigma = LL^{\intercal}$ be the Cholesky decomposition of the covariance matrix.
Perform affine transformations $\vec{R}^t = \vec{\mu}^t +
L^{\intercal}\vec{\bar{R}}^t$ to adjust the center and volatility of each
Gaussian. 
We denote the corresponding return paths and probabilities by $\omega_{R}$ and $p(\omega_{R})$, respectively.

\item If the payoff can be computed directly from the log-returns,
then we directly calculate $\delta(\omega_R)=\arcsin\sqrt{\tilde{f}(\omega_R)}$
into a quantum register
\begin{align}
\sum_{\omega_R}\sqrt{p(\omega_R)}\ket{\omega_R}\ket{\delta(\omega_R)}.
\end{align}
If the payoff is defined in terms of prices and not just log-returns, then we
first compute the price space path $\omega$ for each asset using
$\vec{S}^t = \vec{S}^0 e^{\sum_{j=1}^t\vec{R}^j}$.
This calculation can be done in parallel for each asset.
\item Introduce an ancilla qubit and rotate the value of the $\tilde{f}(\omega_R)$
register into its amplitude:
\begin{align}
\sum_{\omega_R}\sqrt{p(\omega_R)(1-\tilde{f}(\omega_R))}\ket{\omega_R}\ket{0}
    +\sum_{\omega_R}\sqrt{p(\omega_R)\tilde{f}(\omega_R)}\ket{\omega_R}\ket{1}.
\end{align}
\item Use amplitude estimation to extract the probability of the ancilla
being $\ket{1}$.
\end{enumerate}
\OUTPUT The (discretized) expected payoff
$\mathbb{E}(\tilde{f}) = \sum_{\omega_R}p(\omega_R)\tilde{f}(\omega_R)$. We rescale this to obtain
$\mathbb{E}(f) = \fdelta\mathbb{E}(\tilde{f}) +
f_{\text{min}}$.
\end{algorithmic}
\label{alg:reparam}
\end{algorithm}
\end{figure}

\subsubsection{Variationally Trained Gaussian Loaders} \label{sec:var_gaussian}
The standard Gaussian loading operator $\mathcal{G}$ can be pre-computed
because in the re-parameterization method it is problem independent.
In this section, we describe an approach to variationally optimize this
operator. 
Let us consider the problem of preparing a standard normal distribution $g(x_i)$ defined on a discretized mesh of points $x_i = -w + i~\Delta x$, with $i=0,\cdots2^n-1$, and $\Delta x = 2w / 2^n$.
In the following example we fix the domain to $w=5$, so that the full range of values considered is $2w=10$.
This choice leaves outside the domain a probability mass of $\sim 5 \times 10^{-7}$.
We take into account the different metrics used for normalizing a function in real space and a wavefunction in a quantum register (which is normalized in such a way that the sum of its squared elements is one),
and therefore the distribution we aim to load in the quantum register is
\begin{equation}
 g(x_i) \times \Delta x,  \quad \sum_i  g(x_i) \times \Delta x = 1.
\end{equation}
Notice that due to the finite truncation domain, the target distribution is normalized to $1-\alpha$. In principle we can re-normalize the distribution to one in the chosen interval of width $2w$.
Either way this choice provides a negligible difference when compared  with the error we observe in the training.

The variational ansatz of choice is represented by a so-called $Ry$-CNOT ansatz, with linear connectivity (see Appendix~\ref{app:variational}).
In this work we introduce a novel strategy to optimize the circuit in this context, which relies on a energy-based method \cite{peruzzo2014}, and is also detailed in Appendix~\ref{app:variational}.
In short, the target cost function  is  the energy of the associate quantum harmonic oscillator problem,  whose ground state  is
naturally Gaussian \cite{ollitrault2020}.
Here, we must be careful because the squared modulus of the solution of the discretized quantum harmonic oscillator matches a normal distribution only in the limit of $\Delta x \rightarrow 0$.
To fix this we perform a subsequent re-optimization targeting directly the infinity-norm between the two distributions.
\begin{equation}
\label{eq:infnorm}
 L_\infty = \max_i  | g(x_i) \times \Delta x - \tilde{g}(x_i) |,
 \end{equation}
 where the quantum state encoded in the register is defined by coefficients $\sqrt{\tilde{g}(x_i)}$.

We notice that training directly with Eq.~\eqref{eq:infnorm} as a cost function is not sufficient, and pre-training using the energy-based approach is needed to produce accurate results.
We indeed observe how the $L_\infty$ cost-function displays a much more corrugated  landscape in the circuit parameter space compared to the energy of the associate quantum harmonic oscillator problem.

\begin{figure}[th]
  \centering
  \includegraphics[width=0.75\linewidth]{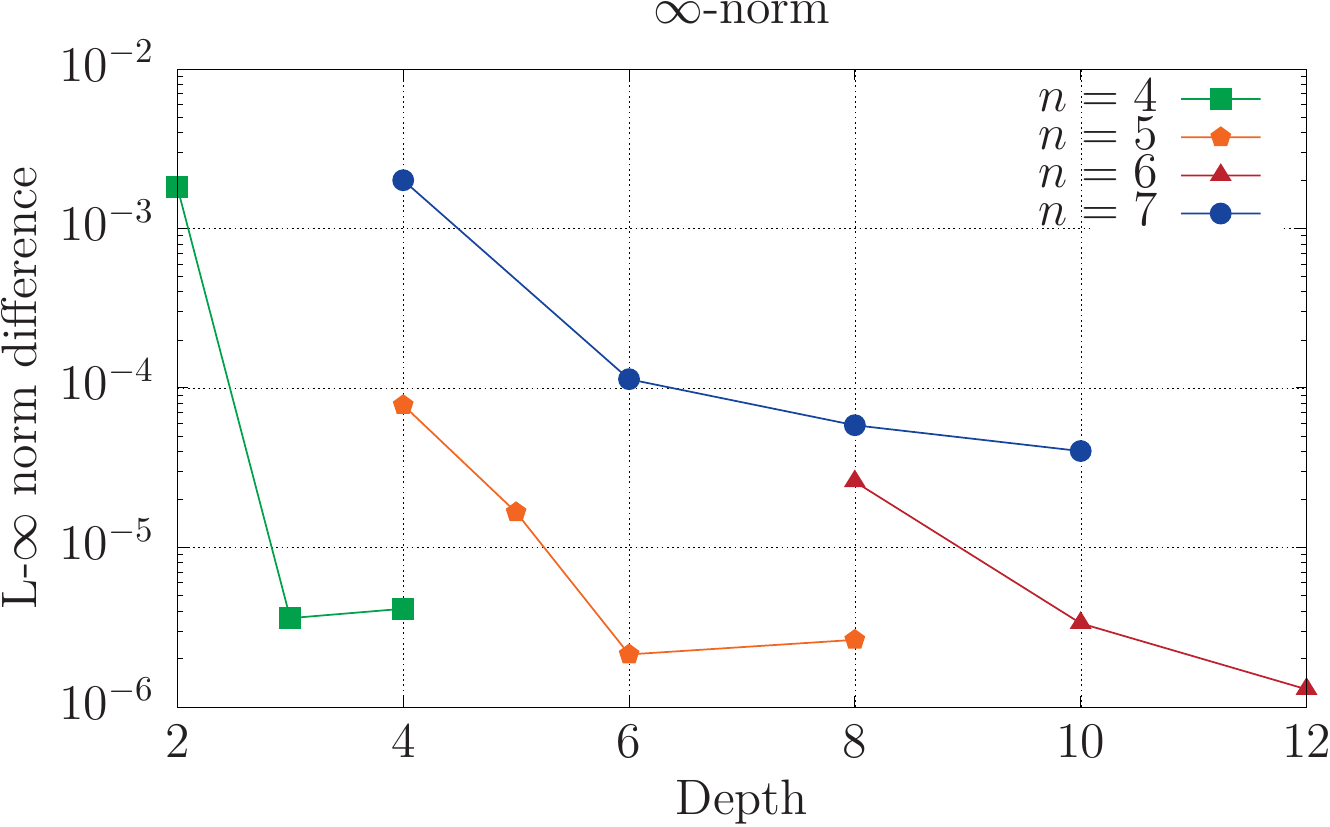}
  \caption{L$_\infty$ errors from training variational Ry-CNOT circuits to
  approximate $\mathcal{G}$ for different register sizes $n$.}
  \label{fig:linf}
\end{figure}

%We find that we can bound the $L_{\infty}$-norm of the prepared state
% by $\epsilon_{\text{dens}}$ of $10^{-4}$ with 8 qubits
%and a depth of 10 CNOTs (Figure~\ref{fig:linf}). These circuits are trained
%using Ry-CNOT ansatzes
%and an optimization function that minimizes
%the energy of the harmonic oscillator Hamiltonian, whose ground state is
%naturally Gaussian.
It is important to note that
the circuits needed to encode these Gaussian states for different choices of the register size $n$ can be
pre-trained and used for any derivative pricing problem, and therefore the training cost is not
included in our overall resource estimations.
We show in Fig.~\ref{fig:linf} results for different register sizes and depths of the circuit ansatz.
More details are provided in Appendix~\ref{app:variational}.

This  numerical study shows that the state we can prepare variationally, approaches
 the target  exponentially  fast in the depth, and hence in the number of gate operations.
This observation is in good agreement with the expected behavior  from the
 Solovay-Kitaev theorem \cite{dawson2005solovay}, that provides an upper bound for the number of gates required to achieve a desired accuracy for a cost function.
Indeed, for any target operation $U \in SU(2^n)$, there is a sequence of operators $S=U_{s_1} U_{s_2} \dots U_{s_D}$ in a dense subset of $SU(2^n)$, such that error in the energy $\epsilon$ decreases exponentially with the depth  $D=\mathcal{O}(log^c(1/\varepsilon))$.
Although the subset of $SU(2^n)$ operations generated by the entangler blocks in our circuit does not generate a dense subset of $SU(2^n)$ arbitrarily close to the exact unitary $U$ (the generator of the target state), we can numerically observe that the exponential decrease of the error with the number of gates still holds.

We end this section investigating the portability of these results in the fault-tolerant regime, which is necessary for the applicability of the derivative pricing algorithm.
While our numerical results provide evidence for a rather efficient Gaussian state preparation in terms of circuit depth for an $Ry$-CNOT ansatz, an additional step has to be made in view of a fault-tolerant implementation of such circuits.
In this new-framework, the continuous rotation $Ry$ gate needs to be expanded as a finite product of discrete operations. Following again the Solovay-Kitaev theorem, or more specialized results \cite{selinger2012efficient}, it is possible to also have an efficient representation of any $SU(2)$ operator with a sequence of Clifford + T gates that scale logarithmically with the threshold error $\epsilon$.
We investigate how the results obtained before can be transferred in this regime where rotation angles can only take discretized values.
We therefore assume that each parameter ${\vartheta}^k_{q_j}$ can only be represented in the format $j * 2\pi/ M_{\text{digit}}$, where $j$ is an integer.
We numerically show in Appendix~\ref{app:variational} that the error introduced by such digitization decreases systematically with the mesh size as $O(1 / M_{\text{digit}})$.

\subsubsection{Error Analysis}
The total error in the re-parameterization approach is
\begin{equation}
\frac{\epsilon_{\text{total}}}{\fdelta} \le \epsilon_{\text{trunc}} +
\epsilon_{\text{disc}} +
\epsilon_{\text{arith}} +
\epsilon_{\text{amp}},
\end{equation}
where $\epsilon_{\text{trunc}}, \epsilon_{\text{disc}}$, and
$\epsilon_{\text{amp}}$ are the truncation, discretization, and amplitude
estimation error bounded in Section~\ref{sec:general-error}.
Here, the term $\epsilon_{\text{arith}}$ arises from the individual errors we
introduce during the preparation of the Gaussians and the calculation of the payoff.
Assuming that each Gaussian $g(x_i)$ we prepare has $L_{\infty}$ error
$\epsilon_{\text{dens}}$ and the payoff calculation introduces a max error of
$\epsilon_{f}$, the total error will be
\begin{equation}
\epsilon_{\text{arith}} = \underbrace{\sum_{x_1=-w}^{w}\cdots \sum_{x_{dT}=-w}^{w}}_{dT}\left[\prod_{i=1}^{dT}(g(x_i)+\epsilon_{\text{dens}})\left(f(\mathbf{x}) + \epsilon_f\right)
- \prod_{i=1}^{dT}g(x_i)f(\mathbf{x})\right],
\end{equation}
where $\mathbf{x}=(x_1, x_2, ..., x_{dT})$.
Expanding the integrand and keeping only the linear error terms, we get
\begin{equation}
    \epsilon_{\text{arith}} \le 2wdT\epsilon_{\text{dens}} + \epsilon_f,
\end{equation}
where we use that $\sum_{-w}^{w}g(x) \le 1$ due to truncation of the probability mass function.

\subsubsection{Resource Estimates}
\label{sec:resource_estimates_reparam}
We calculate the resources required for the same basket autocallable as in Section~\ref{sec:resource_estimates_riemann}, where $d=3$, $T=20$, $\Delta t=1/20$, $\sigma_{\text{max}} = 0.4$, $\sigma_{\text{min}} = 0.1$ and $w=5$, and the contract has $5$ autocall dates.
We further assume that each Gaussian is prepared with $n=5$ qubits, such that $\epsilon_{\text{dens}} = 2 \times 10^{-6}$, $\epsilon_{\text{amp}} = \epsilon_f = 10^{-4}$, which gives us a total error of $\epsilon_{\text{total}}/\fdelta \approx 2 \times 10^{-3}$.
From Fig.~\ref{fig:linf} we observe that we can prepare Gaussian states with $L_{\infty} \sim 2 \times 10^{-6}$ using $5$ qubits and circuit depth $6$, requiring $7$ layers of Ry gates.
With these inputs and using the resource calculations described in Appendix~\ref{app:reparam-resources}, constructing the $\mathcal{Q}$ operator using re-parameterization requires $8$k qubits and has a T-depth of $9.5$k, which includes the computation of prices from log-returns, Eq.~\eqref{eqn:return-to-price}.
For a target confidence level of $1-\alpha = 0.68$, the total T-depth is $5.4 \times 10^7$.
With the re-parameterization method, pricing the TARF of Section~\ref{sec:resource_estimates_riemann} with $d=1$, $T=26$, $\Delta t=1/26$ and $\sigma=0.4$ to accuracy $\epsilon_{\text{total}}/\fdelta \approx 2 \times 10^{-3}$ requires total T-depth of $8.2 \times 10^7$ and $11.5$k qubits.
These resource estimates are summarized in Table~\ref{tbl:resources}.

\section{Payoffs}
\label{sec:payoffs}

The previous sections analyzed methods for performing steps 1-3 in Algorithm~\ref{alg:reparam}. This results in a quantum state representing a superposition
of all
possible paths. In this section, we apply payoff
functions to these superpositions (step 4) so that the normalized expected
discounted payoff is stored in an amplitude of an accumulator qubit. This
allows amplitude estimation to extract this amplitude to complete the pricing
algorithm. We also analyze the additional errors introduced by the payoff function.
We cover two example derivative cases: autocallables and TARFs.
In addition, throughout this section (unlike in the other sections),
we will assume that payoffs are \emph{not} discounted unless explicitly stated.

\subsection{Autocallables}
An \emph{autocallable} contract is typically defined in terms of asset returns relative to predefined reference levels, and includes a \emph{notional} value which is used to calculate the dollar value of the contract.
For a single underlying, an autocallable consists of:
\begin{itemize}
\item a set of $m$ binary options $\{(K_i, t_i, f_i)\}_{i=1...m}$ each with strike returns $K_i$, payment dates $t_i$, and binary payoffs $f_i$. Assume these are sorted so that $t_i < t_{i+1}$.
\item a short knock-in put with strike $K_{put}$, barrier $b$ and notional value $k$, and
\item the condition that if any binary option has a non-zero payoff then all subsequent options at later times including the put option are
knocked out.
\end{itemize}

The payoff $f_i$ of the binary options is equal to a fixed number $p_i$ when $\tilde{R}^{t_i}_c \geq K$ and zero otherwise, where $ \tilde{R}^{t_i}_c$ is the cumulative return of the underlying asset at timestep $t_i$. The payoff of the put option is 
\begin{align}
f_{put} =
\begin{cases}
k(\tilde{R}^T_c-K_{put}) & \mbox{if } \tilde{R}^T_c < K_{put} \mbox{ and  } \tilde{R}^i_c < b, \; \forall i\in\{0, ..., T\} \\
0 & \mbox{otherwise.}
\end{cases}
\end{align}
A more detailed explanation of autocallable options and all the parameters mentioned above can be found in Appendix \ref{app:autocallable}.

\begin{figure}[ht]
\centering
\includegraphics[width=\textwidth]{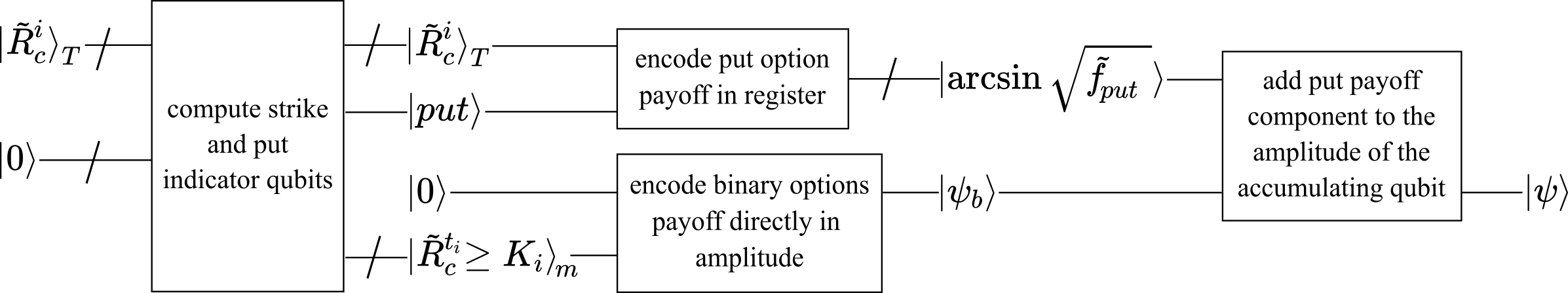}
\caption{A block diagram of the subroutine presented in Algorithm~\ref{alg:autocallable}.
$\ket{\tilde{R}^i_c}_T$ represents the set of all cumulative return vectors for all timesteps, 
$\ket{\tilde{R}^{t_i}_c \geq K_i}_m$ and $\ket{put}$ are the strike and put
qubits respectively,
$\ket{\arcsin\sqrt{\tilde{f}_{put}}}$ is the register containing the value of the arcsine of the normalized discounted payoff of the put option,
$\ket{\psi_b}$ is the accumulator qubit with just the rotations from the binary options applied to it and
$\ket{\psi}$ is the accumulator qubit with the full expected discounted payoff
rotations applied to it.
Junk qubits were omitted from the diagram for clarity.}
\label{fig:AC_block}
\end{figure}

Note that the minimum payoff of the option occurs when the
return of the underlying asset falls to zero. Therefore, we can compute the
normalized discounted payoff $\tilde{f}_i$ from Eq~\eqref{eq:norm_disc_payoff} as
\begin{equation}
\label{eqn:normalized_autocall_payoff}
\tilde{f}_i = \frac{e^{-rt_i}f_i + e^{-rt_{m}}kK_{put}}{f^{\text{disc}}_{\text{max}} + e^{-rt_{m}}kK_{put}},
\end{equation}
where $f^{\text{disc}}_{\text{max}}$ is the maximum possible discounted payoff among the possible discounted payoffs across all the binary options
and $r$ is the risk-free rate.
We now give a sketch of the algorithm used to implement the payoff for an
autocallable shown in Algorithm~\ref{alg:autocallable}. In steps 1-2, we
compute the strike and put qubits that indicate s which of the payoffs $f_1,
f_2,..., f_m, f_{put}$ are non-zero. In step 3, we control on the strike qubits
to apply the appropriate rotation on the accumulator qubit corresponding to the
normalized discounted payoff from the binary options.
In parallel, we execute step 4 which stores the normalized discounted payoff
from the put option into another register. Finally, in step 5, we use the
previously computed register to apply the appropriate rotation on the
accumulator qubit corresponding to the normalized discounted payoff from the
put options.
This procedure is illustrated in Fig.~\ref{fig:AC_block}.

\begin{figure}
\begin{algorithm} [H]
\caption{Autocallable payoff implementation}
\begin{algorithmic}[1]
\REQUIRE An autocallable with parameters $\{(K_i, t_i, f_i)\}_{i=1...m}, K_{put}$, 
$b$ and $k$.
\begin{enumerate}
\item We apply in parallel a set of $T$ comparators to obtain the register $\ket{\tilde{R}^{i}_c < b}_T$, a set of $m$ comparators to obtain the strike register $\ket{\tilde{R}^{t_i}_c \geq K_i}_m$ and a single comparator to obtain the qubit $\ket{\tilde{R}^{T}_c < K_{put}}$ . \label{step:ac_comp}
\item We then apply the necessary AND and OR operations on all the registers from the previous step to obtain the $\ket{put}$ qubit which is set to $\ket{1}$ if all of the conditions for the put option were fulfilled i.e. none of the binary options payed off, the put option was knocked in and $\tilde{R}^{T}_c < K_{put}$.
\item Let $\theta_i = \arcsin(\sqrt{\tilde{f}_i})$. Serially, for each bit of the strike register we apply
a controlled rotation of $\theta_i$ on the accumulator qubit conditioned on all previous bits having been zero.
This is illustrated in Figure~\ref{fig:auto-call_circuits}.
\item We use the register $\ket{\tilde{R}^{T}_c}$ to compute the arcsine of the the normalized expected payoff using quantum arithmetic and obtain the register $\ket{\arcsin(\sqrt{\tilde{f}_{put}})}$ where $\tilde{f}_{put}$ is defined in Eq.~\eqref{eqn:normalized_autocall_payoff}. \label{step:ac_put_reg_compute}
\item Finally we apply a rotation of $\theta_{put} = \arcsin(\sqrt{\tilde{f}_{put}})$ on the accumulator qubit on the condition that the put option is activated. This is done using a series of $R_y(2^i)$ rotations where each rotation is controlled on the $i$th qubit of the $\ket{\arcsin(\sqrt{\tilde{f}_{put}})}$ register and the $\ket{\text{put}}$ qubit.
\end{enumerate}
\end{algorithmic}
\label{alg:autocallable}
\end{algorithm}
\end{figure}

\begin{figure}[ht]
\centering
\includegraphics[width=0.75\textwidth]{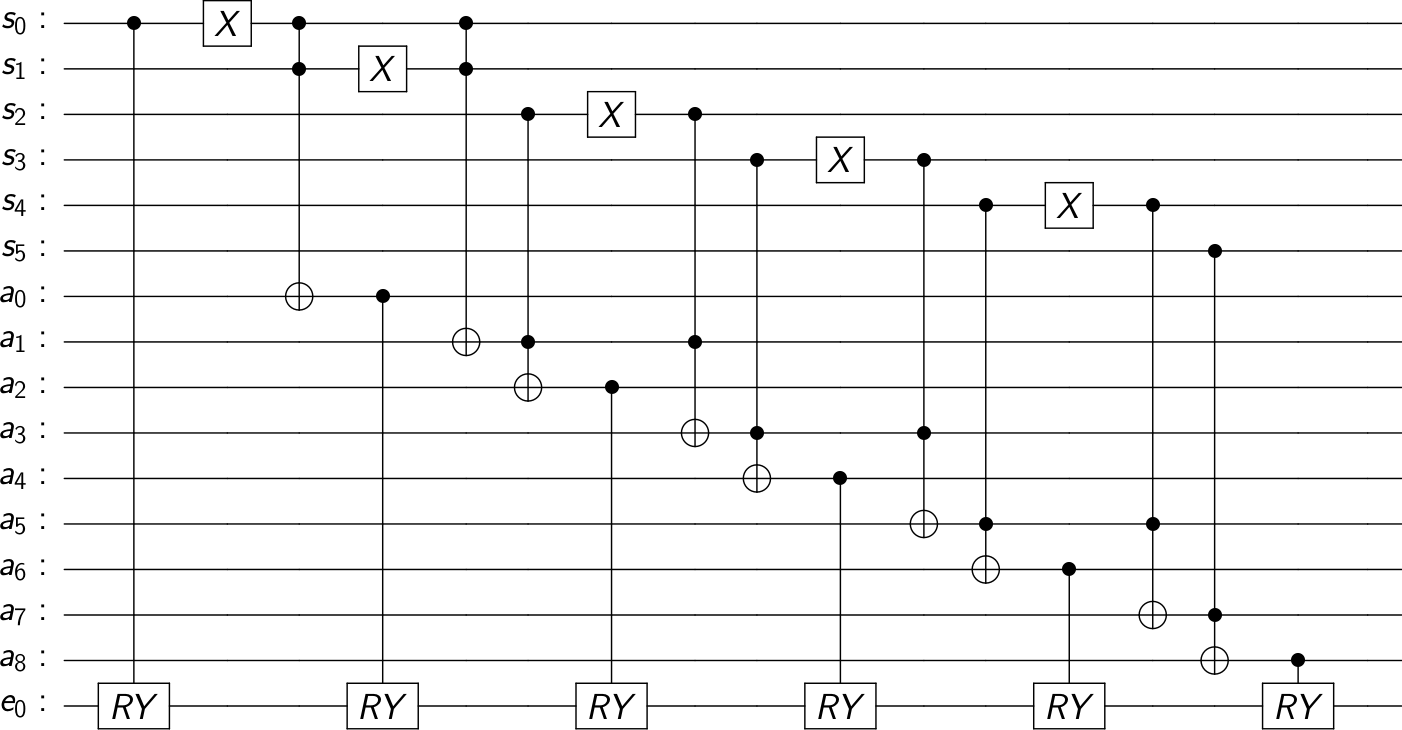}
\caption{Example circuit used in step 4 of Algorithm~\ref{alg:autocallable} to accumulate binary option discounted payoffs in an
autocallable with 6 binary options. Here the qubits $s_0,...,s_5$ represent the
boolean comparisons for the 6 strike values $K_i$. The normalized discounted payoff $\tilde{f}_i$
at time $t_i$ (given by a
particular phase in the RY rotation) is only non-zero if all payoffs at previous times $t_{j<i}$ are zero.
The overall normalized discounted payoff is loaded into the amplitude of qubit $e_0$.}
\label{fig:auto-call_circuits}
\end{figure}

We elaborate on a few subtleties in Algorithm~\ref{alg:autocallable}:
\begin{enumerate}
	\item Throughout the algorithm, we can assume that we have access to
	$\ket{\tilde{R}^i_c}$ for all $i$. This is because these registers are set
	when computing the prices $S^i$ of the underlying when loading in the paths
	 onto the quantum state (Appendix \ref{app:reparam-resources}).
	\item Strictly speaking, the output register in step
	\ref{step:ac_put_reg_compute} is equal to $\ket{\arcsin
	(\sqrt{\tilde{f}_{put}})}$ only if the part of the $\ket{put}$ qubit that
	is entangled to this final register is $\ket{1}$. However, in the next
	step, the $R_y$ rotations that are controlled on this register are also
	always controlled on the $\ket{put}$ qubit, so the instances when the
	register has an unknown value have no effect on the rotations applied to
	the accumulator qubit.
	\item The autocallable can also be defined on a basket assets instead of just one. Typical examples include \emph{BestOf} and \emph{WorstOf}, where the return of the contract is based on the return of the best or the worst performing asset in the basket respectively.
The only difference is that, in step \ref{step:ac_comp}, we would apply the
comparators on all the assets and then compare the return of each asset
$\ket{\tilde{R}^i_{c,j}}_{j=1...d}$ to find the largest or smallest as
necessary. We can then use the fact that if the worst performing asset is above
the strike price, then so are all the other assets. Conversely, if the best
performing asset is below the strike price, then so are all the other assets.
	\item If we want to price an autocallable option that does not have a
	knock-in put, then, not only can skip steps 4-5 of the algorithm, but we
	can also perform
	the algorithm by just using the sum of log-returns instead of the sum of
	returns. Then it would be sufficient to have access to the registers
	$\ket{\sum_{j=1}^{t_i}R^j}$ instead of $\ket{\tilde{R}^{t_i}_c}$ which would
	reduce the resources required in the loading of the paths in superposition
	into the quantum state.
\end{enumerate}

We now discuss the error arising from Algorithm~\ref{alg:autocallable}. Steps
1-2 are performed with logical operation circuits (Comparator, AND, OR) which introduce no error, while steps 3 and 5 require controlled-$Ry$ rotations whose decomposition into T-gates is a function of an additive error $\epsilon$, which we can choose depending on the desired accuracy of the calculation.

Step 4 is the most resource heavy component of the payoff circuit, which requires the computation of the quantum register $\ket{\tilde{R}^T_c - K_{put}}$, the division of that register by the classical constant in the denominator of Eq.~\eqref{eqn:normalized_autocall_payoff}, as well as the computation of the square root and arcsine of the register.
We describe in detail the resource requirements for all the above circuit components in Appendix~\ref{app:arithmetic_resource_estimation}, and the corresponding arithmetic and gate synthesis error in Appendix~\ref{app:arithmetic_error_analysis}.

Consider again the autocallable contract from
Section~\ref{sec:resource_estimates_riemann} and Section~\ref{sec:resource_estimates_reparam} with 5 autocall dates, defined on $d=3$ assets and simulated using $T=20$ timesteps.
We target a total additive payoff error $\epsilon_f$ which when distributed across the operations of steps 3, 5, 6 in Algorithm~\ref{alg:autocallable} determines the resources required by each component.
For $\epsilon_f=10^{-4}$, the circuit computing the autocallable payoff
requires $1.6$k qubits and a T-depth of $3.2$k, assuming we can parallelize
computations wherever possible. These resources are included in the end-to-end
summary estimates of Table~\ref{tbl:resources}.

\subsection{Target Accrual Redemption Forwards}
\label{sec:TARF}
In this section, we consider the payoff implementation for the second example
derivative: TARFs. To simplify the discussion, we describe the TARF
implementation for a single underlying in price space.
TARFs are usually contingent on the price of the underlying asset rather than
the return.

A \emph{TARF} is:
\begin{itemize}
\item A forward price $F$, payment dates denoted chronologically by timesteps $t={1,2,...,T}$, two strike prices $K_{\text{upper}} > F$ and
$K_{\text{lower}} \le F$, a knock-out price $b$, a constant $\alpha$ and an accrual cap $C\in \mathbb{R}_+$.
\item At each timestep $t$ the TARF has a payoff
\begin{align}
f_t =
\begin{cases}
S^t-F \text{ if } S^t > K_{\text{upper}} & \mbox{and the contract is not knocked out (`upper condition')}\\
\alpha(S^t-F) \text{ if } S^t < K_{\text{lower}}&  \mbox{and the contract is not knocked out (`lower condition')}\\
0 & \mbox{otherwise.}
\end{cases}
\end{align}
\item a knock-out condition that if at any time $t$ the price is above $b > F$ all payoffs that haven't been paid yet (including the one for the current time $t$) are knocked out. We will call this the `barrier condition'.
\item an accrual cap condition such that if the total gain accumulated by any payment date exceeds $C$ the contract holder
only receives a payoff such that the total gains equals $C$ and the rest of the forward contracts are knocked out. We will call this the `cap condition'.
\end{itemize}
A more detailed explanation of TARFs and all the parameters mentioned above can be found in Appendix \ref{app:tarf}.

The minimum TARF payoff occurs when the price of the underlying asset falls to
zero, and the maximum payoff occurs when the payoff at every payment date
is $b-F$ until the accrual cap is reached. Thus, we can compute the normalized
discounted payoff $\tilde{f}(\omega)$ using Eq.~\eqref{eq:norm_disc_payoff} to be
\begin{equation}
\tilde{f}(\omega) = \frac{f^{\text{disc}}(\omega) + \sum_{j=1}^{T}e^{-rt_{j}}2TF}{\sum_{j=1}^{C/(b-F)}e^{-rt_j}(b-F) +\sum_{t_j=1}^{T}e^{-rt_{j}}2TF},
\label{eqn:normalized_tarf_payoff}
\end{equation}
where $f^{\text{disc}}(\omega)$ is the discounted payoff of the TARF for the path $\omega$ and $r$ is the risk-free rate.

\begin{figure}[ht]
\centering
\includegraphics[width=\textwidth]{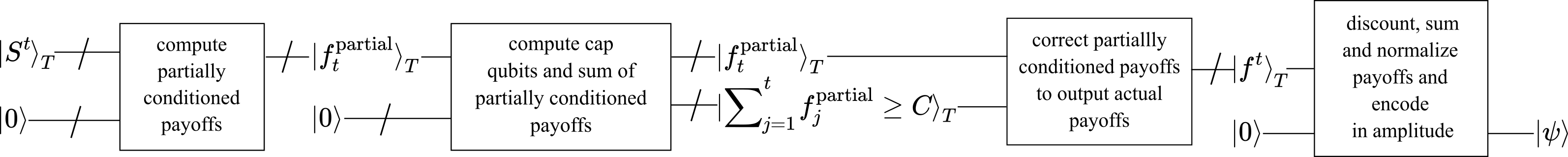}
\caption{A block diagram of the subroutine presented in Algorithm~\ref{alg:tarf}.
$\ket{S^t}_T$ represents the set of all prices for all timesteps, 
$\ket{f^\text{partial}_t}_T$ represent the partial conditional payoffs for all timesteps,
$\ket{\sum_{j=1}^{t}f^\text{partial}_t \geq C}_T$  represent the sum of partial conditional payoffs over each timestep,
$\ket{f_t}_T$ represent the actual payoffs for all timesteps and
$\ket{\psi}$ is the accumulator qubit with the full expected discounted payoff rotations applied to it.
Large amounts of junk qubits were omitted from the diagram to avoid dangling wires.} 
\label{fig:tarf_block}
\end{figure}

Algorithm~\ref{alg:tarf} details an implementation for a TARF payoff.
In steps 1-3, we compute the `partial conditional payoffs' $f_t^{\text{partial}}$ at each timestep $t$,
i.e. the payoff given that we ignore the cap condition.
In steps 4-8, we compute whether the cap condition is fulfilled at each
timestep and correct the partial conditional payoffs to take into account the
cap condition (giving us the actual payoffs).
Finally, in steps 9-11 we discount the payoffs, sum them up, normalize the
result according to Eq.~\eqref{eqn:normalized_tarf_payoff} and apply the
appropriate rotation on the accumulator qubit.
This procedure is illustrated in Fig \ref{fig:tarf_block}.
  
\begin{figure}
\begin{algorithm} [H]
\caption{TARF payoff implementation}
\begin{algorithmic}[1]
\REQUIRE A TARF with parameters $(F, T, K_{\text{upper}}, K_{\text{lower}}, b,\alpha, C)$.
\begin{enumerate}
\item We apply in parallel a set of comparators to obtain the registers $\ket{S^t <  K_{\text{lower}}}_T$, $\ket{S^t >  K_{\text{upper}}}_T$ and $\ket{S^t > b}_T$. \label{step:tarf_comp}
\item We then apply the necessary AND and OR operations on all the registers from the previous step to obtain registers $\ket{upper}_T$ and $\ket{lower}_T$ where the $t$th qubit
in each register represents whether the upper and lower conditions were partially fullfilled (not taking into account the cap condition).
\item We use $\ket{S^{t}}$, $\ket{upper}_T$ and $\ket{lower}_T$ to create $\ket{f^\text{partial}_t}$ registers in parallel using quantum arithmetic and control-copies.
\item For each $t$, we compute $\sum_{j=1}^{t}f^\text{partial}_t$ in series, storing each result in a separate register.
\item We compute the $\ket{\sum_{j=1}^{t}f^\text{partial}_t \geq C}$ qubits in parallel for all $t$.
\item Then applying AND and OR gates in series to the  $\ket{\sum_{j=1}^{t}f^\text{partial}_j \geq C}$ qubits, we compute $\ket{cap}_T$ which has all $T-1$ $\ket{0}$ qubits a single $\ket{1}$ qubit in the $c$th position where $c$ is the timestep at which the cap condition occured. We now have
\begin{enumerate}
	\item the value of $f^\text{partial}_t$ stored in the register $\ket{f^\text{partial}_t}$
	\item the register $\ket{\sum_{j=1}^{t}f^\text{partial}_t \geq C}$ for every timestep $t$ indicating whether the cap condition has been fulfilled at the current or previous timestep
	\item the register $\ket{cap}_T$ indicating which timestep the cap condition was fulfilled.
\end{enumerate}
\item We control-copy all the registers containing $f^\text{partial}_t$ in parallel controlled on the cap conditions to give us $T-1$ registers of $\ket{f_{t \neq c}}$ and an additional register $\ket{\vec{0}}$. We now have a register for each timestep with the correctly encoded payoff at that timestep, except for time $c$ which is the timestep at which the cap condition was fulfilled. \label{step:ctrl-copy-payoff}
\item We then take care of timestep $c$ by adding $C-\sum_{j=1}^{t-1}f_j$ to each register $t$ controlled on the $t$th qubit of $\ket{cap}_T$ such that nothing changes except for the $\ket{\vec{0}}$ register in step~\ref{step:ctrl-copy-payoff} which now turns into $\ket{f_{c}}$
\item We apply the discount factor to each $f_t$ in parallel to obtain $\ket{f^\text{disc}_t}$
\item We then add up all the discounted payoffs, normalize the sum and take the arcsine to obtain $\ket{\arcsin(\sqrt{\sum_{j=1}^T\tilde{f}_{j}})}$
\item Finally we apply a rotation of $\theta = \arcsin(\sqrt{\sum_{j=1}^T\tilde{f}_{j}})$ on the accumulator qubit.
This is done using a series of $R_y(2^i)$ rotations where each rotation is controlled on the $i$th qubit of the $\ket{\arcsin(\sqrt{\sum_{j=1}^T\tilde{f}_{j}})}$.
\end{enumerate}
\end{algorithmic}
\label{alg:tarf}
\end{algorithm}
\end{figure}

An interesting note is that were we to ignore the discount factor, this algorithm would be much simpler: we would add an extra step after step 5 in which we would set the payoffs for all the paths in which the cap condition was fulfilled equal to $C$ and then to skip ahead to steps 10 and 11.

An error analysis for the TARF payoff is very similar to that
described in the previous section for autocallables.
There are 2 main differences. First, when computing the payoff
$f^\text{partial}_t$ for the lower condition, we require a multiplication.
We set the error on this multiplication to be 10 times lower than our final
desired error to make it negligible.
Second, when discounting the payoffs, we have to ensure an error of
$\frac{1}{\sqrt{T}}$ times smaller than in the auto-callable case because we are adding the payoffs after discounting them,
causing the errors to add in quadrature.
For $\epsilon_f=10^{-4}$, the circuit computing the TARF payoff requires $9$k qubits and a T-depth of $6$k, assuming we can parallelize computations wherever possible.

\section{Discussion}
\label{sec:discussion}

We provide a thorough resource and error analysis to price financial
derivatives using quantum computers.
In particular, we investigate autocallables and TARFs which are two important path-dependent options that are relevant in practice and computationally expensive to price classically.
To achieve this we introduce the re-parameterization method: a new method to
load stochastic processes
that overcomes the limitations of existing approaches.
Although we limit our analysis to geometric Brownian motion, our approach can
be straightforwardly extended to other models e.g. to stochastic or local volatility methods
by loading multiple independent stochastic processes and introducing a
conditional or non-stationary re-parametrization.

Our resource estimates give a target performance threshold for quantum computers
capable of demonstrating advantage in derivative pricing.
Assuming a target of 1 second for pricing an autocallable option, the quantum
processor would need
to execute T-gates at a rate of 10MHz at a code distance that can support
$10^{10}$ logical operations. Further improvements in reducing the T-depth for
this algorithm would linearly lessen this requirement.

The resource estimates in this work concur with recent
work~\cite{babbush2020focus} that emphasizes the importance of going beyond
complexity scaling in order to understand thresholds for quantum advantage.
In particular, the quadratic speedup available in amplitude estimation-based
algorithms could be lost in the constant factor overheads of error correction.

Although current estimates target logical clock rates around 10kHz
\cite{Fowler2018} (i.e. orders of magnitudes slower than our requirement),
we are optimistic that future work on algorithms, circuit optimization, error
correction, and
hardware will continue to improve the required resource estimates and runtimes.
For example, in the case of Shor's algorithm, the estimated resource
requirements have reduced by almost three orders of magnitude through
careful analysis across several publications~\cite{gidney2019factor}.
This work represents the first milestone on the journey towards quantum
advantage for pricing financial derivatives and we are looking forward to
future enhancements.

Further, we emphasize that the resource estimation approach here can be
fruitfully applied to analyze thresholds in other financially relevant
applications. As summarized in two recent
reviews~\cite{egger2020quantum,bouland2020prospects} there are many potential
areas for quantum advantage in finance where advantage thresholds would provide
 useful targets for both industry and the research community.

\begin{acknowledgments}
We thank Paul Burchard for guidance on the derivative pricing problem domain
and Thomas H{\"a}ner for useful discussions regarding quantum arithmetic. We thank
Graham Griffiths, Alex Hurst, Dunstan Marris and Elmer Tan for their technical and business
insights on derivative products. We
thank Ryan Babbush for suggesting clarifications to the manuscript. SC
contributed to this work during his internship at Goldman Sachs.
\end{acknowledgments}

\bibliographystyle{apsrev4-1_custom}
\bibliography{derivatives}   % name your BibTeX data base

\begin{thebibliography}{45}%
\makeatletter
\providecommand \@ifxundefined [1]{%
 \@ifx{#1\undefined}
}%
\providecommand \@ifnum [1]{%
 \ifnum #1\expandafter \@firstoftwo
 \else \expandafter \@secondoftwo
 \fi
}%
\providecommand \@ifx [1]{%
 \ifx #1\expandafter \@firstoftwo
 \else \expandafter \@secondoftwo
 \fi
}%
\providecommand \natexlab [1]{#1}%
\providecommand \enquote  [1]{``#1''}%
\providecommand \bibnamefont  [1]{#1}%
\providecommand \bibfnamefont [1]{#1}%
\providecommand \citenamefont [1]{#1}%
\providecommand \href@noop [0]{\@secondoftwo}%
\providecommand \href [0]{\begingroup \@sanitize@url \@href}%
\providecommand \@href[1]{\@@startlink{#1}\@@href}%
\providecommand \@@href[1]{\endgroup#1\@@endlink}%
\providecommand \@sanitize@url [0]{\catcode `\\12\catcode `\$12\catcode
  `\&12\catcode `\#12\catcode `\^12\catcode `\_12\catcode `\%12\relax}%
\providecommand \@@startlink[1]{}%
\providecommand \@@endlink[0]{}%
\providecommand \url  [0]{\begingroup\@sanitize@url \@url }%
\providecommand \@url [1]{\endgroup\@href {#1}{\urlprefix }}%
\providecommand \urlprefix  [0]{URL }%
\providecommand \Eprint [0]{\href }%
\providecommand \doibase [0]{http://dx.doi.org/}%
\providecommand \selectlanguage [0]{\@gobble}%
\providecommand \bibinfo  [0]{\@secondoftwo}%
\providecommand \bibfield  [0]{\@secondoftwo}%
\providecommand \translation [1]{[#1]}%
\providecommand \BibitemOpen [0]{}%
\providecommand \bibitemStop [0]{}%
\providecommand \bibitemNoStop [0]{.\EOS\space}%
\providecommand \EOS [0]{\spacefactor3000\relax}%
\providecommand \BibitemShut  [1]{\csname bibitem#1\endcsname}%
\let\auto@bib@innerbib\@empty
%</preamble>
\bibitem [{\citenamefont {Prabha}\ \emph {et~al.}(2013)\citenamefont {Prabha},
  \citenamefont {Savard},\ and\ \citenamefont
  {Wickramarachi}}]{prabha_deriving_2013}%
  \BibitemOpen
  \bibfield  {author} {\bibinfo {author} {\bibfnamefont {A.}~\bibnamefont
  {Prabha}}, \bibinfo {author} {\bibfnamefont {S.}~\bibnamefont {Savard}}, \
  and\ \bibinfo {author} {\bibfnamefont {H.}~\bibnamefont {Wickramarachi}},\
  }\href
  {https://milkeninstitute.org/sites/default/files/reports-pdf/Derivatives-Report.pdf}
  {\emph {\bibinfo {title} {Deriving the Economic Impact of Derivatives}}},\
  \bibinfo {type} {Tech. Rep.}\ (\bibinfo  {institution} {Milken Institute},\
  \bibinfo {year} {2013})\BibitemShut {NoStop}%
\bibitem [{\citenamefont {Black}\ and\ \citenamefont
  {Scholes}(1973)}]{BlackScholes}%
  \BibitemOpen
  \bibfield  {author} {\bibinfo {author} {\bibfnamefont {F.}~\bibnamefont
  {Black}}\ and\ \bibinfo {author} {\bibfnamefont {M.}~\bibnamefont
  {Scholes}},\ }\bibfield  {title} {\enquote {\bibinfo {title} {The pricing of
  options and corporate liabilities},}\ }\href {\doibase 10.1086/260062}
  {\bibfield  {journal} {\bibinfo  {journal} {Journal of Political Economy}\
  }\textbf {\bibinfo {volume} {81}},\ \bibinfo {pages} {637} (\bibinfo {year}
  {1973})}\BibitemShut {NoStop}%
\bibitem [{\citenamefont {Montanaro}(2015)}]{montanaro2015quantum}%
  \BibitemOpen
  \bibfield  {author} {\bibinfo {author} {\bibfnamefont {A.}~\bibnamefont
  {Montanaro}},\ }\bibfield  {title} {\enquote {\bibinfo {title} {Quantum
  speedup of monte carlo methods},}\ }\href {\doibase 10.1098/rspa.2015.0301}
  {\bibfield  {journal} {\bibinfo  {journal} {Proceedings of the Royal Society
  of London A: Mathematical, Physical and Engineering Sciences}\ }\textbf
  {\bibinfo {volume} {471}} (\bibinfo {year} {2015}),\
  10.1098/rspa.2015.0301}\BibitemShut {NoStop}%
\bibitem [{\citenamefont {Brassard}\ \emph {et~al.}(2002)\citenamefont
  {Brassard}, \citenamefont {Hoyer}, \citenamefont {Mosca},\ and\ \citenamefont
  {Tapp}}]{brassard2002quantum}%
  \BibitemOpen
  \bibfield  {author} {\bibinfo {author} {\bibfnamefont {G.}~\bibnamefont
  {Brassard}}, \bibinfo {author} {\bibfnamefont {P.}~\bibnamefont {Hoyer}},
  \bibinfo {author} {\bibfnamefont {M.}~\bibnamefont {Mosca}}, \ and\ \bibinfo
  {author} {\bibfnamefont {A.}~\bibnamefont {Tapp}},\ }\bibfield  {title}
  {\enquote {\bibinfo {title} {{Quantum Amplitude Amplification and
  Estimation}},}\ }\href {\doibase 10.1090/conm/305/05215} {\bibfield
  {journal} {\bibinfo  {journal} {Contemporary Mathematics}\ }\textbf {\bibinfo
  {volume} {305}} (\bibinfo {year} {2002}),\
  10.1090/conm/305/05215}\BibitemShut {NoStop}%
\bibitem [{\citenamefont {Rebentrost}\ \emph {et~al.}(2018)\citenamefont
  {Rebentrost}, \citenamefont {Gupt},\ and\ \citenamefont
  {Bromley}}]{rebentrost2018quantum}%
  \BibitemOpen
  \bibfield  {author} {\bibinfo {author} {\bibfnamefont {P.}~\bibnamefont
  {Rebentrost}}, \bibinfo {author} {\bibfnamefont {B.}~\bibnamefont {Gupt}}, \
  and\ \bibinfo {author} {\bibfnamefont {T.~R.}\ \bibnamefont {Bromley}},\
  }\bibfield  {title} {\enquote {\bibinfo {title} {Quantum computational
  finance: Monte carlo pricing of financial derivatives},}\ }\href {\doibase
  10.1103/PhysRevA.98.022321} {\bibfield  {journal} {\bibinfo  {journal} {Phys.
  Rev. A}\ }\textbf {\bibinfo {volume} {98}},\ \bibinfo {pages} {022321}
  (\bibinfo {year} {2018})}\BibitemShut {NoStop}%
\bibitem [{\citenamefont {Stamatopoulos}\ \emph {et~al.}(2020)\citenamefont
  {Stamatopoulos}, \citenamefont {Egger}, \citenamefont {Sun}, \citenamefont
  {Zoufal}, \citenamefont {Iten}, \citenamefont {Shen},\ and\ \citenamefont
  {Woerner}}]{Stamatopoulos_2020}%
  \BibitemOpen
  \bibfield  {author} {\bibinfo {author} {\bibfnamefont {N.}~\bibnamefont
  {Stamatopoulos}}, \bibinfo {author} {\bibfnamefont {D.~J.}\ \bibnamefont
  {Egger}}, \bibinfo {author} {\bibfnamefont {Y.}~\bibnamefont {Sun}}, \bibinfo
  {author} {\bibfnamefont {C.}~\bibnamefont {Zoufal}}, \bibinfo {author}
  {\bibfnamefont {R.}~\bibnamefont {Iten}}, \bibinfo {author} {\bibfnamefont
  {N.}~\bibnamefont {Shen}}, \ and\ \bibinfo {author} {\bibfnamefont
  {S.}~\bibnamefont {Woerner}},\ }\bibfield  {title} {\enquote {\bibinfo
  {title} {Option pricing using quantum computers},}\ }\href {\doibase
  10.22331/q-2020-07-06-291} {\bibfield  {journal} {\bibinfo  {journal}
  {Quantum}\ }\textbf {\bibinfo {volume} {4}},\ \bibinfo {pages} {291}
  (\bibinfo {year} {2020})}\BibitemShut {NoStop}%
\bibitem [{\citenamefont {Carrera~Vazquez}\ and\ \citenamefont
  {Woerner}(2021)}]{vazquez2020}%
  \BibitemOpen
  \bibfield  {author} {\bibinfo {author} {\bibfnamefont {A.}~\bibnamefont
  {Carrera~Vazquez}}\ and\ \bibinfo {author} {\bibfnamefont {S.}~\bibnamefont
  {Woerner}},\ }\bibfield  {title} {\enquote {\bibinfo {title} {Efficient state
  preparation for quantum amplitude estimation},}\ }\href {\doibase
  10.1103/physrevapplied.15.034027} {\bibfield  {journal} {\bibinfo  {journal}
  {Physical Review Applied}\ }\textbf {\bibinfo {volume} {15}} (\bibinfo {year}
  {2021}),\ 10.1103/physrevapplied.15.034027}\BibitemShut {NoStop}%
\bibitem [{\citenamefont {Egger}\ \emph
  {et~al.}(2020{\natexlab{a}})\citenamefont {Egger}, \citenamefont {Gutierrez},
  \citenamefont {Mestre},\ and\ \citenamefont {Woerner}}]{egger2019credit}%
  \BibitemOpen
  \bibfield  {author} {\bibinfo {author} {\bibfnamefont {D.~J.}\ \bibnamefont
  {Egger}}, \bibinfo {author} {\bibfnamefont {R.~G.}\ \bibnamefont
  {Gutierrez}}, \bibinfo {author} {\bibfnamefont {J.~C.}\ \bibnamefont
  {Mestre}}, \ and\ \bibinfo {author} {\bibfnamefont {S.}~\bibnamefont
  {Woerner}},\ }\bibfield  {title} {\enquote {\bibinfo {title} {Credit risk
  analysis using quantum computers},}\ }\href {\doibase
  10.1109/TC.2020.3038063} {\bibfield  {journal} {\bibinfo  {journal} {IEEE
  Transactions on Computers}\ } (\bibinfo {year} {2020}{\natexlab{a}}),\
  10.1109/TC.2020.3038063}\BibitemShut {NoStop}%
\bibitem [{\citenamefont {Woerner}\ and\ \citenamefont
  {Egger}(2019)}]{Woerner_2019}%
  \BibitemOpen
  \bibfield  {author} {\bibinfo {author} {\bibfnamefont {S.}~\bibnamefont
  {Woerner}}\ and\ \bibinfo {author} {\bibfnamefont {D.~J.}\ \bibnamefont
  {Egger}},\ }\bibfield  {title} {\enquote {\bibinfo {title} {Quantum risk
  analysis},}\ }\href {\doibase 10.1038/s41534-019-0130-6} {\bibfield
  {journal} {\bibinfo  {journal} {npj Quantum Information}\ }\textbf {\bibinfo
  {volume} {5}} (\bibinfo {year} {2019}),\
  10.1038/s41534-019-0130-6}\BibitemShut {NoStop}%
\bibitem [{\citenamefont {Grover}\ and\ \citenamefont
  {Rudolph}(2002)}]{grover2002creating}%
  \BibitemOpen
  \bibfield  {author} {\bibinfo {author} {\bibfnamefont {L.}~\bibnamefont
  {Grover}}\ and\ \bibinfo {author} {\bibfnamefont {T.}~\bibnamefont
  {Rudolph}},\ }\bibfield  {title} {\enquote {\bibinfo {title} {{Creating
  superpositions that correspond to efficiently integrable probability
  distributions}},}\ }\href@noop {} {\  (\bibinfo {year} {2002})},\ \Eprint
  {http://arxiv.org/abs/quant-ph/0208112} {arXiv:quant-ph/0208112} \BibitemShut
  {NoStop}%
\bibitem [{\citenamefont {Rubinstein}(1981)}]{rubinstein2016simulation}%
  \BibitemOpen
  \bibfield  {author} {\bibinfo {author} {\bibfnamefont {R.~Y.}\ \bibnamefont
  {Rubinstein}},\ }\href {\doibase 10.1002/9780470316511} {\emph {\bibinfo
  {title} {Simulation and the Monte Carlo Method}}},\ Wiley Series in
  Probability and Statistics\ (\bibinfo  {publisher} {Wiley},\ \bibinfo {year}
  {1981})\BibitemShut {NoStop}%
\bibitem [{\citenamefont {Suzuki}\ \emph {et~al.}(2020)\citenamefont {Suzuki},
  \citenamefont {Uno}, \citenamefont {Raymond}, \citenamefont {Tanaka},
  \citenamefont {Onodera},\ and\ \citenamefont
  {Yamamoto}}]{suzuki2020amplitude}%
  \BibitemOpen
  \bibfield  {author} {\bibinfo {author} {\bibfnamefont {Y.}~\bibnamefont
  {Suzuki}}, \bibinfo {author} {\bibfnamefont {S.}~\bibnamefont {Uno}},
  \bibinfo {author} {\bibfnamefont {R.}~\bibnamefont {Raymond}}, \bibinfo
  {author} {\bibfnamefont {T.}~\bibnamefont {Tanaka}}, \bibinfo {author}
  {\bibfnamefont {T.}~\bibnamefont {Onodera}}, \ and\ \bibinfo {author}
  {\bibfnamefont {N.}~\bibnamefont {Yamamoto}},\ }\bibfield  {title} {\enquote
  {\bibinfo {title} {Amplitude estimation without phase estimation},}\ }\href
  {\doibase 10.1007/s11128-019-2565-2} {\bibfield  {journal} {\bibinfo
  {journal} {Quantum Information Processing}\ }\textbf {\bibinfo {volume}
  {19}},\ \bibinfo {pages} {75} (\bibinfo {year} {2020})}\BibitemShut {NoStop}%
\bibitem [{\citenamefont {Aaronson}\ and\ \citenamefont
  {Rall}(2020)}]{aaronson2020quantum}%
  \BibitemOpen
  \bibfield  {author} {\bibinfo {author} {\bibfnamefont {S.}~\bibnamefont
  {Aaronson}}\ and\ \bibinfo {author} {\bibfnamefont {P.}~\bibnamefont
  {Rall}},\ }\bibfield  {title} {\enquote {\bibinfo {title} {Quantum
  approximate counting, simplified},}\ }\href {\doibase
  10.1137/1.9781611976014.5} {\bibfield  {journal} {\bibinfo  {journal}
  {Symposium on Simplicity in Algorithms}\ ,\ \bibinfo {pages} {24–32}}
  (\bibinfo {year} {2020})}\BibitemShut {NoStop}%
\bibitem [{\citenamefont {Grinko}\ \emph {et~al.}(2021)\citenamefont {Grinko},
  \citenamefont {Gacon}, \citenamefont {Zoufal},\ and\ \citenamefont
  {Woerner}}]{grinko2019iterative}%
  \BibitemOpen
  \bibfield  {author} {\bibinfo {author} {\bibfnamefont {D.}~\bibnamefont
  {Grinko}}, \bibinfo {author} {\bibfnamefont {J.}~\bibnamefont {Gacon}},
  \bibinfo {author} {\bibfnamefont {C.}~\bibnamefont {Zoufal}}, \ and\ \bibinfo
  {author} {\bibfnamefont {S.}~\bibnamefont {Woerner}},\ }\bibfield  {title}
  {\enquote {\bibinfo {title} {Iterative quantum amplitude estimation},}\
  }\href {\doibase 10.1038/s41534-021-00379-1} {\bibfield  {journal} {\bibinfo
  {journal} {npj Quantum Information}\ }\textbf {\bibinfo {volume} {7}}
  (\bibinfo {year} {2021}),\ 10.1038/s41534-021-00379-1}\BibitemShut {NoStop}%
\bibitem [{\citenamefont {{Nakaji}}(2020)}]{nakaji2020faster}%
  \BibitemOpen
  \bibfield  {author} {\bibinfo {author} {\bibfnamefont {K.}~\bibnamefont
  {{Nakaji}}},\ }\bibfield  {title} {\enquote {\bibinfo {title} {{Faster
  Amplitude Estimation}},}\ }\href@noop {} {\  (\bibinfo {year} {2020})},\
  \Eprint {http://arxiv.org/abs/2003.02417} {arXiv:2003.02417 [quant-ph]}
  \BibitemShut {NoStop}%
\bibitem [{\citenamefont {Tanaka}\ \emph {et~al.}(2020)\citenamefont {Tanaka},
  \citenamefont {Suzuki}, \citenamefont {Uno}, \citenamefont {Raymond},
  \citenamefont {Onodera},\ and\ \citenamefont
  {Yamamoto}}]{tanaka2020amplitude}%
  \BibitemOpen
  \bibfield  {author} {\bibinfo {author} {\bibfnamefont {T.}~\bibnamefont
  {Tanaka}}, \bibinfo {author} {\bibfnamefont {Y.}~\bibnamefont {Suzuki}},
  \bibinfo {author} {\bibfnamefont {S.}~\bibnamefont {Uno}}, \bibinfo {author}
  {\bibfnamefont {R.}~\bibnamefont {Raymond}}, \bibinfo {author} {\bibfnamefont
  {T.}~\bibnamefont {Onodera}}, \ and\ \bibinfo {author} {\bibfnamefont
  {N.}~\bibnamefont {Yamamoto}},\ }\bibfield  {title} {\enquote {\bibinfo
  {title} {Amplitude estimation via maximum likelihood on noisy quantum
  computer},}\ }\href@noop {} {\  (\bibinfo {year} {2020})},\ \Eprint
  {http://arxiv.org/abs/2006.16223} {arXiv:2006.16223 [quant-ph]} \BibitemShut
  {NoStop}%
\bibitem [{\citenamefont {Giurgica-Tiron}\ \emph {et~al.}(2020)\citenamefont
  {Giurgica-Tiron}, \citenamefont {Kerenidis}, \citenamefont {Labib},
  \citenamefont {Prakash},\ and\ \citenamefont {Zeng}}]{giurgicatiron2020low}%
  \BibitemOpen
  \bibfield  {author} {\bibinfo {author} {\bibfnamefont {T.}~\bibnamefont
  {Giurgica-Tiron}}, \bibinfo {author} {\bibfnamefont {I.}~\bibnamefont
  {Kerenidis}}, \bibinfo {author} {\bibfnamefont {F.}~\bibnamefont {Labib}},
  \bibinfo {author} {\bibfnamefont {A.}~\bibnamefont {Prakash}}, \ and\
  \bibinfo {author} {\bibfnamefont {W.}~\bibnamefont {Zeng}},\ }\bibfield
  {title} {\enquote {\bibinfo {title} {Low depth algorithms for quantum
  amplitude estimation},}\ }\href@noop {} {\  (\bibinfo {year} {2020})},\
  \Eprint {http://arxiv.org/abs/2012.03348} {arXiv:2012.03348 [quant-ph]}
  \BibitemShut {NoStop}%
\bibitem [{\citenamefont {Nielsen}\ and\ \citenamefont
  {Chuang}(2010)}]{mikeandike}%
  \BibitemOpen
  \bibfield  {author} {\bibinfo {author} {\bibfnamefont {M.~A.}\ \bibnamefont
  {Nielsen}}\ and\ \bibinfo {author} {\bibfnamefont {I.~L.}\ \bibnamefont
  {Chuang}},\ }\href {\doibase 10.1017/CBO9780511976667} {\emph {\bibinfo
  {title} {Cambridge University Press}}}\ (\bibinfo {year} {2010})\ p.\
  \bibinfo {pages} {702}\BibitemShut {NoStop}%
\bibitem [{\citenamefont {Bouland}\ \emph {et~al.}(2020)\citenamefont
  {Bouland}, \citenamefont {van Dam}, \citenamefont {Joorati}, \citenamefont
  {Kerenidis},\ and\ \citenamefont {Prakash}}]{bouland2020prospects}%
  \BibitemOpen
  \bibfield  {author} {\bibinfo {author} {\bibfnamefont {A.}~\bibnamefont
  {Bouland}}, \bibinfo {author} {\bibfnamefont {W.}~\bibnamefont {van Dam}},
  \bibinfo {author} {\bibfnamefont {H.}~\bibnamefont {Joorati}}, \bibinfo
  {author} {\bibfnamefont {I.}~\bibnamefont {Kerenidis}}, \ and\ \bibinfo
  {author} {\bibfnamefont {A.}~\bibnamefont {Prakash}},\ }\bibfield  {title}
  {\enquote {\bibinfo {title} {Prospects and challenges of quantum finance},}\
  }\href@noop {} {\  (\bibinfo {year} {2020})},\ \Eprint
  {http://arxiv.org/abs/2011.06492} {arXiv:2011.06492 [q-fin.CP]} \BibitemShut
  {NoStop}%
\bibitem [{\citenamefont {Plesch}\ and\ \citenamefont
  {Brukner}()}]{plesch2011quantum}%
  \BibitemOpen
  \bibfield  {author} {\bibinfo {author} {\bibfnamefont {M.}~\bibnamefont
  {Plesch}}\ and\ \bibinfo {author} {\bibfnamefont {{\v{C}}.}~\bibnamefont
  {Brukner}},\ }\bibfield  {title} {\enquote {\bibinfo {title} {Quantum-state
  preparation with universal gate decompositions},}\ }\href {\doibase
  10.1103/physreva.83.032302} {\bibfield  {journal} {\bibinfo  {journal}
  {Physical Review A}\ }\textbf {\bibinfo {volume} {83}},\
  10.1103/physreva.83.032302}\BibitemShut {NoStop}%
\bibitem [{\citenamefont {Zoufal}\ \emph {et~al.}(2019)\citenamefont {Zoufal},
  \citenamefont {Lucchi},\ and\ \citenamefont {Woerner}}]{zoufal2019quantum}%
  \BibitemOpen
  \bibfield  {author} {\bibinfo {author} {\bibfnamefont {C.}~\bibnamefont
  {Zoufal}}, \bibinfo {author} {\bibfnamefont {A.}~\bibnamefont {Lucchi}}, \
  and\ \bibinfo {author} {\bibfnamefont {S.}~\bibnamefont {Woerner}},\
  }\bibfield  {title} {\enquote {\bibinfo {title} {Quantum generative
  adversarial networks for learning and loading random distributions},}\ }\href
  {\doibase 10.1038/s41534-019-0223-2} {\bibfield  {journal} {\bibinfo
  {journal} {npj Quantum Information}\ }\textbf {\bibinfo {volume} {5}},\
  \bibinfo {pages} {1} (\bibinfo {year} {2019})}\BibitemShut {NoStop}%
\bibitem [{\citenamefont {Peruzzo}\ \emph {et~al.}(2014)\citenamefont
  {Peruzzo}, \citenamefont {McClean}, \citenamefont {Shadbolt}, \citenamefont
  {Yung}, \citenamefont {Zhou}, \citenamefont {Love}, \citenamefont
  {Aspuru-Guzik},\ and\ \citenamefont {O’Brien}}]{peruzzo2014}%
  \BibitemOpen
  \bibfield  {author} {\bibinfo {author} {\bibfnamefont {A.}~\bibnamefont
  {Peruzzo}}, \bibinfo {author} {\bibfnamefont {J.}~\bibnamefont {McClean}},
  \bibinfo {author} {\bibfnamefont {P.}~\bibnamefont {Shadbolt}}, \bibinfo
  {author} {\bibfnamefont {M.-H.}\ \bibnamefont {Yung}}, \bibinfo {author}
  {\bibfnamefont {X.-Q.}\ \bibnamefont {Zhou}}, \bibinfo {author}
  {\bibfnamefont {P.~J.}\ \bibnamefont {Love}}, \bibinfo {author}
  {\bibfnamefont {A.}~\bibnamefont {Aspuru-Guzik}}, \ and\ \bibinfo {author}
  {\bibfnamefont {J.~L.}\ \bibnamefont {O’Brien}},\ }\bibfield  {title}
  {\enquote {\bibinfo {title} {A variational eigenvalue solver on a photonic
  quantum processor},}\ }\href {\doibase 10.1038/ncomms5213} {\bibfield
  {journal} {\bibinfo  {journal} {Nature Communications}\ }\textbf {\bibinfo
  {volume} {5}} (\bibinfo {year} {2014}),\ 10.1038/ncomms5213}\BibitemShut
  {NoStop}%
\bibitem [{\citenamefont {Ollitrault}\ \emph {et~al.}(2020)\citenamefont
  {Ollitrault}, \citenamefont {Mazzola},\ and\ \citenamefont
  {Tavernelli}}]{ollitrault2020}%
  \BibitemOpen
  \bibfield  {author} {\bibinfo {author} {\bibfnamefont {P.~J.}\ \bibnamefont
  {Ollitrault}}, \bibinfo {author} {\bibfnamefont {G.}~\bibnamefont {Mazzola}},
  \ and\ \bibinfo {author} {\bibfnamefont {I.}~\bibnamefont {Tavernelli}},\
  }\bibfield  {title} {\enquote {\bibinfo {title} {Nonadiabatic molecular
  quantum dynamics with quantum computers},}\ }\href {\doibase
  10.1103/physrevlett.125.260511} {\bibfield  {journal} {\bibinfo  {journal}
  {Physical Review Letters}\ }\textbf {\bibinfo {volume} {125}} (\bibinfo
  {year} {2020}),\ 10.1103/physrevlett.125.260511}\BibitemShut {NoStop}%
\bibitem [{\citenamefont {Dawson}\ and\ \citenamefont
  {Nielsen}(2006)}]{dawson2005solovay}%
  \BibitemOpen
  \bibfield  {author} {\bibinfo {author} {\bibfnamefont {C.~M.}\ \bibnamefont
  {Dawson}}\ and\ \bibinfo {author} {\bibfnamefont {M.~A.}\ \bibnamefont
  {Nielsen}},\ }\bibfield  {title} {\enquote {\bibinfo {title} {The
  solovay-kitaev algorithm},}\ }\href@noop {} {\bibfield  {journal} {\bibinfo
  {journal} {Quantum Info. Comput.}\ }\textbf {\bibinfo {volume} {6}},\
  \bibinfo {pages} {81–95} (\bibinfo {year} {2006})},\ \Eprint
  {http://arxiv.org/abs/quant-ph/0505030} {arXiv:quant-ph/0505030} \BibitemShut
  {NoStop}%
\bibitem [{\citenamefont {Selinger}(2015)}]{selinger2012efficient}%
  \BibitemOpen
  \bibfield  {author} {\bibinfo {author} {\bibfnamefont {P.}~\bibnamefont
  {Selinger}},\ }\bibfield  {title} {\enquote {\bibinfo {title} {Efficient
  clifford+t approximation of single-qubit operators},}\ }\href@noop {}
  {\bibfield  {journal} {\bibinfo  {journal} {Quantum Info. Comput.}\ }\textbf
  {\bibinfo {volume} {15}},\ \bibinfo {pages} {159–180} (\bibinfo {year}
  {2015})},\ \Eprint {http://arxiv.org/abs/1212.6253} {arXiv:1212.6253
  [quant-ph]} \BibitemShut {NoStop}%
\bibitem [{\citenamefont {Babbush}\ \emph {et~al.}(2021)\citenamefont
  {Babbush}, \citenamefont {McClean}, \citenamefont {Newman}, \citenamefont
  {Gidney}, \citenamefont {Boixo},\ and\ \citenamefont
  {Neven}}]{babbush2020focus}%
  \BibitemOpen
  \bibfield  {author} {\bibinfo {author} {\bibfnamefont {R.}~\bibnamefont
  {Babbush}}, \bibinfo {author} {\bibfnamefont {J.~R.}\ \bibnamefont
  {McClean}}, \bibinfo {author} {\bibfnamefont {M.}~\bibnamefont {Newman}},
  \bibinfo {author} {\bibfnamefont {C.}~\bibnamefont {Gidney}}, \bibinfo
  {author} {\bibfnamefont {S.}~\bibnamefont {Boixo}}, \ and\ \bibinfo {author}
  {\bibfnamefont {H.}~\bibnamefont {Neven}},\ }\bibfield  {title} {\enquote
  {\bibinfo {title} {Focus beyond quadratic speedups for error-corrected
  quantum advantage},}\ }\href {\doibase 10.1103/prxquantum.2.010103}
  {\bibfield  {journal} {\bibinfo  {journal} {PRX Quantum}\ }\textbf {\bibinfo
  {volume} {2}} (\bibinfo {year} {2021}),\
  10.1103/prxquantum.2.010103}\BibitemShut {NoStop}%
\bibitem [{\citenamefont {Fowler}\ and\ \citenamefont
  {Gidney}(2018)}]{Fowler2018}%
  \BibitemOpen
  \bibfield  {author} {\bibinfo {author} {\bibfnamefont {A.~G.}\ \bibnamefont
  {Fowler}}\ and\ \bibinfo {author} {\bibfnamefont {C.}~\bibnamefont
  {Gidney}},\ }\bibfield  {title} {\enquote {\bibinfo {title} {Low overhead
  quantum computation using lattice surgery},}\ }\href
  {https://arxiv.org/abs/1808.06709} {\bibfield  {journal} {\bibinfo  {journal}
  {arXiv:1808.06709}\ } (\bibinfo {year} {2018})}\BibitemShut {NoStop}%
\bibitem [{\citenamefont {Gidney}\ and\ \citenamefont
  {Ekerå}(2021)}]{gidney2019factor}%
  \BibitemOpen
  \bibfield  {author} {\bibinfo {author} {\bibfnamefont {C.}~\bibnamefont
  {Gidney}}\ and\ \bibinfo {author} {\bibfnamefont {M.}~\bibnamefont
  {Ekerå}},\ }\bibfield  {title} {\enquote {\bibinfo {title} {How to factor
  2048 bit rsa integers in 8 hours using 20 million noisy qubits},}\ }\href
  {\doibase 10.22331/q-2021-04-15-433} {\bibfield  {journal} {\bibinfo
  {journal} {Quantum}\ }\textbf {\bibinfo {volume} {5}},\ \bibinfo {pages}
  {433} (\bibinfo {year} {2021})}\BibitemShut {NoStop}%
\bibitem [{\citenamefont {Egger}\ \emph
  {et~al.}(2020{\natexlab{b}})\citenamefont {Egger}, \citenamefont {Gambella},
  \citenamefont {Marecek}, \citenamefont {McFaddin}, \citenamefont {Mevissen},
  \citenamefont {Raymond}, \citenamefont {Simonetto}, \citenamefont {Woerner},\
  and\ \citenamefont {Yndurain}}]{egger2020quantum}%
  \BibitemOpen
  \bibfield  {author} {\bibinfo {author} {\bibfnamefont {D.~J.}\ \bibnamefont
  {Egger}}, \bibinfo {author} {\bibfnamefont {C.}~\bibnamefont {Gambella}},
  \bibinfo {author} {\bibfnamefont {J.}~\bibnamefont {Marecek}}, \bibinfo
  {author} {\bibfnamefont {S.}~\bibnamefont {McFaddin}}, \bibinfo {author}
  {\bibfnamefont {M.}~\bibnamefont {Mevissen}}, \bibinfo {author}
  {\bibfnamefont {R.}~\bibnamefont {Raymond}}, \bibinfo {author} {\bibfnamefont
  {A.}~\bibnamefont {Simonetto}}, \bibinfo {author} {\bibfnamefont
  {S.}~\bibnamefont {Woerner}}, \ and\ \bibinfo {author} {\bibfnamefont
  {E.}~\bibnamefont {Yndurain}},\ }\bibfield  {title} {\enquote {\bibinfo
  {title} {Quantum computing for finance: State-of-the-art and future
  prospects},}\ }\href {\doibase 10.1109/tqe.2020.3030314} {\bibfield
  {journal} {\bibinfo  {journal} {IEEE Transactions on Quantum Engineering}\
  }\textbf {\bibinfo {volume} {1}},\ \bibinfo {pages} {1–24} (\bibinfo {year}
  {2020}{\natexlab{b}})}\BibitemShut {NoStop}%
\bibitem [{\citenamefont {Herbert}(2021)}]{herbert2021problem}%
  \BibitemOpen
  \bibfield  {author} {\bibinfo {author} {\bibfnamefont {S.}~\bibnamefont
  {Herbert}},\ }\bibfield  {title} {\enquote {\bibinfo {title} {The problem
  with grover-rudolph state preparation for quantum monte-carlo},}\ }\href@noop
  {} {\  (\bibinfo {year} {2021})},\ \Eprint {http://arxiv.org/abs/2101.02240}
  {arXiv:2101.02240 [quant-ph]} \BibitemShut {NoStop}%
\bibitem [{\citenamefont {Kaneko}\ \emph {et~al.}(2020)\citenamefont {Kaneko},
  \citenamefont {Miyamoto}, \citenamefont {Takeda},\ and\ \citenamefont
  {Yoshino}}]{kaneko2020quantum}%
  \BibitemOpen
  \bibfield  {author} {\bibinfo {author} {\bibfnamefont {K.}~\bibnamefont
  {Kaneko}}, \bibinfo {author} {\bibfnamefont {K.}~\bibnamefont {Miyamoto}},
  \bibinfo {author} {\bibfnamefont {N.}~\bibnamefont {Takeda}}, \ and\ \bibinfo
  {author} {\bibfnamefont {K.}~\bibnamefont {Yoshino}},\ }\href@noop {}
  {\enquote {\bibinfo {title} {Quantum pricing with a smile: Implementation of
  local volatility model on quantum computer},}\ } (\bibinfo {year} {2020}),\
  \Eprint {http://arxiv.org/abs/2007.01467} {arXiv:2007.01467 [quant-ph]}
  \BibitemShut {NoStop}%
\bibitem [{\citenamefont {Selinger}(2013)}]{Selinger_2013}%
  \BibitemOpen
  \bibfield  {author} {\bibinfo {author} {\bibfnamefont {P.}~\bibnamefont
  {Selinger}},\ }\bibfield  {title} {\enquote {\bibinfo {title} {Quantum
  circuits of t-depth one},}\ }\href {\doibase 10.1103/physreva.87.042302}
  {\bibfield  {journal} {\bibinfo  {journal} {Physical Review A}\ }\textbf
  {\bibinfo {volume} {87}} (\bibinfo {year} {2013}),\
  10.1103/physreva.87.042302}\BibitemShut {NoStop}%
\bibitem [{\citenamefont {H{\"a}ner}\ \emph {et~al.}(2018)\citenamefont
  {H{\"a}ner}, \citenamefont {Roetteler},\ and\ \citenamefont
  {Svore}}]{hner2018optimizing}%
  \BibitemOpen
  \bibfield  {author} {\bibinfo {author} {\bibfnamefont {T.}~\bibnamefont
  {H{\"a}ner}}, \bibinfo {author} {\bibfnamefont {M.}~\bibnamefont
  {Roetteler}}, \ and\ \bibinfo {author} {\bibfnamefont {K.~M.}\ \bibnamefont
  {Svore}},\ }\href@noop {} {\enquote {\bibinfo {title} {Optimizing quantum
  circuits for arithmetic},}\ } (\bibinfo {year} {2018}),\ \Eprint
  {http://arxiv.org/abs/1805.12445} {arXiv:1805.12445 [quant-ph]} \BibitemShut
  {NoStop}%
\bibitem [{\citenamefont {Draper}\ \emph {et~al.}(2006)\citenamefont {Draper},
  \citenamefont {Kutin}, \citenamefont {Rains},\ and\ \citenamefont
  {Svore}}]{draper2006logarithmic}%
  \BibitemOpen
  \bibfield  {author} {\bibinfo {author} {\bibfnamefont {T.~G.}\ \bibnamefont
  {Draper}}, \bibinfo {author} {\bibfnamefont {S.~A.}\ \bibnamefont {Kutin}},
  \bibinfo {author} {\bibfnamefont {E.~M.}\ \bibnamefont {Rains}}, \ and\
  \bibinfo {author} {\bibfnamefont {K.~M.}\ \bibnamefont {Svore}},\ }\bibfield
  {title} {\enquote {\bibinfo {title} {{A logarithmic-depth quantum
  carry-lookahead adder}},}\ }\href@noop {} {\bibfield  {journal} {\bibinfo
  {journal} {Quantum Information and Computation}\ }\textbf {\bibinfo {volume}
  {6}},\ \bibinfo {pages} {351} (\bibinfo {year} {2006})},\ \Eprint
  {http://arxiv.org/abs/quant-ph/0406142} {arXiv:quant-ph/0406142} \BibitemShut
  {NoStop}%
\bibitem [{\citenamefont {Maslov}\ and\ \citenamefont
  {Saeedi}(2011)}]{maslov2011reversible}%
  \BibitemOpen
  \bibfield  {author} {\bibinfo {author} {\bibfnamefont {D.}~\bibnamefont
  {Maslov}}\ and\ \bibinfo {author} {\bibfnamefont {M.}~\bibnamefont
  {Saeedi}},\ }\bibfield  {title} {\enquote {\bibinfo {title} {Reversible
  circuit optimization via leaving the boolean domain},}\ }\href {\doibase
  10.1109/TCAD.2011.2105555} {\bibfield  {journal} {\bibinfo  {journal} {IEEE
  Transactions on Computer-Aided Design of Integrated Circuits and Systems}\
  }\textbf {\bibinfo {volume} {30}},\ \bibinfo {pages} {806} (\bibinfo {year}
  {2011})}\BibitemShut {NoStop}%
\bibitem [{\citenamefont {Takahashi}\ \emph {et~al.}(2009)\citenamefont
  {Takahashi}, \citenamefont {Tani},\ and\ \citenamefont
  {Kunihiro}}]{takahashi2009quantum}%
  \BibitemOpen
  \bibfield  {author} {\bibinfo {author} {\bibfnamefont {Y.}~\bibnamefont
  {Takahashi}}, \bibinfo {author} {\bibfnamefont {S.}~\bibnamefont {Tani}}, \
  and\ \bibinfo {author} {\bibfnamefont {N.}~\bibnamefont {Kunihiro}},\
  }\href@noop {} {\enquote {\bibinfo {title} {Quantum addition circuits and
  unbounded fan-out},}\ } (\bibinfo {year} {2009}),\ \Eprint
  {http://arxiv.org/abs/0910.2530} {arXiv:0910.2530 [quant-ph]} \BibitemShut
  {NoStop}%
\bibitem [{\citenamefont {Mu\~{n}oz Coreas}\ and\ \citenamefont
  {Thapliyal}(2018)}]{Munoz2018}%
  \BibitemOpen
  \bibfield  {author} {\bibinfo {author} {\bibfnamefont {E.}~\bibnamefont
  {Mu\~{n}oz Coreas}}\ and\ \bibinfo {author} {\bibfnamefont {H.}~\bibnamefont
  {Thapliyal}},\ }\bibfield  {title} {\enquote {\bibinfo {title} {T-count and
  qubit optimized quantum circuit design of the non-restoring square root
  algorithm},}\ }\href {\doibase 10.1145/3264816} {\bibfield  {journal}
  {\bibinfo  {journal} {J. Emerg. Technol. Comput. Syst.}\ }\textbf {\bibinfo
  {volume} {14}} (\bibinfo {year} {2018}),\ 10.1145/3264816}\BibitemShut
  {NoStop}%
\bibitem [{\citenamefont {Ross}\ and\ \citenamefont
  {Selinger}(2016)}]{ross_2016}%
  \BibitemOpen
  \bibfield  {author} {\bibinfo {author} {\bibfnamefont {N.~J.}\ \bibnamefont
  {Ross}}\ and\ \bibinfo {author} {\bibfnamefont {P.}~\bibnamefont
  {Selinger}},\ }\bibfield  {title} {\enquote {\bibinfo {title} {Optimal
  ancilla-free clifford+t approximation of z-rotations},}\ }\href@noop {}
  {\bibfield  {journal} {\bibinfo  {journal} {Quantum Info. Comput.}\ }\textbf
  {\bibinfo {volume} {16}},\ \bibinfo {pages} {901} (\bibinfo {year} {2016})},\
  \Eprint {http://arxiv.org/abs/1403.2975} {arXiv:1403.2975 [quant-ph]}
  \BibitemShut {NoStop}%
\bibitem [{\citenamefont {Bocharov}\ \emph {et~al.}(2015)\citenamefont
  {Bocharov}, \citenamefont {Roetteler},\ and\ \citenamefont
  {Svore}}]{Bocharov_2015}%
  \BibitemOpen
  \bibfield  {author} {\bibinfo {author} {\bibfnamefont {A.}~\bibnamefont
  {Bocharov}}, \bibinfo {author} {\bibfnamefont {M.}~\bibnamefont {Roetteler}},
  \ and\ \bibinfo {author} {\bibfnamefont {K.~M.}\ \bibnamefont {Svore}},\
  }\bibfield  {title} {\enquote {\bibinfo {title} {Efficient synthesis of
  universal repeat-until-success quantum circuits},}\ }\href {\doibase
  10.1103/physrevlett.114.080502} {\bibfield  {journal} {\bibinfo  {journal}
  {Physical Review Letters}\ }\textbf {\bibinfo {volume} {114}} (\bibinfo
  {year} {2015}),\ 10.1103/physrevlett.114.080502}\BibitemShut {NoStop}%
\bibitem [{\citenamefont {Kim}\ and\ \citenamefont
  {Choi}(2018)}]{Kim_Efficient_2018}%
  \BibitemOpen
  \bibfield  {author} {\bibinfo {author} {\bibfnamefont {T.}~\bibnamefont
  {Kim}}\ and\ \bibinfo {author} {\bibfnamefont {B.}~\bibnamefont {Choi}},\
  }\bibfield  {title} {\enquote {\bibinfo {title} {Efficient decomposition
  methods for controlled-r n using a single ancillary qubit},}\ }\href
  {\doibase 10.1038/s41598-018-23764-x} {\bibfield  {journal} {\bibinfo
  {journal} {Scientific Reports}\ }\textbf {\bibinfo {volume} {8}} (\bibinfo
  {year} {2018}),\ 10.1038/s41598-018-23764-x}\BibitemShut {NoStop}%
\bibitem [{\citenamefont {Zalka}(1998)}]{zalka1998}%
  \BibitemOpen
  \bibfield  {author} {\bibinfo {author} {\bibfnamefont {C.}~\bibnamefont
  {Zalka}},\ }\bibfield  {title} {\enquote {\bibinfo {title} {Simulating
  quantum systems on a quantum computer},}\ }\href {\doibase
  10.1098/rspa.1998.0162} {\bibfield  {journal} {\bibinfo  {journal}
  {Proceedings of the Royal Society of London. Series A: Mathematical, Physical
  and Engineering Sciences}\ }\textbf {\bibinfo {volume} {454}},\ \bibinfo
  {pages} {313} (\bibinfo {year} {1998})}\BibitemShut {NoStop}%
\bibitem [{\citenamefont {Wiesner}(1996)}]{wiesner1996}%
  \BibitemOpen
  \bibfield  {author} {\bibinfo {author} {\bibfnamefont {S.}~\bibnamefont
  {Wiesner}},\ }\bibfield  {title} {\enquote {\bibinfo {title} {Simulations of
  many-body quantum systems by a quantum computer},}\ }\href@noop {} {\
  (\bibinfo {year} {1996})},\ \Eprint {http://arxiv.org/abs/quant-ph/9603028}
  {arXiv:quant-ph/9603028 [quant-ph]} \BibitemShut {NoStop}%
\bibitem [{\citenamefont {Barkoutsos}\ \emph {et~al.}(2018)\citenamefont
  {Barkoutsos}, \citenamefont {Gonthier}, \citenamefont {Sokolov},
  \citenamefont {Moll}, \citenamefont {Salis}, \citenamefont {Fuhrer},
  \citenamefont {Ganzhorn}, \citenamefont {Egger}, \citenamefont {Troyer},
  \citenamefont {Mezzacapo}, \citenamefont {Filipp},\ and\ \citenamefont
  {Tavernelli}}]{Barkoutsos2018}%
  \BibitemOpen
  \bibfield  {author} {\bibinfo {author} {\bibfnamefont {P.~K.}\ \bibnamefont
  {Barkoutsos}}, \bibinfo {author} {\bibfnamefont {J.~F.}\ \bibnamefont
  {Gonthier}}, \bibinfo {author} {\bibfnamefont {I.}~\bibnamefont {Sokolov}},
  \bibinfo {author} {\bibfnamefont {N.}~\bibnamefont {Moll}}, \bibinfo {author}
  {\bibfnamefont {G.}~\bibnamefont {Salis}}, \bibinfo {author} {\bibfnamefont
  {A.}~\bibnamefont {Fuhrer}}, \bibinfo {author} {\bibfnamefont
  {M.}~\bibnamefont {Ganzhorn}}, \bibinfo {author} {\bibfnamefont {D.~J.}\
  \bibnamefont {Egger}}, \bibinfo {author} {\bibfnamefont {M.}~\bibnamefont
  {Troyer}}, \bibinfo {author} {\bibfnamefont {A.}~\bibnamefont {Mezzacapo}},
  \bibinfo {author} {\bibfnamefont {S.}~\bibnamefont {Filipp}}, \ and\ \bibinfo
  {author} {\bibfnamefont {I.}~\bibnamefont {Tavernelli}},\ }\bibfield  {title}
  {\enquote {\bibinfo {title} {Quantum algorithms for electronic structure
  calculations: Particle-hole hamiltonian and optimized wave-function
  expansions},}\ }\href {\doibase 10.1103/PhysRevA.98.022322} {\bibfield
  {journal} {\bibinfo  {journal} {Phys. Rev. A}\ }\textbf {\bibinfo {volume}
  {98}},\ \bibinfo {pages} {022322} (\bibinfo {year} {2018})}\BibitemShut
  {NoStop}%
\bibitem [{\citenamefont {Stokes}\ \emph {et~al.}(2020)\citenamefont {Stokes},
  \citenamefont {Izaac}, \citenamefont {Killoran},\ and\ \citenamefont
  {Carleo}}]{stokes2020quantum}%
  \BibitemOpen
  \bibfield  {author} {\bibinfo {author} {\bibfnamefont {J.}~\bibnamefont
  {Stokes}}, \bibinfo {author} {\bibfnamefont {J.}~\bibnamefont {Izaac}},
  \bibinfo {author} {\bibfnamefont {N.}~\bibnamefont {Killoran}}, \ and\
  \bibinfo {author} {\bibfnamefont {G.}~\bibnamefont {Carleo}},\ }\bibfield
  {title} {\enquote {\bibinfo {title} {Quantum natural gradient},}\ }\href
  {\doibase 10.22331/q-2020-05-25-269} {\bibfield  {journal} {\bibinfo
  {journal} {Quantum}\ }\textbf {\bibinfo {volume} {4}},\ \bibinfo {pages}
  {269} (\bibinfo {year} {2020})}\BibitemShut {NoStop}%
\bibitem [{\citenamefont {McArdle}\ \emph {et~al.}(2019)\citenamefont
  {McArdle}, \citenamefont {Jones}, \citenamefont {Endo}, \citenamefont {Li},
  \citenamefont {Benjamin},\ and\ \citenamefont
  {Yuan}}]{mcardle2019variational}%
  \BibitemOpen
  \bibfield  {author} {\bibinfo {author} {\bibfnamefont {S.}~\bibnamefont
  {McArdle}}, \bibinfo {author} {\bibfnamefont {T.}~\bibnamefont {Jones}},
  \bibinfo {author} {\bibfnamefont {S.}~\bibnamefont {Endo}}, \bibinfo {author}
  {\bibfnamefont {Y.}~\bibnamefont {Li}}, \bibinfo {author} {\bibfnamefont
  {S.~C.}\ \bibnamefont {Benjamin}}, \ and\ \bibinfo {author} {\bibfnamefont
  {X.}~\bibnamefont {Yuan}},\ }\bibfield  {title} {\enquote {\bibinfo {title}
  {Variational ansatz-based quantum simulation of imaginary time evolution},}\
  }\href {\doibase 10.1038/s41534-019-0187-2} {\bibfield  {journal} {\bibinfo
  {journal} {npj Quantum Information}\ }\textbf {\bibinfo {volume} {5}}
  (\bibinfo {year} {2019}),\ 10.1038/s41534-019-0187-2}\BibitemShut {NoStop}%
\end{thebibliography}%

\appendix
\section{Background on Derivatives}
\label{app:derivatives}
This section presents some examples of commonly used derivatives in the financial sector. Unlike in most other sections of this paper where all payoffs are assumed to be discounted payoffs, in this section they are by default \emph{not} discounted unless explicitly stated. 

\subsection{Forwards}
\label{app:forwards}

An example of a derivative is a forward contract, often simply called a
forward. Here, the holder promises to buy or sell a certain asset to the
issuer on a specified date in the future at a fixed price $F$ known as the
\emph{forward price}. A simple path-independent example is where the holder
promises to buy $x$ amount of an asset at $F$ dollars per asset $m$ months
from now. Forwards are typically \emph{settled in cash} i.e. instead of the
money and asset exchanging hands on the expiration date, a payoff is
determined based on the value of the asset and there is only an exchange of
money determined by this payoff. For example, if the price at the expiration
date $T$ of the asset is $S^T$, the payoff is given by $f(S^T)=x(S^T-F)$,
where if $f(S^T) > 0$, the contract holder makes a profit (and the issuer
a loss) and the opposite if  $f(S^T) < 0$.

%A typical use case of a forward would be to hedge currency risk e.g. a
%corporation that is headquartered in the US may need to periodically buy some
%resources from Tunisia. In this case, the corporation may want to take out a
%currency forward with a bank, where it promises to buy $x$ dinars to the bank
%at a price of $F$ dollars per dinar on some payment date to remove any
%currency speculation when planning its cash flow. Again, instead of an actual
%exchange in currency, the bank will give $xf(S^T)=x(S^T-F)$ dollars to
%the corporation on the payment date (where $S^T$ is the price of a dinar in
%dollars on the payment date).

\subsection{Options}
\label{app:options}

Another example of a derivative is an option. Options can be viewed as
conditional forwards. With an option contract, the holder has
\emph{the option} to buy or sell a certain asset to the issuer on some
future date at a pre-determined price (unlike the foward where the issuer is
obliged to buy or sell the asset). If the holder chooses to buy or sell the
asset, we say that they have \emph{exercised the option}. Similarly to the
forwards, option contracts are usually settled in cash based on the value of
the asset on the exercise date. An example of a path-independent option with
a single underlying asset is a \emph{European call option}, where the issuer
has the option of buying an asset at a strike price $K$ on expiration date.
The payoff on expiration date can then be written as $f(S^T)=\max(S^T-K, 0)$.
A European \emph{put} option is where the issuer has the option of
\emph{selling} an asset at a strike price $K$ on expiration date, which gives
a payoff of  $f(S^T)=\max(K-S^T, 0)$. Another example of a path-independent
option is a \emph{binary option} which has a fixed payoff if the underlying
asset is above (or below) the strike at time $T$.

\subsection{Path-dependence and Discounted Payoffs}
\label{app:path-dependent}

An example of a path\emph{-dependent} derivative is a \emph{knock-out}
European call option. This is the same as a European call option, but with an
additional knock-out price (or \emph{barrier}) $b$. If at any time from $0$ to $T$ the
underlying asset goes above this value, then the contract is worth nothing.
This path-dependent payoff function has the form
\begin{align}
f(S^0, S^1, ..., S^T) =
\begin{cases}
S^T-K & \mbox{if } S^T > K \mbox{ and  } S^i < b, \; \forall i\in\{0, ..., T\} \\
0 & \mbox{otherwise.}
\end{cases}
\end{align}
The inclusion of the value of the underlying at times other than $T$ is what
introduces path dependence. Another example is a \emph{knock-in} put option
which has payoff
\begin{align}
f(S^0, S^1, ..., S^T) =
\begin{cases}
K-S^T & \mbox{if } S^T < K \mbox{ and  } S^i < b, \; \forall i\in\{0, ..., T\} \\
0 & \mbox{otherwise.}
\end{cases}
\end{align}
Here the contract is knock-in because it only has non-zero payoff if the asset
goes below some value $b$.

In the examples discussed so far, there has only been one \emph{payment} date
where an exchange takes place between the contract issuer and holder, at time
$T$. It is possible (as we will see later) for some path-dependent options to
have several payment dates, i.e. where several payments are made at different
times throughout the course of the contract duration.

We now introduce the notion of a discounted payoff. As expected, the price
today for any derivative is related to its expected payoff in the future.
However we also want to take into account the time delay for the payoff to
account for the opportunity cost of investing in a risk-free asset with
interest rate $r$. If a contract has a payoff $f_i$ at time $t_i$ from today,
we define the discounted payoff as $e^{-rt_i}f_i$.

The price of a derivatives contract is given by the expected value of the
discounted payoff under the stochastic process for the underlying assets.
In practice, path-dependent derivatives are much more difficult to price
computationally and are often priced using Monte Carlo simulations of the
paths. This is in contrast to some models for path-independent derivatives
that can even have analytic solutions, such as the Black-Scholes model for
European call options~\cite{BlackScholes}. Path-dependent options present an
opportunity to use quantum speedups for Monte Carlo to gain advantage.
In this work, we will consider two specific examples of path-dependent
derivatives: autocallables and target accrual redemption forwards (TARFs).

\subsection{Auto-callable Options}
\label{app:autocallable}

A typical example of an auto-callable (`automatically callable') option is a
set of binary options, each of which pays different binary payouts at different
payment dates and then knocks out the whole product (i.e. voids all future
payoffs) if it makes a non-zero payout at any of the payment dates. Autocallables are typically
contingent on the returns of the underlying asset (as opposed to directly on the price). More formally,
let $(K_i, t_i, f_i)$ be a binary option that has payoff $f_i$ defined as
\begin{align}
f_i =
\begin{cases}
p_i & \mbox{if } \tilde{R}^{t_i}_c > K_i \\
0 & \mbox{otherwise.}
\end{cases}
\end{align}
where $p_i$ is a fixed dollar value and $\tilde{R}^{t_i}_c$ is the cumulative return of the underlying asset at time $t_i$ defined as the product of the returns $\tilde{R}^{t_j}$ 
for $j=\{1, 2,..., i\}$. We have used the notation $\tilde{R}$ to represent the return to differentiate from 
$R$ which we have used previously to represent the log return. An autocallable is then a set
\begin{align}
\{(K_1, t_1, f_1), (K_2, t_2, f_2), ..., (K_3, t_m, f_m)\},
\end{align}
where $\{t_i\}$ and $\{p_i\}$ typically increase linearly. If any of the
binary options $(K_i, t_i, f_i)$ pays out a non-zero dollar amount (i.e. is \emph{in the money}), then all
subsequent options $\{(K_j, t_j, f_j)\}_{j>i}$ are knocked out i.e. voided.

In practice, these binary options are often bundled with a \emph{short}
knock-in put option i.e. a knock-in put option given to the \emph{issuer} by
the holder, which mitigates risk for the issuer and decreases the price for
the holder. This put option is also typically contingent on the return space of the underlying asset.
More formally, the payoff (to the holder) from the put option is defined as
\begin{align}
f_{put} =
\begin{cases}
k(\tilde{R}^T_c-K_{put}) & \mbox{if } \tilde{R}^T_c < K_{put} \mbox{ and  } \tilde{R}^i_c < b, \; \forall i\in\{0, ..., T\} \\
0 & \mbox{otherwise.}
\end{cases}
\end{align}
where $K_{put}$ and $b$ are the dimensionless put strike and barrier parameters respectively
and $k$ is a constant notional value.
We note that in the case where $\tilde{R}^T_c < K_{put}$, the payoff is negative, 
implying that the contract \emph{holder} has to pay the contract \emph{issuer}.
As with the set of binary options, this put option is also knocked
out if any of the binary options $(K_i, t_i, f_i)$ is in the money.
An example of the full payoff structure for an auto-callable option with a
single underlying and 3 payment dates is given
in Algorithm~\ref{alg:autocall_example}.

\begin{algorithm} [H]
\caption{Auto-callable example} \label{alg:autocall_example}
\begin{algorithmic}[1]
\REQUIRE The following displays the payoff scheme of a 3-year auto-callable with yearly payment dates, binary option strike return of 1.1, a
knock-in barrier of 0.7, a put option strike return of 1 and a notional value of \$18. The timestep values $i$ represents the elapsed time in years and the non-zero payoffs $p_i$ at each year from the binary option are $\$2i$.
Finally the cumulative return of the underlying asset at timestep $i$ is denoted by $\tilde{R}^i_c$.
\begin{enumerate}
\item For $i$ in $[1,2,3]$:
	\begin{itemize}
	\item if $\tilde{R}^i_c \geq 1.1$: the contract holder receives $\$2i$ from the issuer and the contract is immediately ended.
	\end{itemize}
\item After 3 years, if the contract was not previously ended and at any point in time in the last 3 years the cumulative return of the asset was less than 0.7:
	\begin{itemize}
	\item if $\tilde{R}_c^3 \leq 1$: the contract holder pays the issuer $\$18(1-\tilde{R}_c^3)$
	\end{itemize}
\end{enumerate}
\end{algorithmic}
\end{algorithm}

It is possible (and simpler) to describe the autocallable option with a single underlying to be contingent directly on the price of the underlying asset
as opposed to the return.
However defining it as such does not allow us to to trivially generalize it to the case of multiple
underlying assets. This is because it is typical to tie the overall option payoff to the best or worst performing asset, where performance is defined in
terms of returns. In principle the different underlying assets could have
independent put strike returns but this in not common.

The contingent payoffs and the knock-in put mean that autocallables have a
payoff that is strongly path dependent. This means that they are
computationally expensive to price in practice, sometimes taking five to ten
seconds using classical Monte Carlo methods with at least forty thousand paths.

\subsection{Target Accrual Redemption Forwards}
\label{app:tarf}

A target accrual redemption note (TARN) is any derivative whose payoff is
capped at a specified target amount.\footnote{The term historically referred
only to notes (hence the name) but has now come to include any derivative with
an accrual cap} For the purposes of this paper, we will focus on a commonly
used TARN called a target accrual redemption forward (TARF). A TARF is a
set of forwards with a couple of knock-out conditions. Specifically, it is a
derivative with a single underlying with several (typically 20-60) payment
dates and a forward price $F$. Throughout the contract, we have two fixed
strike prices $K_{\text{upper}} \geq F$ and $K_{\text{lower}} < F$. At each
payment date $t$, the payoff is defined as:
\begin{align}
f_t =
\begin{cases}
S^{t} - F & \mbox{if } S^{t}>K_{\text{upper}} \\
0 & \mbox{if } K_{\text{lower}} \leq S^{t} \leq K_{\text{upper}} \\
\alpha(S^{t} - F) & \mbox{if } S^{t}<K_{\text{lower}}
\end{cases}
\end{align}
where $S^{t}$ is the price of the underlying at the payment date $t$ and
$\alpha$ is a positive constant. We note that when $S^{t}<K_{\text{lower}}$,
the payoff is negative and hence the holder of derivative makes a loss. The
constant $\alpha$ makes this loss asymmetric if it happens and is often one
or two.

In addition, a TARF will have two knock-out conditions based on a knock-out
threshold $b$ and accrual cap $C$. The first condition states that if at
any payment date the price of the underlying is greater or equal to $b$,
the derivative contract is immediately knocked out (without payment for that
date). The second condition is if at any payment date $t$ the total gains
of the holder are going to exceed the accrual cap $C$ due to the payoff
$f_t$, the contract holder instead only receives the amount such that
their total gains sum up to $C$ and the contract is then knocked out.
 An example of the full payoff structure for TARF with 52 payment dates is
 given in Algorithm~\ref{alg:TARF_example}.

\begin{algorithm} [H]
\caption{TARF example} \label{alg:TARF_example}
\begin{algorithmic}[1]
\REQUIRE The following displays the payoff scheme of a 1-year TARF with weekly payment dates, a forward price of \$20, strike prices of $K_{\text{upper}}= \$20$ and $K_{\text{lower}}= \$15$, an $\alpha = 2$, a knock-out threshold of \$30 and an accrual cap of \$5. The timestep values $i$ represents the elapsed time weeks..
Finally the price of the underlying asset at timestep $t$ is denoted by $S^t$.
\begin{enumerate}
\item Total := \$0
\item For $t$ in $[1,2,...,52]$:
\begin{itemize}
	\item if $S^t \geq \$30$: the contract is immediately ended.
	\item if \$20 $\leq S^t < \$30$: set $f_t = S^t - \$20$.
	\item if $S^t < \$15$: set $f_t = 2(S^t - \$20)$ (this will be  negative number).
	\item if Total + $f_t \geq \$5$:
	\begin{itemize}
		\item set $f_t =$ \$5 - Total
		\item the contract holder receives $f_t$ from the issuer and the contract is ended.
	\end{itemize}
	\item else:
	\begin{itemize}
		\item the contract holder receives $f_t$ from the issuer (if $f_t$ is negative then the holder effectively gives money to the issuer)
		\item Total $:=$ Total + $f_t$
	\end{itemize}
\end{itemize}
\end{enumerate}
\end{algorithmic}
\end{algorithm}

\section{Insufficiency of Grover-Rudolph Loading}
\label{app:grover-rudolph}
The Grover-Rudolph algorithm~\cite{grover2002creating} is often cited as a
method to efficiently create quantum superpositions that correspond to
classical distributions. For a given probability distribution $\{p_i\}$ of a
random variable $x$, the algorithm creates a quantum superposition of the form
\begin{equation}
\ket{\psi(\{p_i\})} = \sum_i \sqrt{p_i}\ket{i}.
\end{equation}
The algorithm is inductive in nature and starts by assuming that there is a
way to divide the probability distributions into some number $2^m$ of regions
 in the domain of interest and create the state
\begin{equation}
\label{eqn:GR_state}
\ket{\psi_m} = \sum_{i=0}^{2^m-1}\sqrt{p_i^{(m)}}\ket{i},
\end{equation}
where $p_i^{(m)}$ is the probability for the random variable to lie in region $i$.
Then it aims to add one qubit to the state of Eq.~\eqref{eqn:GR_state}, to
further subdivide the $2^m$ regions into a $2^{m+1}$ discretization of the
probability distribution with an evolution of the form
\begin{equation}
\sqrt{p_i^{(m)}}\ket{i} \rightarrow \sqrt{\alpha_i}\ket{i}\ket{0} +
\sqrt{\beta_i}\ket{i}\ket{1},
\end{equation}
where $\alpha_i$ ($\beta_i$) is the probability for the random variable to
lie in the left (right) half of region $i$.
Letting $x_L^i$ and $x_R^i$ denote the left and right boundaries of region
$i$, the function
\begin{equation}
\label{eqn:f_i_GR}
f(i)=\frac{\int_{x_L^i}^{\frac{x_R^i-x_L^i}{2}} p(x) dx}{\int_{x_L^i}^{x_R}p(x) dx}
\end{equation}
is the probability that, given $x$ lies in region $i$, it also lies in the
left half of the region.
If we can construct a circuit which performs the computation
\begin{equation}
\sqrt{p_i^{(m)}}\ket{i}\ket{0 \cdots 0} \rightarrow \sqrt{p_i^{(m)}}\ket{i}\ket{\theta_i},
\end{equation}
with $\theta_i=\arccos\sqrt{f(i)}$, then a controlled rotation of angle
$\theta_i$ on the $m+1$th qubit yields
\begin{equation}
\sqrt{p_i^{(m)}}\ket{i}\ket{\theta_i}\ket{0} \rightarrow
\sqrt{p_i^{(m)}}\ket{i}\ket{\theta_i}(\cos\theta_i\ket{0} + \sin\theta_i\ket{1}).
\end{equation}
After uncomputing $\ket{\theta_i}$, we are left with
\begin{equation}
\ket{\psi_{m+1}} = \sum_{i=0}^{2^{m+1}-1}\sqrt{p_i^{(m+1)}}\ket{i},
\end{equation}
which is the extension of the state in Eq.~\eqref{eqn:GR_state} to one extra qubit.
Performing this iteration $n=\log_2 N$ times, we will have a discretization
of the distribution over $N$ total number of points across $n$ qubits.

In practice, the efficiency of the Grover-Rudolph method relies on the
ability to perform the integrals in Eq.~\eqref{eqn:f_i_GR} in superposition.
The argument in the original formulation in \cite{grover2002creating} is
that probability distributions that can be integrated efficiently classically
using probabilistic methods (e.g using Monte Carlo) can be equivalently
efficiently integrated quantumly.
However, since the ultimate goal in quantum derivative pricing is to provide
a faster alternative to Monte Carlo integration over a probability distribution,
performing this integral as part of our initial state preparation without any
corresponding quantum speedup, nullifies the advantage offered by amplitude
estimation as an alternative to Monte Carlo. While efficient from a complexity
point of view, this means that Grover-Rudolph is insufficient as a method for
quantum advantage in derivative pricing.
\footnote{During revisions of the manuscript, a new pre-print \cite{herbert2021problem} rigorously demonstrated the insufficiency of the Grover-Rudolph method for quantum-accelerated Monte Carlo. }

More recently, an approximate method to implement the Grover-Rudolph algorithm
for standard normal probability distributions was presented in
\cite{kaneko2020quantum}, where the authors suggest the expression in
Eq.~\eqref{eqn:f_i_GR}, written as
\begin{equation}
g(x, \delta)=\frac{\int_{x}^{x+\delta/2} p(x) dx}{\int_{x}^{x+\delta}p(x) dx},
\end{equation}
can be approximated as
\begin{equation}
\label{eqn:kaneko_approx}
g(x, \delta) \approx \frac{1}{2} + \frac{1}{8}\delta x + \mathcal{O}(\delta^2),
\end{equation}
for small $\delta$.
As the $\delta$ parameter decreases with each iteration of the Grover-Rudolph
algorithm adding a qubit to the discretization, the authors highlight that
for $m \geq 7$ the approximation in Eq.~\eqref{eqn:kaneko_approx} becomes
sufficiently accurate.
However, because the Grover-Rudolph construction is iterative, the $m < 7$
terms need to be computed before the above approximation becomes possible.
As such, the integrals in Eq.~\eqref{eqn:f_i_GR} are computed classically and
then loaded into the corresponding quantum registers.
While this approximation allows the simplification of the general
Grover-Rudolph algorithm for standard normal distributions after a certain
point in the iteration, it does not change the fact that it requires computing
integrals over the entire domain of the probability distribution, thus making
it practically infeasible for the same reason as the original Grover-Rudolph
method.

\section{Fixed-point Quantum Arithmetic Resources}
\label{app:arithmetic}
This section reviews preliminaries
for common quantum arithmetic operations and the synthesis of arbitrary rotations.
These operations are used in resource estimation and error analysis.
Quantum arithmetic is required for path loading using the Riemann
summation method (Section~\ref{sec:riemann-sum}) and the re-parameterization
method (Section~\ref{sec:re-parameterization}), as well as the payoff
calculation
described in Section~\ref{sec:payoffs}.
For the Riemann sum method, we need to perform all the arithmetic
operations involved in Eq.~\eqref{eqn:multivariate_normal_pdf_T} as well as compute the arcsine and square root of a quantum register for the payoff calculation in Eq.~\eqref{eqn:payoff_amplitude_encoding}.
We identify algorithms for performing individual arithmetic operations
efficiently, where resources are usually reported as a number of Toffoli gates
or T-gates.
In cases where we employ arithmetic algorithms from previous work in the
literature, we report the gate cost in terms of the gate set reported by the authors.

As we are working in the fault-tolerant setting, we estimate the T-depth of the
circuits in a Clifford + T gate set decomposition and assume Toffoli gates can
be constructed with a T-depth of one using ancilla qubits \cite{Selinger_2013}.
For each operation we assume that we can parallelize the resulting circuits wherever possible.

\subsection{Resource Estimation}
\label{app:arithmetic_resource_estimation}
We perform all calculations in fixed-point arithmetic similarly to
\cite{hner2018optimizing}, which allows us to use the quantum algorithms for reversible function evaluation described therein.
An $n$-bit representation of a number $x$ is

\begin{equation}
\label{eqn:fixed_point_repr}
x=\underbrace{x_{n-1}\cdots x_{n-p}}_p.\underbrace{x_{n-p-1} \cdots x_0}_{n-p},
\end{equation}
where $x_i \in {0,1}$ denotes the $i$-th bit of the binary representation of
$x$ and $p$ denotes the number of bits to the left of the binary decimal point.
The choice of $n$ and $p$ controls the error that we allow in each calculation
as well as the resources required to perform arithmetic on the registers.
Once we choose the values of $(n, p)$ so that the overall arithmetic error is acceptable for the problem under consideration, we keep them constant throughout the analysis.
It is possible that we can tailor these values for different components of the circuit and reduce the overall resources required, but for simplicity in this paper we ignore this potential optimization.

Let $\text{TF}_{f}$ and $T_{f}$ denote the number of Toffoli gates and the T-depth required to compute an arithmetic function or logical operation $f$.
The estimates for the operations are functions of the
fixed-point register size $(n,p)$ that will be used to represent the underlying quantum states involved in the computations.

\paragraph*{Addition/Subtraction}
Using the algorithm described in~\cite{draper2006logarithmic}, we can perform addition of two
$n$-qubit registers in place with a Toffoli cost of $10n-3w(n)-3w(n-1)-3\log_2n - 3\log_2(n-1) -7$ where $w(n)$ denotes the number of ones in the binary expansion of $n$,
and a Toffoli depth of $\lfloor \log_2(n)\rfloor + \lfloor\log_2(n-1)\rfloor + \lfloor\log_2\left(\frac{n}{3}\right)\rfloor +\lfloor\log_2\left(\frac{n-1}{3}\right)\rfloor + 8$.
Note that subtraction is given by $a - b = \,\sim(\sim a + b)$ and so can be implemented as an addition with $2n$ extra $X$ gates, which does not change the Toffoli count.

We can turn an addition gate into a
controlled addition gate by using the method shown in Figure 3 in
\cite{maslov2011reversible}. This requires an additional $n$-qubit ancilla
register, along with two sets of $n$ parallel controlled swap gates. Each individual
controlled swap gate is comprised of 3 Toffoli gates in series.

\paragraph*{Multiplication} For multiplication we follow the method
from \cite{hner2018optimizing}, which uses the controlled addition circuit in \cite{takahashi2009quantum} and requires a Toffoli count of

\begin{equation}
\text{TF}_{\text{mul}}(n,p) = \frac{3}{2}n^2 + 3np + \frac{3}{2}n - 3p^2 + 3p.
\end{equation}
This method can also be used for division of a quantum register by a classical value, which we do by inverting the classical value and employing the multiplication algorithm.

The controlled additions in the fixed-point multiplication method from \cite{hner2018optimizing} require ancilla qubits proportional to the register size, but the circuits include uncomputing the ancillas, meaning that they can be reused for each subsequent addition that is not done in parallel.
Because we parallelize the computations across the $d$ assets and $T$ timesteps, we include an additional $T*d*n$ qubits when we count the total to account for these required ancilla qubits.

We can additionally parallelize each multiplication circuit, by considering the register of one factor as $z \geq 1$ independent registers of size $n/z$, and each controlled addition can happen in parallel for the $z$ subregisters.
This requires $n \cdot (z-1)$ extra qubits and $z-1$ additions to accumulate the
$z$ sub-results into the final result.
$z=1$ denotes that no extra parallelization is employed.
If we can parallelize the pairwise accumulation additions as well, we arrive at
a total T-depth cost of parallelized fixed-point multiplication given by
\begin{equation}
\label{eqn:t_depth_parallel_mult}
\text{T}_{\text{mul}}(n, z) = \lceil\frac{n}{z}\rceil \cdot (\text{T}_{\text{add}} + 6) + \lceil\log_2z\rceil \cdot \text{T}_{\text{add}}.
\end{equation}
$(\text{T}_{\text{add}} + 6)$ is the T-depth of a controlled addition discussed in the Addition/Subtraction section.

\paragraph*{Square Root} We employ the square root algorithm described in \cite{Munoz2018}, which we extend for quantum registers in fixed-point representation.
For an $(n, p)$-sized number $x$, we can compute $\sqrt{x}$ by treating $x$ as
an $n$-digit integer, and then shifting the result to the right $(n-p)/2$ times.
This amounts to performing
\begin{equation}
\label{eqn:sqrt_mapping}
\sqrt{x} \mapsto \sqrt{\frac{x * 2^{n-p}}{2^{n-p}}}.
\end{equation}
The Toffoli count of this square root algorithm is \cite{Munoz2018}
\begin{equation}
\text{TF}_{\text{sq}}(n,p) = \frac{n^2}{2} + 3n - 4.
\end{equation}
The T-depth of this algorithm as reported by the authors is given $\text{T}_{\text{sq}}(n)= 5n + 3$ and requires $2n+1$ qubits.

\paragraph*{Logical Operations} For comparisons between quantum registers or between a quantum register and a constant, we use the logarithmic comparator from \cite{draper2006logarithmic} with Toffoli/T-depth of $2\lfloor\log_2(n-1)\rfloor + 5$, which includes uncomputing the intermediate
ancillas.
The logical \emph{OR} operation for a 2-qubit input can be performed with a Toffoli/T-depth of one \cite{Stamatopoulos_2020}.

\paragraph*{Exponential} In \cite{hner2018optimizing}, the authors
introduce a generic quantum algorithm to calculate smooth classical functions
using a parallel piecewise polynomial approximation. We apply this to estimate
the resources of computing exponentials.
The algorithm takes parameters $k$ and $M$, which control the polynomial degree chosen for the piecewise approximations and the number of domain subintervals respectively.
The total number of Toffolis is given by

\begin{equation}
\text{TF}_{\text{exp}}(n, p, k, M) = \frac{3}{2}n^2k + 3npk + \frac{7}{2}nk - 3p^2d + 3pk - d + 2Md(4\lceil\log_2M\rceil - 8) + 4Mn.
\end{equation}

This algorithm, which we also use to compute the arcsine function,
requires $k$ iterations of a multiplication and an addition, where $k$-degree polynomials are used for the approximation.
Additionally, for $M$ chosen subintervals, it requires $M$ comparison circuits between the $n$-qubit input register and a classical value.
Using the comparator from \cite{draper2006logarithmic} with T-depth of
$2\lfloor\log_2(n-1)\rfloor+5$, the T-depth of a parallel polynomial evaluation circuit is
\begin{equation}
\label{eqn:t_depth_pp}
\text{T}_{\text{pp}}(n, z) = k\left(\text{T}_{\text{mul}}(n, z) + \text{T}_{\text{add}}\right) + M(2\lfloor\log_2(n-1)\rfloor+5),
\end{equation}
where $z$ is the optional parallelization factor introduced in the resource estimation above for the multiplication circuit.

The qubit count for the
parallel polynomial evaluation scheme for choices of the polynomial degree $k$
and number of subintervals $M$ is given by \cite{hner2018optimizing}
\begin{equation}
\label{eqn:q_pp}
q_{\text{pp}}(n, k, M) = n(d+1) + \lceil\log_2M\rceil + 1.
\end{equation}

\paragraph*{Arcsine} To calculate the arcsine we employ the algorithm from \cite{hner2018optimizing} just as we do for the exponential.
However, because the derivative
\begin{equation}
\frac{d\arcsin(x)}{x}=\frac{1}{\sqrt{1-x^2}}
\end{equation}
diverges near $\pm 1$, the authors use the transformation
\begin{equation}
\arcsin(x)=\frac{\pi}{2}-2\arcsin\left(\sqrt{\frac{1-x}{2}}\right)
\end{equation}
to handle the interval $x \in [0.5, 1]$.
Since the computation of the arcsine requires a conditional square root
evaluation of the argument and, whenever we need to calculate an arcsine, we have to calculate the square root as well (e.g Eq.~\eqref{eqn:payoff_amplitude_encoding}), we can instead use the transformation
\begin{equation}
\arcsin(\sqrt{x})=\frac{\pi}{2}-\arcsin(\sqrt{1-x}).
\end{equation}
The resource estimation considerations then follow similarly to those in Appendix D.1/D.2 of \cite{hner2018optimizing}.
We need:
\begin{itemize}
\item A comparator to check if $x < 0.25$ ($\sqrt{x} < 0.5$) that indicates
whether we need to apply the above transformation, which would require two
Toffoli gates assuming the value in the quantum register is normalized.
\item A conditional subtraction and conditional copy depending on the comparator value above to either prepare $\sqrt{x}$ or $\sqrt{1-x}$. A conditional copy requires $n$ Toffolis, a conditional subtraction requires $\text{TF}_{\text{add}} + n$ Toffoli gates \cite{hner2018optimizing}.
\item $\text{TF}_{\text{sq}}$ Toffoli gates for the square root computation \cite{Munoz2018}.
\item The Toffoli gates required for the polynomial evaluation from \cite{hner2018optimizing} to compute the arcsine.
\item A conditional copy and conditional subtraction depending again on the comparator result from the first step, to get either $\arcsin(\sqrt{x})$ for $x < 0.25$ or $\pi/2 - \arcsin(\sqrt{1-x})$ otherwise.
\end{itemize}
With the above considerations and the Toffoli count for the polynomial approximation of $\arcsin(x)$ from \cite{hner2018optimizing}, the total Toffoli count for computing $\ket{\arcsin\sqrt{x}}$ is
\begin{equation}
\text{TF}_{\text{arcsq}}(n,p, k, M) = k\left(\frac{3}{2}n^2 + n(3p + \frac{7}{2}) -3(p-1)p -1\right) + \frac{n^2}{2}  + 11n + 2Md(4\lceil\log_2M\rceil - 8) + 4Mn - 2.
\end{equation}

The T-depth for computing $\arcsin(\sqrt{x})$ of a number $x$ represented in a register of size $(n,p)$, calculated similarly to the exponential is
\begin{equation}
\text{T}_{\text{arcsq}}(n, p, z) = \text{T}_{\text{sq}}(n) + \text{T}_{\text{pp}}(n, z) + 8n + 6,
\end{equation}
where $\text{T}_{\text{sq}}(n)= 5n + 3$ is the T-depth for the square root algorithm from \cite{Munoz2018}.

The operation will require $q_{\text{arcsq}}$ qubits, where the qubit requirements for the arcsine will be given by Eq.~\eqref{eqn:q_pp} for a choice of $k$ and $M$, and $2n+1$ for the square root operation

\begin{equation}
\label{eqn:q_arcsq}
	q_{\text{arcsq}}(n, k, M) = q_{\text{pp}}(n, k, M) + 2n + 1
\end{equation}

\paragraph*{$R_y$} We use $R_y(\theta)$ rotations in the variational preparation of Gaussians discussed in Sec.~\ref{sec:var_gaussian} and controlled-$R_y$ rotations to encode the payoff into the
amplitude of an ancilla in Eq.~\eqref{eqn:payoff_ancilla_rotation}
as well as the transition probabilities in the Riemann summation method in Eq.~\eqref{eqn:riemann_payoff_rotation}.
Using the method described in \cite{ross_2016}, an arbitrary single-qubit unitary can be performed within precision $\epsilon$ with a T-depth of approximately $3\log_2(1/\epsilon)$.
\footnote{A possible optimization could be to use the Repeat-Until-Success method described in \cite{Bocharov_2015}. This method shows that an arbitrary single-qubit unitary can be performed within precision $\epsilon$ with a T-depth of approximately $1.15\log_2(1/\epsilon)$ using one ancilla qubit and measurement. However the method includes a probability of failure that complicates the analysis. Thus we leave the inclusion of this method to future work.}

When the angle $\theta$ we wish to rotate is stored in a separate register $\ket{\theta}$,
 we require a series of $R_y(\theta_k)$ rotations, each controlled on the $k$th qubit of $\ket{\theta}$ where
\begin{align}
	\theta_k = \frac{2^k}{2^{n-p}}.
	\label{eq:theta_k}
\end{align}
A single controlled-$R_n$ can be performed with an $R_n$-depth of one, $R_n$-count of 3 and with a single ancilla qubit using the decomposition from \cite{Kim_Efficient_2018}. However each rotation contributes an error $\epsilon$ so if $\ket{\theta}$ is an $n$-qubit register (with $p$ bits to the left of the binary point), the end-to-end operation can be performed to precision $\epsilon$ with T-depth of at most $3n\log_2(n/\epsilon)$.
We can reduce this depth slightly by noticing that the amplitude increase due to any controlled-$R_n$ rotation where $\theta_k < \arcsin(\epsilon)$ is less than $\epsilon$ and hence is unnecessary.
Therefore using that observation and Eq.~\eqref{eq:theta_k}, we compute the total number of rotations required to be $n-\max(\lfloor\log_2(\arcsin(\epsilon)\rfloor+(n-p),0)$.
This gives us a final T-depth for a controlled-$Ry(\theta)$ operation of
\begin{equation}
\text{T}_{Ry}(n,p,\epsilon) = 3\tilde{n}\log_2(\tilde{n}/\epsilon)
\end{equation}
where $\tilde{n} = n-\max(\lfloor\log_2(\arcsin(\epsilon)\rfloor+(n-p),0)$.

\subsection{Error Analysis}
\label{app:arithmetic_error_analysis}
Given the fixed-point representation of Eq.~\eqref{eqn:fixed_point_repr}, each
arithmetic operation involving registers results in some approximation error, depending on the specific method used.
Here we outline the arithmetic error associated with each of the operations described in the previous section.

\paragraph*{Addition/Multiplication} We use the fixed-point addition and multiplication methods described in \cite{hner2018optimizing}, where the addition of two $(n,p)$-sized registers introduces an error bounded by $\epsilon_A=\frac{1}{2^{n-p}}$, and the error associated with multiplication is at most
\begin{equation}
\epsilon_M(n, p)=\frac{n}{2^{n-p}}.
\end{equation}
For $(n, p)$-sized registers $X$ and $Y$, where each register already contains additive errors $\epsilon_X$, $\epsilon_Y$ and each factor $X,Y$ is bounded above by $b$, the error in the computation of $X \cdot Y$ is given by
\begin{equation}
\label{eqn:mult_error}
\epsilon_{\text{mul}}=b*(\epsilon_X + \epsilon_Y) + \epsilon_X\epsilon_Y + \epsilon_M(n, p)
\end{equation}

\paragraph*{Exponential} We employ the parallel polynomial evaluation methods from \cite{hner2018optimizing} to estimate the resources and associated error in computing exponentials.
The error associated with the algorithm depends on choices for the degree of
the polynomial approximation and the number of subintervals chosen, but the
authors provide explicit error estimates and corresponding required resources
in Table II for errors ranging from $10^{-5}$ to $10^{-9}$. We use these
in our overall error estimate.
In our case, we compute the exponential of a register that itself contains
arithmetic error $\xi$.
Denoting the error in computing the exponential of a register
$\epsilon_{\text{exp}}$, the total arithmetic error in computing the exponential of a register can be approximated to first order in $\xi$ in the Taylor expansion of $\text{exp}(-x+\xi)$ as
\begin{equation}
\label{eqn:num_error}
\bar{\epsilon}_{\text{exp}} \lesssim \epsilon_{\text{exp}} + \xi.
\end{equation}

\paragraph*{Square root} As discussed in the previous section, for square root computations we consider the square root algorithm described in \cite{Munoz2018}, extended for quantum registers in fixed-point representation.
The mapping in Eq.~\eqref{eqn:sqrt_mapping} introduces a maximum error of
\begin{equation}
\epsilon_{\text{sq}}=\frac{1}{2^{(n-p)/2}}.
\end{equation}
When computing the square root of a register $x$ which already contains
(positive) additive error $\xi$, the total additive error from the square root
operation is bounded by $\epsilon_{\text{sq}} + \sqrt{\xi}$.
This is easily seen by observing that if we have a square root algorithm which
gives us an estimate $\hat{x}$ with $|\sqrt{x}-\hat{x}| \leq
\epsilon_{\text{sq}}$, then
\begin{eqnarray}
|\sqrt{x+\xi} - \hat{x}| & \leq & |\sqrt{x} - \hat{x}| + \sqrt{\xi} \nonumber \\
& \leq & \epsilon_{\text{sq}} + \sqrt{\xi} \nonumber
\end{eqnarray}
where the first inequality follows from $(\sqrt{x} + \sqrt{\xi})^2 = x + \xi + 2 \sqrt{x\xi} \geq x + \xi$, for positive $x$ and $\xi$.

\paragraph*{Arcsine} For the arcsine calculation we again use the polynomial
evaluation method from \cite{hner2018optimizing},
where the authors give sample resource estimates for error rates ranging from
$10^{-5}$ to $10^{-9}$.
We want to bound the error from the computation of arcsine on a register containing an arithmetic error $\xi$ to begin with.
As discussed in Appendix~\ref{app:arithmetic_resource_estimation} we only need to compute $\arcsin(x)$ for $x \leq 0.5$. 
In addition, whenever we are computing the function $\arcsin(x)$  in our algorithms presented in the paper, we are only doing it for $x \geq 0$. 
This gives us a domain of $0 \leq x \leq 0.5$ for our $\arcsin(x)$ error calculation. Given this domain, we notice that the slope of $\arcsin(x)$ is always monotonically increasing with a maximum at $x = 0.5$.
Therefore computing the error when $x=0.5$ gives us the upper bound:
\begin{equation}
\bar{\epsilon}_{\text{arcsin}} \leq \left|\arcsin(0.5) - \arcsin(0.5-\xi)\right| + \epsilon_{\text{arcsin}},
\end{equation}
where $\epsilon_{\text{arcsin}}$ is the error from the computation of the arcsine from \cite{hner2018optimizing}, given a choice of polynomial degree and number of subintervals.

\paragraph*{Sine} As discussed in the previous section, we compute the $\sin(\theta)$ function with a series of controlled-$Ry$ rotations
controlled on qubits from a register containing the angle $\theta$.
We can bound the error from the computation of $\sin(\theta)$ when the register that is supposed to represent $\theta$ is actually representing $\theta+\xi$ due to an arithmetic error.
To quantify the upper bound, we notice that in the domain of $0\leq \theta \leq \pi/2$, the slope of $\sin(\theta)$ is monotonically decreasing, and therefore has a maximum slope at $\theta = 0$.
Therefore computing the error when $\theta=0$ gives us the upper bound:
\begin{align}
\bar{\epsilon}_{\text{sin}} &\leq \left|\sin(0+\xi) - \sin(0) \right| + \epsilon_{\text{sin}} \notag \\
 &\leq \xi + \epsilon_{\text{sin}}
\end{align}
where we have used the inequality $\sin(a+b) \leq \sin(a) + b$ for $b\geq 0$ and where $\epsilon_{\text{sin}}$ is the error arising from the gate decomposition of the $Ry$ operator discussed in Appendix~\ref{app:arithmetic_resource_estimation}.

\section{Riemann Summation}

\subsection{Riemann Summation Path Loading Resource Estimates}
\label{app:riemann-resource}
In this section, we examine the T-depth and qubit count required to compute
Eq.~\eqref{eqn:multivariate_normal_pdf_T} in a quantum register, and encode that value into the amplitude of an ancilla qubit as described in Algorithm~\ref{alg:riemann-sum}.
The calculation is done in log-return space (see Sec.~\ref{sec:spaces}) and it involves the resource estimates for the operations that we introduced in
Appendix~\ref{app:arithmetic_resource_estimation}.

Let $T_f$ and $q_f$ denote the T-depth and qubit count required for an operation $f$ respectively.
Assuming we can parallelize the computation across the $d$ assets and $T$ timesteps wherever possible, the contributions to the resources for computing $\ket{\arcsin\sqrt{P(\vec{R})}}$ with $P(\vec{R})$ given by Eq.~\eqref{eqn:multivariate_normal_pdf_T} are
\begin{itemize}
\item $\text{T}_{\text{add}}$ for computing the terms $(R - \mu)$ which can be done in parallel for $d$ assets and $T$ timesteps, where $T*d*n$ qubits are used to hold the log-returns $R$ for all assets and timesteps.
\item $\text{T}_{\text{mul}}$ for all $R^2$ terms in the expansion of Eq.~\eqref{eqn:normal_pdf_numerator_expansion} (in parallel for all $d$ and $T$), requiring $T*d*n$ additional qubits.
\item $\text{T}_{\text{mul}} * {d \choose 2}/(d/2)$ for all $R_iR_j$ terms  in the expansion of Eq.~\eqref{eqn:normal_pdf_numerator_expansion} (parallel in $d$, $T$) and $T*{d \choose 2}*n$ qubits.
\item $\text{T}_{\text{add}} * \lceil{\log{\left({d \choose 2} + d \right)}}\rceil$ to sum all the terms in Eq.~\eqref{eqn:normal_pdf_numerator_expansion} in parallel. The qubits from the previous step can be reused here.
\item $\text{T}_{\text{exp}}$ to calculate the exponential in Eq.~\eqref{eqn:multivariate_normal_pdf}, requiring $q_{\text{exp}}$ extra qubits with $q_{\text{exp}}$ given by Eq.~\eqref{eqn:q_pp} for a choice of parameter values determined by the desired approximation accuracy.
\item $\text{T}_{\text{arcsq}}$ to calculate the $\arcsin$ and square root in $\ket{\arcsin\sqrt{P(\vec{R})}}$, with qubit resources given by Eq.~\eqref{eqn:q_arcsq}.
\item $\text{T}_{\text{add}} * (T-1)$ and $(T-1)*d*n$ qubits to calculate all the sums $R_j^{t=1} + R_j^{t=2} + \dotsc + R_j^{t=t'}$ for $t' \in[2,T]$ in Eq.~\eqref{eqn:return-to-price}.
\item $\text{T}_{\text{exp}}$ to calculate the prices across all assets and all timesteps in Eq.~\eqref{eqn:return-to-price} in parallel, using $q_{\text{exp}}*d*T$ more qubits.
\item A T-depth of $3n\log_2(n/\epsilon)$ to perform the ancilla rotation in Eq.~\eqref{eqn:riemann_payoff_rotation} to precision $\epsilon$, controlled on the register where $\ket{\arcsin\sqrt{P(\vec{R})}}$ is computed
This requires $n$ ancilla qubits using the controlled-$R_y$ decomposition from \cite{Kim_Efficient_2018}.
\end{itemize}

Moreover, we will need an additional register of size $T*d*n$ to implement the addition circuit used in \cite{Munoz2018} with constant T-depth and $(z-1)*T*d$ extra qubits if we use the parallel multiplication scheme described in Appendix~\ref{app:arithmetic_resource_estimation} during the calculation of prices across assets and timesteps, where $z \geq 1$ is the optional parallelization factor we choose.
Note that we have not included extra qubit counts to compute the $(R - \mu)$ terms and the sum in Eq.~\eqref{eqn:normal_pdf_numerator_expansion} because we can do these in place using the existing registers we have to hold each $R_i$.
This is possible because after we compute the sums and exponentials in Eq.~\eqref{eqn:return-to-price} (which we can do before computing the sums) we do not need the values of $R_i$ again.

The total T-depth of the Riemann summation path loading process to precision $\epsilon$ for $d$ assets and $T$ timesteps using registers of size $(n,p)$ is then
\begin{equation}
    T_{\text{RS}}(n,p,d,T,\epsilon) = n^2 +2n^2{d \choose 2}/d + 10\left( {d \choose 2} +d  \right) + 10T + 9n + 5 + 3n\log_2(n/\epsilon) + 2T_{\text{exp}}(n, p, \epsilon) + T_{\text{arcsin}}(n, p, \epsilon),
\end{equation}
where the dependency of $T_{\text{exp}}$ and $T_{\text{arcsin}}$ on $\epsilon$ denotes that the polynomial approximation parameters $k$ and $M$ in Eq.~\eqref{eqn:t_depth_pp} for each function will depend on the target accuracy of the process.
The total number of qubits required is
\begin{equation}
    q_{\text{RS}}(n,p,d,T,\epsilon) = Tn \left(4d + {d \choose 2} \right) + 3n+1+ q_{\text{exp}}(n, p, \epsilon)(1+dT) + q_{\text{arcsin}}(n,p,\epsilon).
\end{equation}

\subsection{Importance Sampling for Normalization in Riemann Summation}
\label{app:importance}
Within this section, we introduce a technique closely related to classical
importance sampling to overcome the problem of the exponentially increasing
scaling shown in~\ref{alg:riemann-sum}.
The main idea is to approximate the target distribution by another
distribution that can be loaded efficiently and then use quantum arithmetic
only to adjust for the (multiplicative) error.

% univariate case
Suppose a univariate probability density function
$f: [0, 1] \rightarrow [0, P]$, with $P > 1$ and $\int_{x=0}^1 f(x) dx = 1$
and a payoff function $g: [0, 1] \rightarrow [0, 1]$\footnote{In the
considered context, $g$ will be applied only once. Thus, we can assume it
to take values in $[0, 1]$ without changing the changing the overall
complexity of our approach.}.
As introduced before, we can consider the scaled function $f(x)/P$ and a
corresponding operator $\mathcal{F}$, as well as a corresponding operator
$\mathcal{G}$ to prepare a state on $n+2$ qubits given by
\begin{eqnarray}
\frac{1}{\sqrt{N}} \sum_{i=0}^{N - 1} \ket{i}_n \left( \sqrt{1 - f(x_i)/P}
\ket{0} + \sqrt{f(x_i)/P}  \ket{1} \right)\left(\sqrt{1 - g(x_i)}\ket{0} +
\sqrt{g(x_i)}\ket{1}  \right),
\end{eqnarray}
where we set $x_i = i/N$.
Then, the probability of measuring $\ket{11}$ in the last two qubits is
given by
\begin{eqnarray}
\frac{1}{PN} \sum_{i=0}^{N-1} f(x_i) g(x_i),
\end{eqnarray}
and when multiplied with $P$ corresponds to the Riemann sum approximating
$\int_{x=0}^1 f(x)g(x) dx = \mathbb{E}[g(X)]$ for $X \sim f$.

Further, let us consider a probability distribution $h(x_i) \in [0, 1]$ that
can be efficiently loaded into a quantum state, i.e., where we know how to
efficiently construct a quantum operator $\mathcal{H}$ such that
\begin{eqnarray}
\mathcal{H} \ket{0}_n &=& \sum_{i=0}^{N-1} \sqrt{h(x_i)} \ket{i}_n.
\end{eqnarray}
Suppose now that we have $h$ such that $f(x) / (h(x)  N) \in [0, 1]$ for all
$x$, then we can construct a new operator $\mathcal{F}_h$ defined as
\begin{eqnarray}
\mathcal{F}_h : \ket{i}_n\ket{0} \mapsto \ket{i}_n\left(\sqrt{1-f(x_i)/
(h(x_i) N)}\ket{0} + \sqrt{f(x_i)/ (h(x_i) N)}\ket{1}\right).
\end{eqnarray}
Combining $\mathcal{H}$ and $\mathcal{F}_h$ leads to
\begin{eqnarray}
\mathcal{F}_h \mathcal{H} \ket{0}_n \ket{0} &=&  \sum_{i=0}^{N-1}
\sqrt{h(x_i)} \ket{i}_n \left(\ldots + \sqrt{f(x_i)/ (h(x_i) N)}
\ket{1}\right) \left(\ldots + \sqrt{g(x_i)}\ket{1}  \right),
\end{eqnarray}
which implies a probability of measuring $\ket{11}$ in the last two qubits
given by
\begin{eqnarray}
\frac{1}{N} \sum_{i=0}^{N-1} f(x_i)g(x_i),
\end{eqnarray}
i.e., the Riemann sum approximating
$\int_{x=0}^1 f(x)g(x) dx = \mathbb{E}[g(X)]$ for $X \sim f$.
Thus, if we can find such a probability distribution $h$, we can construct a
state that directly corresponds to $\mathbb{E}[g(X)]$ without the need to
rescale by multiplying $P$.
It can be easily seen that for $P \leq 1$ we can set $h(x) = 1/N$ to recover
the original approach without importance sampling.

In case of multivariate probability density functions, we distinguish three
cases.
First, separable functions that we can write as a product of univariate
functions $f_t$ for $t=0, \ldots, T$.
In this case, the univariate approach can be applied directly and we need to
find a corresponding $h_t$ for each $f_t$.
Second, non-separable multivariate probability density functions
$f: [0, 1]^d \rightarrow [0, P]$, with $P > 1$ and $\int_{x \in [0, 1]^d} f(x) dx = 1$.
Suppose we discretize each dimension using $n$ qubits, i.e., we have in
total $N^d$ grid points.
Then, we need to find a probability distribution $h$ such that
$f(x) / (h(x)  N^d) \in [0, 1]$ for all $x$, and the analysis is analog to
the univariate case.
Last, we consider the case of a multivariate probability density function
coming from a stochastic process and given by
\begin{eqnarray}
f(x_0, \ldots x_T) &=& f_0(x_0) \prod_{t=1}^T f_t(x_t \mid x_{t-1}),
\end{eqnarray}
where $x_t \in [0, 1]^d$ and $f_0(x_0), f_t(x_t \mid x_{t-1}) \in [0, P]$
for $t = 0, \ldots, T$.
Suppose a separable probability distribution
\begin{eqnarray}
h(x_0, \ldots, x_T) &=& \prod_{t=0}^T h_t(x_t),
\end{eqnarray}
that can be loaded efficiently as well as a corresponding decomposition
$h_t(x_t) = h_t^{t}(x_t)h_t^{t+1}(x_t)$, with $h_T^{T+1}(x) = 1$.
Then, we can write
\begin{eqnarray}
\frac{f(x_0, \ldots, x_T)}{N^{(T+1)d}} &=&
\frac{f_0(x_0)}{h_0^0(x_0) N^d} h_0^0(x_0)h_0^1(x_0)
\prod_{t=1}^T
\frac{
f_t(x_t \mid x_{t-1})
}{
h_{t-1}^t(x_{t-1}) h_t^t(x_t) N^d
}
h_t^{t}(x_t) h_t^{t+1}(x_t).
\end{eqnarray}
Thus, if we find $h$ such that the individual $h_t$ can be efficiently loaded and
\begin{eqnarray}
\frac{f_0(x_0)}{h_0^0(x_0) N^d} \in [0, 1], && \forall x_0 \\
\frac{
f_t(x_t \mid x_{t-1})
}{
h_{t-1}^t(x_{t-1}) h_t^t(x_t) N^d
} \in [0, 1],
&&\forall x_{t-1}, x_t,  \quad t=1, \ldots, T,
\end{eqnarray}
then, we can efficiently load the stochastic processes without the
exponential scaling overhead $P^{T+1}$.
Again, it can be easily seen that for
$P \leq 1$ we can set $h_t(x_t) = h_t^t(x_t) = 1/N^{d}$ and
$h_t^{t+1}(x_t) = 1$ to recover the original approach without importance sampling.

Note that even though we may not always find an $h$ that satisfies
all requirements, this approach can still help to lower the overhead
coming from scaling.

\section{Re-parameterization Path Loading Resource Estimates}
\label{app:reparam-resources}
\begin{figure}[ht]
\centering
\includegraphics[width=0.5\textwidth]{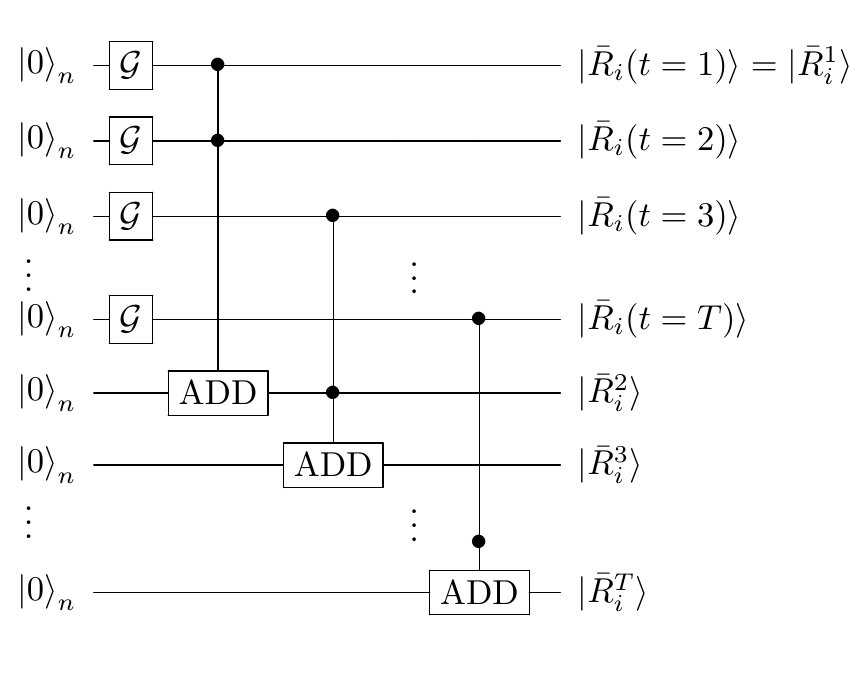}
\caption{Circuit that computes $T$ registers containing the cumulative log-returns $\bar{R}_i^{t'}$ from Eq.~\eqref{eqn:reparam_prices} for one asset at each timestep $t' \in [1, T]$.
We first apply $T$ $\mathcal{G}$ operators (see Sec.~\ref{sec:var_gaussian}) to generate states corresponding to standard Gaussian probability distributions for each timestep, and then serially apply ADD operators which perform $\ket{x}\ket{y}\ket{0} \rightarrow \ket{x}\ket{y}\ket{x+y}$. The ADD operator is discussed in more detail in Appendix~\ref{app:arithmetic}.
The circuit has $\mathcal{G}$-depth of 1 and $T_{\text{add}}$-depth of $T-1$ and can be applied in parallel for each asset in the derivative pricing calculation.}
\label{fig:reparam_sum_circuit_1}
\end{figure}

\begin{figure*}[htbp!]
\centering
\begin{tabular}{c}
\includegraphics[width=1\textwidth]{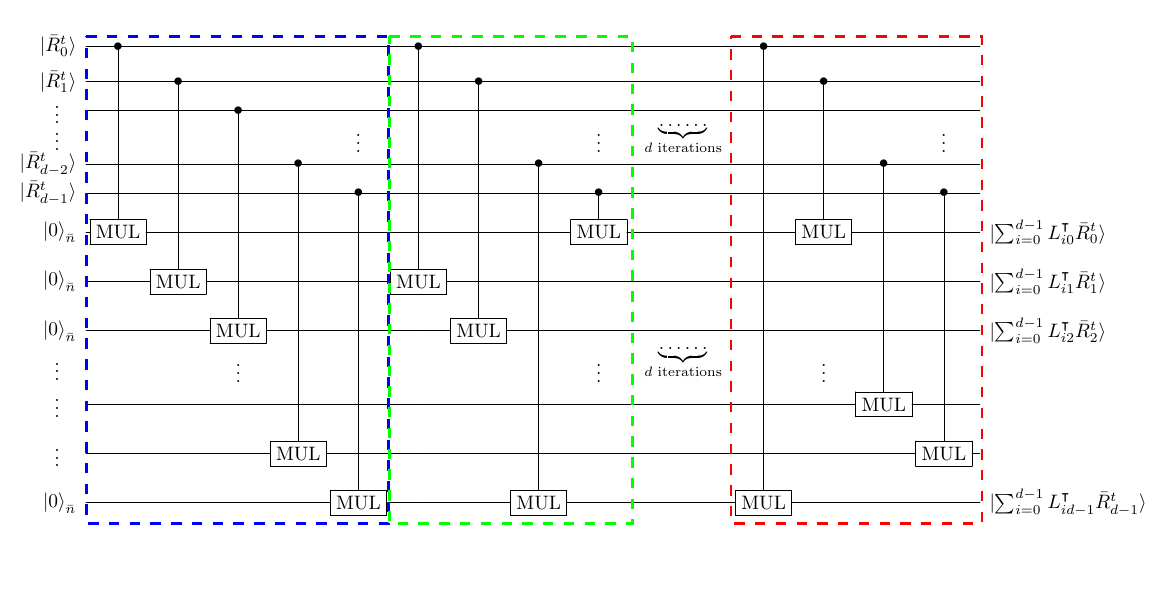} \\
\includegraphics[width=0.4\textwidth]{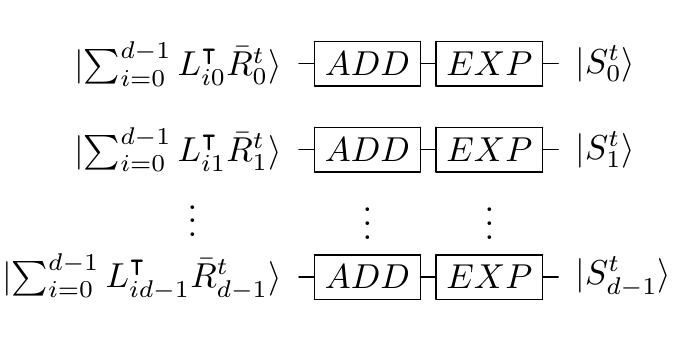}
\end{tabular}
\caption{Circuits that compute asset prices $\ket{S_i^t}$ in separate quantum registers for $d$ assets at timestep $t$ using Eq.~\eqref{eqn:reparam_prices}.
The top figure shows the circuit which takes the cumulative log-returns for each asset created by the circuit in Fig.~\ref{fig:reparam_sum_circuit_1} as input states and computes $\ket{\sum_{i=0}^{d-1}L^{\intercal}_{ji}\bar{R}_i^{t}}$ for each asset $j \in [0, d-1]$. There are $d$ layers of multiplications where each MUL operator performs $\ket{x}\ket{y}\ket{z} \rightarrow \ket{x}\ket{y}\ket{z+xy}$.
Note that the MUL operators in each box can be performed in parallel and therefore the entire circuit has depth of $T_{\text{mul}} * d$.
Then, each computed register $\ket{\sum_{i=0}^{d-1}L^{\intercal}_{ji}\bar{R}_i^{t}}$ requires one extra addition and exponentiation to compute the asset price $\ket{S_i^t}$ (bottom figure), which can also be applied in parallel for each asset.
Both circuits can be applied in parallel across all timesteps $T$ of the calculation.
For more details on the ADD/MUL/EXP operators, see Appendix~\ref{app:arithmetic}.
}
\label{fig:reparam_sum_circuit_2}
\end{figure*}

To prepare the standard normal distributions that we require in the re-parameterization loading approach, we can employ the variational method described in Sec.~\ref{sec:var_gaussian} and the corresponding gate/qubit cost depending on the desired accuracy of the approximation.
In addition to that, we will also have to incur the cost of computing the affine transformation $\vec{R}^t = \vec{\mu}^t +
L^{\intercal}\vec{\bar{R}}^t$ as described in Algorithm~\ref{alg:reparam}.
Note that the affine transformation is required when we need to calculate the asset prices from the log-returns, which for asset $j$ at time $t'$ will be

\begin{equation}
\label{eqn:reparam_prices}
S_j^{t'}=S_j^{t=0}e^{\mu_jt' + \sum_{i=0}^d L^{\intercal}_{ji}\bar{R}_i^{t'}} = e^{\ln S_j^{t=0} + \mu_jt' + \sum_{i=0}^{d-1}L^{\intercal}_{ji}\bar{R}_i^{t'}},
\end{equation}
where $\bar{R}_i^{t'}$ is the $i$th component of the sum $\left(\vec{\bar{R}}(t=1) + \vec{\bar{R}}(t=2) + \cdots + \vec{\bar{R}}(t=t')\right)$.
The graphical representation of the circuit that performs this calculation is shown in Fig.~\ref{fig:reparam_sum_circuit_1}
One complication in Eq.~\eqref{eqn:reparam_prices} is that we cannot compute each asset price fully in parallel across the $d$ assets, because the log-returns of any correlated assets will contribute to the computation of each other's price.
In the case where all assets are pairwise correlated, we will need to compute the contributions to each asset's price from the log-returns of all $d$ assets at that timestep, requiring in total $d^2$ additions to compute all asset prices per timestep.
We can however perform $d$ additions in parallel where the contribution of asset $j$'s return to the price of asset $(j+i) \%d$ is computed for a choice of $i \in [0, d-1]$, since all $d$ such operations have distinct source and target registers.
Then $d$ rounds of additions will compute the term $\sum_{i=0}^{d-1}L^{\intercal}_{ji}\bar{R}_i^{t'}$ for all assets, and if we compute $\left(\vec{\bar{R}}(t=1) + \vec{\bar{R}}(t=2) + \cdots + \vec{\bar{R}}(t=t')\right)$ in a separate register for each $t'$ and each asset, the above calculation can be also parallelized across all timesteps.
This procedure is illustrated in Fig.~\ref{fig:reparam_sum_circuit_2}.

The arithmetic error in computing Eq.~\eqref{eqn:reparam_prices} can be minimized by increasing the qubit register sizes to accommodate the largest values possible for the sums over the timesteps $T$ and assets $d$.
If each gaussian prepared in Eq.~\eqref{eqn:gaussian_preparation} is discretized using $n$ qubits, then $n+\lceil\log_2T\rceil$ qubits will be enough to hold the largest value of the sum represented by $\bar{R}_i^{t'}$.
An additional $\lceil\log_2d\rceil$ qubits will achieve the same for $\sum_{i=0}^{d-1}L^{\intercal}_{ji}\bar{R}_i^{t'}$, assuming the coefficients $|L^{\intercal}_{ji}| \leq 1$ for all $i,j$.
This condition is not hard to satisfy for typical situations of practical interest, which we can argue by looking at the elements of the covariance matrix $\Sigma_{ij}=\Delta t\rho_{ij}\sigma_i\sigma_j$ (where by definition $|\rho_{ij}| \leq 1$).
Typically, annualized volatilities are smaller than $100\%$ (i.e. $\sigma_i < 1$) and the timestep usually satisfies $\Delta t < 1$, meaning the price of the underlying assets needs to be sampled more frequently than just yearly.
If neither condition is satisfied however, we can choose a smaller $\Delta t$ to ensure $|\Sigma_{ij}|<1$, at the cost of increasing the number of timesteps in the calculation.

The contributions to the T-depth and qubit count for loading the paths and computing the asset prices in the re-parameterization approach for a derivative defined on $d$ assets $T$ timesteps are

\begin{itemize}
\item $T_{R_y}(n) \cdot (L+1)$ T-depth for loading the gaussian states in Eq.~\eqref{eqn:gaussian_preparation} using the variational method from Sec.~\ref{sec:var_gaussian}, where each Gaussian is prepared in parallel and the variational ansatz has depth $L$. This step requires $T*d*n$ qubits where $n$ qubits are used to prepare each Gaussian state.
\item $\text{T}_{\text{add}} * (T-1)$ for calculating all the component-wise sums $\left(\vec{\bar{R}}(t=1) + \vec{\bar{R}}(t=2) + \cdots + \vec{\bar{R}}(t=t')\right)$ for $t' \in[2,T]$ in Eq.~\eqref{eqn:reparam_prices}, requiring an extra $T*d*(n+\lceil\log_2T\rceil)$ qubits (see Fig.~\ref{fig:reparam_sum_circuit_1}).
\item $\text{T}_{\text{mul}} * d$ to compute all contributions to $\sum_{i=0}^{d-1}L^{\intercal}_{ji}\bar{R}_i^{t'}$ in Eq.~\eqref{eqn:reparam_prices} and $T*d*\lceil\log_2d\rceil$ more qubits.
\item $\text{T}_{\text{add}}$ to compute the $\mu_jt' + \ln S_j^{t=0}$ contribution in Eq.~\eqref{eqn:reparam_prices} across assets and timesteps.
\item $\text{T}_{\text{exp}}$ to compute the exponential in Eq.~\eqref{eqn:reparam_prices} across assets and timesteps, and $q_{\text{exp}}*d*T$ additional qubits with $q_{\text{exp}}$ given by Eq.~\eqref{eqn:q_pp}.
\end{itemize}
All in all, the total T-depth for path loading using the re-parameterization method to precision $\epsilon$ for $d$ assets and $T$ timesteps is

\begin{equation}
    T_{RP}(n, d, T, L, \epsilon) = 3n\log_2(n/\epsilon)(L+1) + 10T + d\bar{n}^2 + T_{\text{exp}}(\bar{n}, \epsilon),
\end{equation}
with qubit count
\begin{equation}
    q_{RP}(n, d, T) = \left(n + \bar{n} + q_{\text{exp}}(\bar{n}, \epsilon)\right)dT,
\end{equation}
where $\bar{n} = n + \lceil\log_2T\rceil + \lceil\log_2d\rceil$.

\section{Method for Gaussian Loader Training}
\label{app:variational}
In this section we illustrate an approximate method to initialize the quantum register using
 the Variational Quantum Eigensolver (VQE) approach~\cite{peruzzo2014}.
This algorithm features a parametrized circuit which in turn produces a parametrized state $|\psi (\{\theta\}) \rangle$ that approximately represents the target state $|\phi_0 \rangle$  and updates its parameters $\{\theta\}$ to optimize the expectation value of a suitable cost function. Here we show that the choice of the cost function to optimize is crucial for the success of the training.

{\bf Energy based training} As a first method, we adopt a physics-based approach and define an operator $H$,  such that its expectation value $E$ assumes its lowest possible value, $E_0$, when evaluated on the target state,
\begin{equation}
  E_0= \langle \psi_0| H |\psi_0 \rangle .
\end{equation}
In physics applications, the operator $H$ is usually called the Hamiltonian,  $E$ the energy, and $|\phi_0 \rangle$ the ground state.
It is well known the Gaussian function
\begin{equation}
\label{eq:harmgs}
    \phi_0(x) = \left( \frac{m}{\pi} \right)^{1/4} e^{-m ( x -x_0)^2}
\end{equation}
is the ground state of the quantum harmonic oscillator Hamiltonian
\begin{equation}
\label{eq:H}
    H = { P^2 \over 2 m} + { m ~ (X-x_0)^2 \over 2},
\end{equation}
where $X$ is the position operator in real space, and $P= -i { d \over dx}$  is the momentum operator \cite{zalka1998,wiesner1996}.  $m$ is a parameter that determines the \emph{variance} of the desired Gaussian distribution, and $x_0$ is the center of gaussian distribution.
In this case, as we seek to find a state $\phi_0(x)$ such that $\phi_0^2(x)=\mathcal{N}(x_0,\sigma)$, we have to set $m = 1/ (2 \sigma^2)$.
We notice that it is always possible to find a generating Hamiltonian function such that its ground state is the square root of the smooth distribution function that we aim to load.

To translate these considerations into an operational workflow we just have to define a way to compute the expectation value of Eq.~\eqref{eq:H} using a quantum computer.
To this end we observe that the operator $X^2$ is diagonal in the computational basis, so it can be measured directly from the bit-string histogram counts  $N_{\text{counts}}(j)$ generated by the repeated wavefunction collapses.
The operator $P^2$ is diagonal in the momentum basis. This implies the addition of a centered Quantum Fourier Transform (QFT) circuit after the state preparation block.
We use the centered Fourier transform to allow for negative momenta \cite{ollitrault2020}.
As introduced in the main text, we work in discrete position space $x_i = -w + i~\Delta x$, with $i=0,\cdots2^n-1$, and $\Delta x = 2w / 2^n$.
Without loss of generality we choose the domain to be centered at zero.
The energy, $E = E_{X^2} + E_{P^2}$, can be computed in the following way,
\begin{align}
    &  E_{X^2} = \frac{1}{N_{\text{shots}}}\sum_{j=0}^{\mathcal{N}}\frac{m}{2} N_{\text{counts}}(j) (j\times\Delta x - x_0)^2 \\
    & E_{P^2} = \frac{1}{N_{\text{shots}}}\sum_{j=0}^{\mathcal{N}}\frac{1}{2m} N_{\text{counts}}(j) (j\times\Delta p )^2
\end{align}
where $N_{\text{shots}}$ is the total number circuit repetitions for the spacial and momentum basis. $N_{\text{counts}}(j)$ (with $0 \le N_{\text{counts}}(j) \le N_{\text{shots}} $, $\sum_j N_{\text{counts}}(j) = N_{\text{shots}}$) is the number of measurements that collapsed onto the qubit basis state corresponding to the binary representation of integer $j$.
This strategy bypasses the need to obtain a Pauli representation of Eq.~\eqref{eq:H}, which would include an exponentially increasing number of Pauli strings to be measured with the qubit register size.

The first step of our program is to  verify numerically the possibility to prepare a state that systematically converges to Eq.~\eqref{eq:harmgs}, using a quantum circuit.
Adopting a variational approach will circumvent the need of costly quantum arithmetic operations  at the expense of introducing  sources of error which are always present in  numerical variational approaches.
The most trivial one concerns the possibility of getting trapped in local minima during the (classical) optimization procedure.
The second, and more profound one, is linked with the representational power of trial states produced by the (shallow) quantum circuits.

Our main choice for the ansatz is the so-called $R_y$-CNOT circuit \cite{Barkoutsos2018}.
 The initial state, defined on an $n$-qubit register which we set to $\ket{0}^{\otimes n}$,
 is evolved under the action of a unitary $U(\vec{\theta})$ to give the trial wave function $\ket{\psi(\vec{\theta})}$.

The circuit is made of a series of $L$ blocks built from single-qubit rotations $U_R(\vec{\theta}^k)$, followed by an entangler $U_{\text{ENT}}$, that spans the required length of the qubit register.
In our tests, we used the simplest choice of a ladder of CNOT gates with linear connectivity, such that qubit $q_i$ is target of qubit $q_{i-1}$ and controls qubit $q_{i+1}$, with $i=1,\cdots,n-2$.
One additional layer of $U_R$ gates is applied at the end, such that the number of variational parameters is $n \times (L+1)$.

Since the single-qubit rotations are all local operations, $U_R(\vec{\theta}^k)$ can be written as a tensor product of rotations of a single qubit:
\begin{equation}
U_R (\vec{\theta}^k) = \bigotimes_{i=0}^{n-1} R_Y({\vartheta}_{q_i}^k),
\end{equation}
where $R_y({\vartheta}^k_{q_i})$ is a rotation on the Y-axis on the Bloch sphere of qubit $q_i$, and $k = 1,\cdots,L+1$.
The full unitary circuit operation is described by
\begin{equation}
U(\vec{\theta})=  U_R(\vec{\theta}^{ L+1})~ \overbrace{
U_{\rm ENT} U_R(\vec{\theta}^{ L}) \ldots U_{\rm ENT} U_R(\vec{\theta}^{ 1})}^{\rm{L-times}},
\label{eq:trial_ansatz_u_theta}
\end{equation}
and the parametrized state is
\begin{equation}
\ket{\psi(\vec{\theta})} = U(\vec{\theta}) \ket{0}^{\otimes n}.
\label{eq:trial_ansatz}
\end{equation}
%The quantum circuit corresponding to this unitary is depicted in Fig.~\ref{fig:Trial_circuit}.
Note that the unitary $U(\vec{\theta})$ describes the full circuit, but not the pre-measurement change of basis required to collapse the wavefunction in momentum space as explained above.

%We numerically show that this circuit, which provides a real-valued trial state, is able to produce with high fidelity the desired quantum state (cfn. Fig~\ref{fig:ovsd}).

For each value of parameters $n$ and $L$, we repeat the optimization runs eight times in order to gather sufficient statistics, as it may happen that the optimizations remain stuck in suboptimal minima.
Since we use classical emulation of the quantum circuits the only source of error in the optimizations originates from the classical optimizer.
In our runs we first perform a warm up run with the COBYLA optimizer, followed by a longer run using the BFGS optimizer.
To enhance the efficiency of the optimizations, the starting point for the VQE run at depth $L$, uses the optimal parameters found at previous optimization at depth of $L-2$ or $L-1$ when available.
We notice that the part of the algorithm that concerns the classical optimization feedback can be greatly improved, for example using gradient based methods \cite{stokes2020quantum} or imaginary-time inspired update schemes \cite{mcardle2019variational}.

\begin{figure*}
    \centering
    \includegraphics[width = 1 \textwidth]{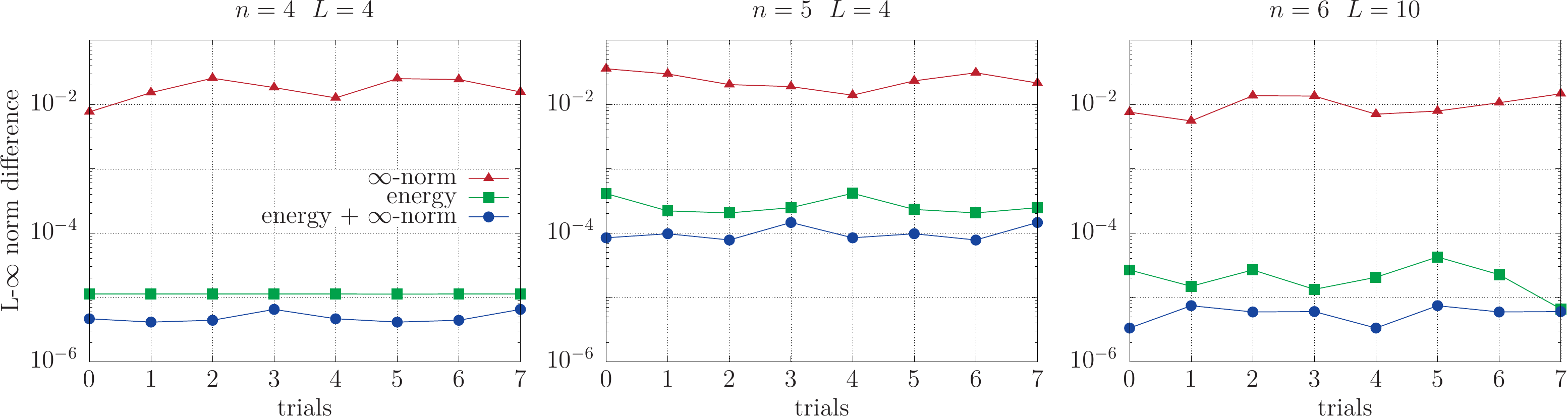}
    \caption{Optimization runs  obtained with the energy-based method (green), the direct  $L_\infty$  optimization (red) and the mixed strategy where the energy based optimization is further refined using the $L_\infty$  optimization (blue).   
     We run eight independent runs for different values of the parameters $n$ (number of qubits) and $L$ (ansatz depth).}
    \label{fig:optifail}
\end{figure*}

{\bf $L_\infty$ training refinements} As discussed in the main text we use pre-optimized circuits obtained using the energy optimization method as a starting guess, and then re-optimize using the $L_\infty$ as the cost function.
In Fig.~\ref{fig:optifail} we show indeed how the direct $L_\infty$ optimization consistently fails to provide accurate results.

\begin{figure*}
    \centering
    \includegraphics[width = 1 \textwidth]{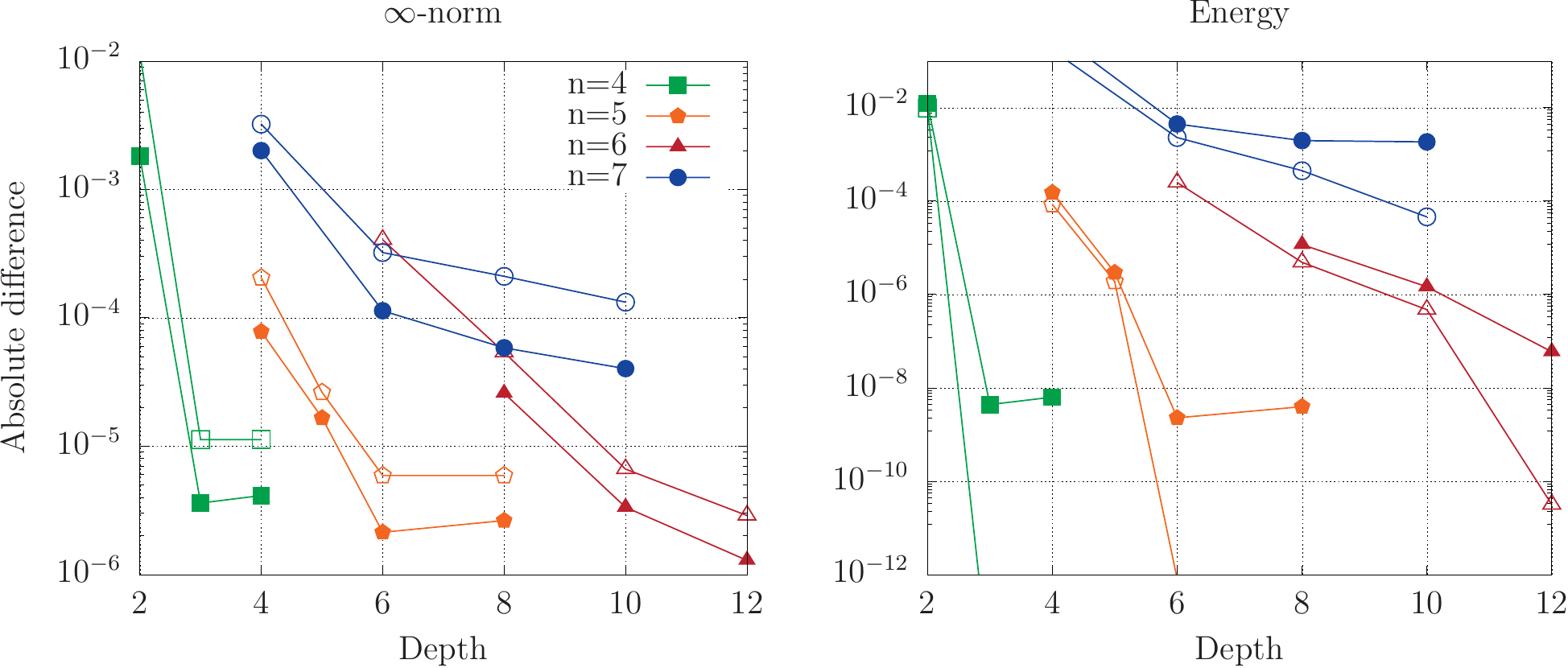}
    \caption{Left figure:  $L_\infty$ norm difference between the prepared and the target distribution as a function of the circuit depth $L$ for different qubit register sizes $n$. We plot here the best of the eight independent optimizations for each parameter. Empty symbols correspond to optimizations performed using the energy of the quantum harmonic oscillator as a cost function, while solid symbols denote the refined optimizations using the  $L_\infty$ as cost function. Right figure: we plot the difference in energy of the associated quantum harmonic oscillator model. As expected the refinement targeting the  $L_\infty$ does not improve this quantity.}
    \label{fig:evsd}
\end{figure*}

We show the complete outcome of the optimizations in Fig.~\ref{fig:evsd}.
This careful numerical study shows that
 the convergence to the exact ground state is exponential in the depth, and therefore the number of gate operations.\\
%This observation is in good agreement with the expected behaviour  from the
 %Solovay-Kitaev theorem\cite{dawson2005solovay}, that provides an upper bound for the number of gates required to achieve a desired accuracy for the energy. Indeed, for any target operation $U \in SU(2^n)$, there is a sequence $S=U_{s_1} U_{s_2} \dots U_{s_D}$ of operators in a dense subset of $SU(2^N)$, such that error in the energy $\epsilon$ decreases exponentially with the depth  $D=\mathcal{O}(log^c(1/\varepsilon))$.
%Although the subset of $SU(2^n)$ operations generated by the entangler blocks in our circuit does not generate a dense subset of $SU(2^N)$ arbitrarily close to the exact unitary $U$ (generator of the exact ground state), we can numerically observe that the exponential decrease of the error with the number of gates still hold.

{\bf Failure of the $L_\infty$ norm direct optimization} We provide an empirical explanation concerning the observed failure of the direct norm optimization technique.
To this end we probe the cost function landscape for both methods, the energy-based and the direct $L_\infty$ optimization,
We start from an optimized parameter configuration $\vec{\theta}_0$ and we perform a \emph{cut} in the parameter space, using
\begin{equation}
\vec{\theta} = \vec{\theta}_0  + \lambda  \vec{\eta}
\end{equation}
where $\vec{\eta}$ is an vector containing uniformly distributed random numbers in the range $[-1,1]$, and $\lambda \in  [-\pi,\pi]$ is a scalar which parametrizes the deviation from the optimal solution.

 \begin{figure*}
    \centering
    \includegraphics[width = 0.6 \textwidth]{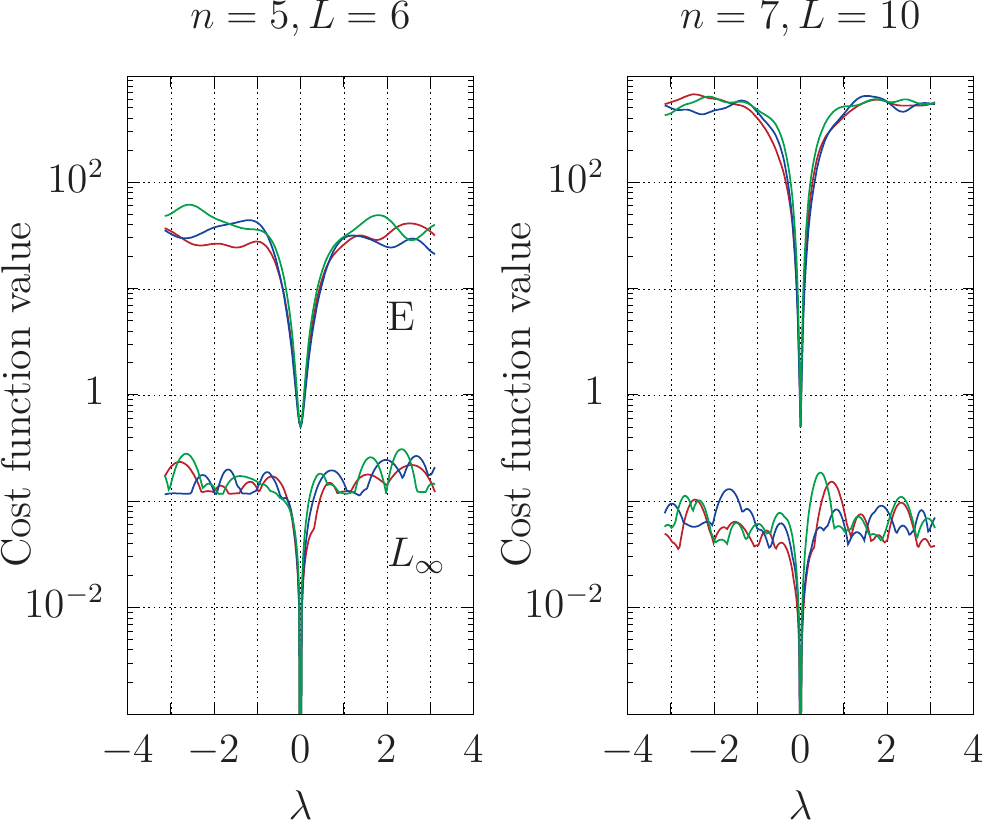}
    \caption{Cost function landscapes for the energy $E$ (above), and the  $L_\infty$ norm (below)  computed using three different cuts (red, blue and green colors) along the parameter space, and for two different setup $n=5,7$ and depths $L=6,10$ respectively.}
    \label{fig:landscape}
\end{figure*}
 In Fig.~\ref{fig:landscape} we probe the cost function landscape for three different cut direction (e.g. three different realizations of the vector $\vec{\eta}$).
We observe indeed that the cost function defined by the  $L_\infty$ norm is much more corrugated than the one defined by the energy $E$ of the associate quantum mechanical toy problem, which instead displays a smoother surface.
 Crucially the basins of attraction of the energy cost-function and the $L_\infty$ cost-function are overlapping (this happens because the ground state of the physical problem is very close to the Gaussian function we want to achieve), therefore the second optimization with the $L_\infty$ norm does not remain stuck in high-cost local minima outside such basin.\\

%Most importantly we observe that the scaling of the depth $L$ to reach a fixed accuracy, i.e. $1- \mathcal{F} = 10^{-4}$, does not seem to increase exponentially with $n$.
%This observation holds in the relevant - large $n$- limit, such that $L < 2^n / n - 1$, namely the depth $L$ is such that the circuits contains a number of variational parameters $n (L+1)$ much smaller than the Hilbert space size $2^n$.
%Indeed we \todo{ spell out a concrete case from the figure once the data is completed}

{\bf Variational parameters digitization.} While our numerical results provide evidence for a rather efficient Gaussian state preparation in terms of circuit depths for a parametrized circuit, an additional step has to be made in view of a fault-tolerant implementation of such circuits.
In this new-framework, the continuous rotation $R_y$ gate needs to be expanded as a finite product of discrete operations.
Following again the Solovay-Kitaev theorem, or more specialized results \cite{selinger2012efficient}, it is possible to also have an efficient representation of any $SU(2)$ operator with a sequence of Clifford + T gates  that scale logarithmically with the threshold error $\epsilon$.
We investigate how the results obtained before can be transferred in this regime where rotation angles can only take discretized values.
We therefore assume that each parameter ${\vartheta}^k_{q_i}$ can only be represented in the format $i * 2\pi/ M_{digit}$, where $i$ is an integer.

We adopt a simple protocol to optimize the parameters on an a grid.
First we project the original continuous parameter values on the grid, choosing the closest grid point for each parameter.
Subsequently, we perform a local search on the grid to find a better combination of the digitized parameters which minimize the $L_\infty$ norm difference compared to the target distribution.
We numerically show that the error introduced by such digitization decreases systematically with the mesh size.
Interestingly, if we consider the error in the $L_\infty$ norm difference introduced by this digitization, it decreases as $O(1 / M_{digit})$.
We observe that in all cases, we are able to obtain values comparable, or better, with the continuous solution, when the mesh size reaches $M_{digit} \sim 10^5$, which is equivalent to discretizing the space using $ 2\pi/ M_{digit} \approx 0.0001$ rad.
%Interestingly if we consider the error in the energy of the trial state introduced by said digitization, it decreases as $c_L / M_{digit}^2$. That is, its scaling is independent by $L$ and $D$, and only its prefactor seems to be $L$ dependent (cfn. Fig.~\ref{fig:digit_en_varform} ).

\begin{figure*}
    \centering
    \includegraphics[width = 0.6 \textwidth]{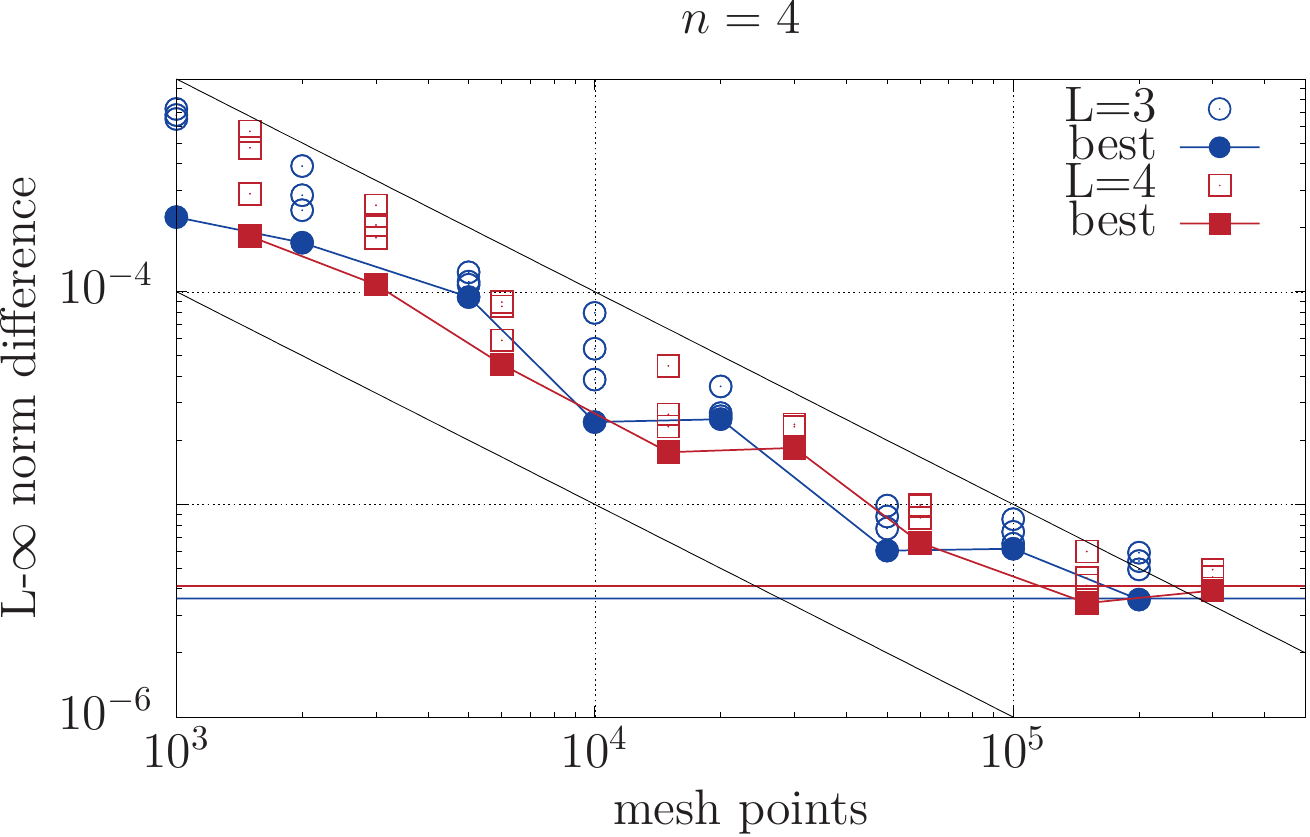}
    \caption{$L_\infty$ norm difference between the prepared and the target distribution as a function of the digitization mesh size $M_{digit}$ for two different circuit depths $L$ (blue and red), for the $n=4$ qubit case.
    For each $M_{digit}$ we \emph{digitize} the eight parameter sets obtained by the previous independent optimizations (which were performed considering a continuous domain for the values of the rotation angles). Empty symbols refer to the full dataset, while the solid symbols highlight only minimum values in the set.
    Colored horizontal lines denote the best values obtained in the previous optimizations with a continuous domain of rotation angles for each $L$ parameter. Interestingly, in some cases the digitization helps in escaping local minima and achieve slightly better solutions. Black diagonal lines are a guide-to-the-eye and represent the functions $1 / M_{digit}$ and  $0.1 / M_{digit}$.}
    \label{fig:m4}
\end{figure*}

\begin{figure*}
    \centering
    \includegraphics[width = 0.6 \textwidth]{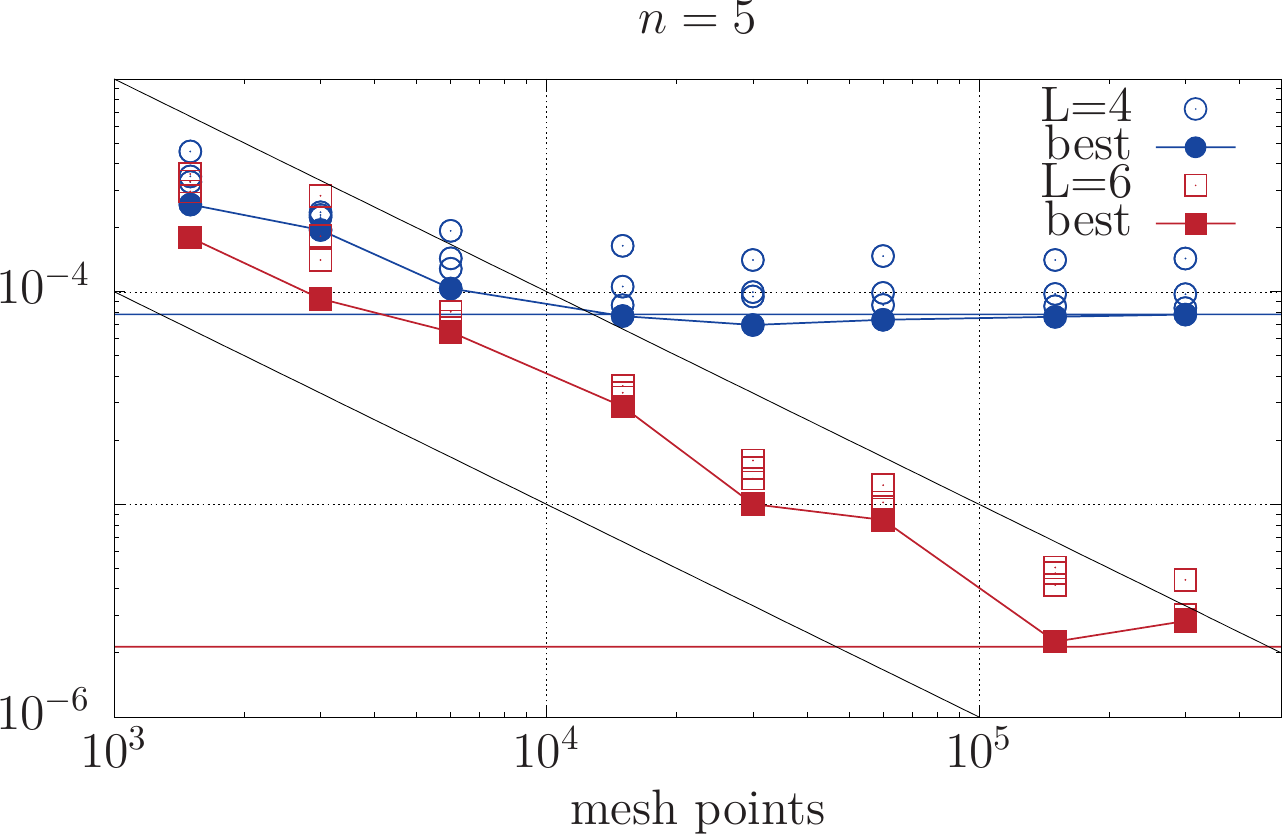}
    \caption{Same as Fig.~\ref{fig:m4} but with $n=5$}
    \label{fig:m5}
\end{figure*}

\begin{figure*}
    \centering
    \includegraphics[width = 0.6 \textwidth]{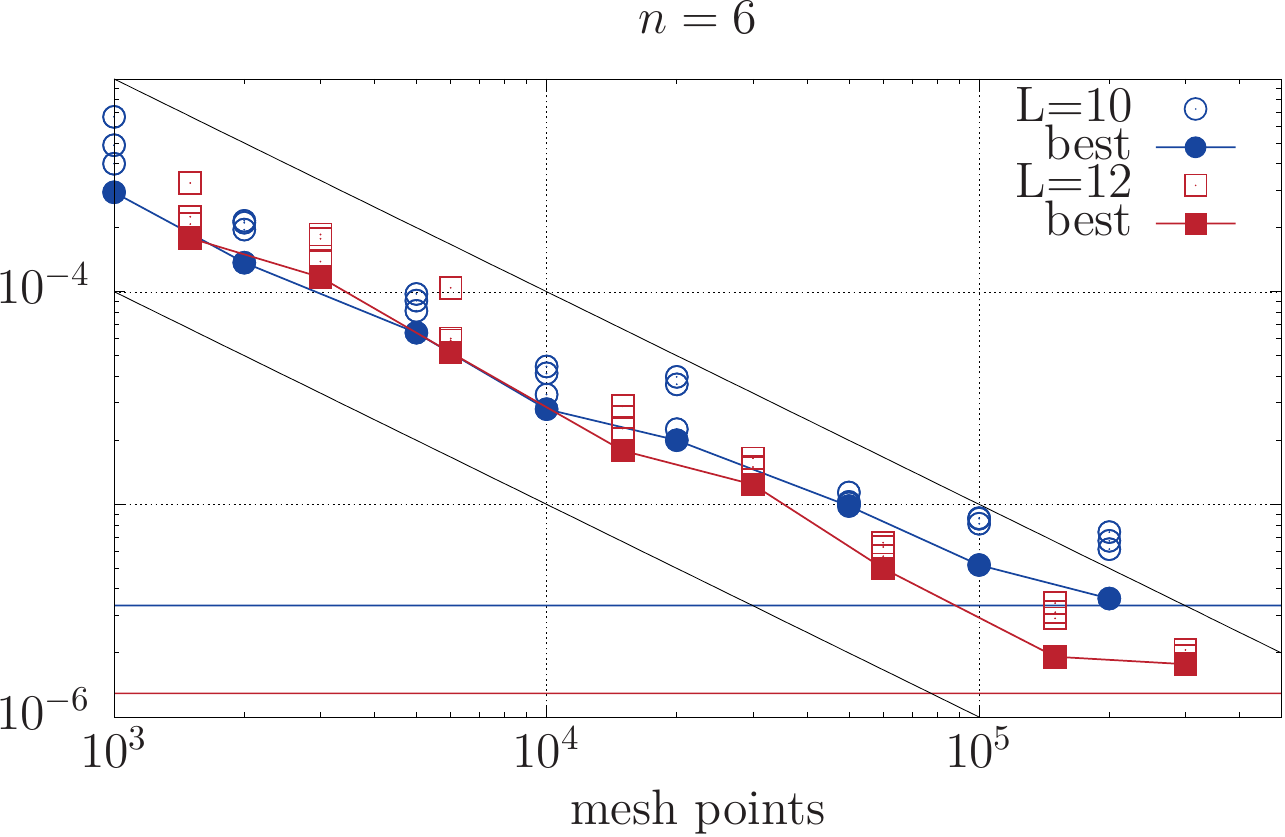}
    \caption{Same as Fig.~\ref{fig:m4} but with $n=6$}
    \label{fig:m6}
\end{figure*}

%\begin{figure*}
  %  \centering
   % \includegraphics[width = 0.7 \textwidth]{graphics/digit_en.pdf}
    %\caption{Energy error resulting from the digitization of the angle values as a function of the mesh size $M_{digit}$. Purple, red, green and blue data series correspond to $L=4,5,6,7$ respectively. The different shapes correspond to different depths $D$, if available we employ three depths values for each $$. Straight lines are guides for the eye and represent $c_L / M_{digit}^2$ functions with different prefactors $c_L$.
    %The unconstrained parameters are taken from the dataset harvested for previous optimization runs at each $L$ and $D$ (between 8 and 16 indipendent optimizations). Error bars are statistical.   }
    %\label{fig:digit_en_varform}
%\end{figure*}

\end{document}